\documentclass[structabstract]{aa}  
\usepackage{graphicx}
\usepackage[authoryear]{natbib}
\usepackage{lscape}
\usepackage{lmodern}
\usepackage{afterpage}
\usepackage{txfonts}
\usepackage{color}
\usepackage{textcomp}

\DeclareFontFamily{U}{euc}{}% I chose euc because the chart is called Euler cursive 
\DeclareFontShape{U}{euc}{m}{n}{<-6>eurm5<6-8>eurm7<8->eurm10}{}% 
\DeclareSymbolFont{AMSc}{U}{euc}{m}{n} % I chose AMSc because AMSa and AMSb are defined in the amsfonts-package 
\DeclareMathSymbol{\umu}{\mathord}{AMSc}{"16} 
 
\newcommand{\hii}{H\,{\scriptsize II}}

\newcommand{\at}{{\sl ATLASGAL}}
\newcommand{\ATLASGAL}{{\sl ATLASGAL}}
\newcommand{\gc}{{\sl Gaussclumps}}
\newcommand{\mum}{$\umu$m}
\newcommand{\msol}{M$_{\odot}$}
\newcommand{\uchii}{{UC-H\,\scriptsize II}}

\newcommand{\nsou}{10565} 
\newcommand{\nsouext}{296}

    \newcommand{\compvar}{7}

    \newcommand{\distlimitmsx}{13} 
    \newcommand{\totalmsxmatch}{1669} 
    \newcommand{\totalmsxmatchpercent}{16} 
    \newcommand{\distlimitmsxeight}{13} 
    \newcommand{\totalmsxmatcheight}{2238} 
    \newcommand{\totalmsxmatchpercenteight}{21} 
    \newcommand{\totalmsxmatchtwoband}{1647} 

    \newcommand{\distlimit}{10} 
    \newcommand{\totalwisematch}{3151} 
    \newcommand{\totalwisematchpercent}{30} 

    \newcommand{\overlap}{1308}
    \newcommand{\totalmirmatch}{3483}
    \newcommand{\totalmirmatchpercent}{33}

\begin{document}
   \title{The ATLASGAL survey: a catalog of dust condensations in the Galactic plane}
   \authorrunning{T. Csengeri et al.}
   \titlerunning{The ATLASGAL survey: a catalog of dust condensations in the Galactic plane}

 \author{T. Csengeri
          \inst{1}
          \and
          J. S. Urquhart
          \inst{1}
          \and
          F. Schuller
          \inst{2}
          \and
          F. Motte
          \inst{3}
          \and
          S. Bontemps
          \inst{4}
          \and
          F. Wyrowski 
          \inst{1}
          \and
          K. M. Menten 
          \inst{1}
          \and
          L. Bronfman
          \inst{5}
          \and
          H. Beuther
          \inst{6}
          \and
          Th. Henning
          \inst{6}
          \and
          L. Testi
          \inst{7,8}
          \and
          A. Zavagno
          \inst{9}
          \and
          M. Walmsley
          \inst{8,10}
      %    \and
          %\inst{2}\fnmsep\thanks{Just to show the usage
          %of the elements in the author field}
          }

   \institute{Max Planck Institute for Radioastronomy,
              Auf dem H\"ugel 69, 53121 Bonn, Germany\\
              \email{ctimea@mpifr-bonn.mpg.de}
         \and
             European Southern Observatory, Alonso de Cordova 3107, Vitacura, Santiago, Chile
         \and
          Laboratoire AIM Paris Saclay, CEA-INSU/CNRS-Universit\'e Paris Diderot, IRFU/SAp CEA-Saclay, 91191 Gif-sur-Yvette, France
          \and
           OASU/LAB-UMR5804, CNRS, Universit\'e Bordeaux 1, 33270 Floirac, France    
          \and
          Departamento de Astronom\'{i}a, Universidad de Chile, Casilla 36-D, Santiago, Chile
           \and
          Max Planck Institute for Astronomy, K\"onigstuhl 17, 69117 Heidelberg, Germany
           \and
         European Southern Observatory, Karl-Schwarzschild-Strasse 2, D-85748 Garching, Germany
          \and
          INAF-Osservatorio Astrofisico di Arcetri, Largo E. Fermi 5, I-50125 Firenze, Italy
           \and
          Aix Marseille Universit\'e, CNRS, LAM (Laboratoire d'Astrophysique de Marseille), UMR 7326, 13388 Marseille, France
           \and
           Dublin Institute for Advanced Studies, Burlington Road 10, Dublin 4, Ireland
             }

   \date{Received 2 August 2013 ; accepted 27 February, 2014}

% \abstract{}{}{}{}{} 
% 5 {} token are mandatory
 
  \abstract
  % context heading (optional)
  % {} leave it empty if necessary  
   {
      The formation processes and the evolutionary stages of high-mass stars are poorly 
      understood compared to low-mass stars. Large-scale surveys are needed to provide 
      an unbiased census of
      high column density sites which can potentially host precursors to high-mass stars.
   }
  % aims heading (mandatory)
   {
      The \at\ survey covers 420 sq.\,degree of the Galactic plane, between $-80^\circ<\ell<+60^\circ$ at 870~{\mum}. 
      Here we identify the population of embedded
       sources throughout the inner Galaxy. With this catalog we first investigate the general statistical properties 
       of dust condensations in terms of their observed 
       parameters, such as flux density and angular size.
      Then using mid-IR surveys we aim to investigate their star-formation activity
      and the Galactic distribution of star-forming and quiescent clumps. 
      Our ultimate goal is to determine the statistical properties of quiescent and star-forming
      clumps within the Galaxy and to constrain the star-formation processes.
   }
  % methods heading (mandatory)
   {
      We optimized the source extraction method, referred to as \textsl{MRE-GCL}, 
        for the \at\ maps in order to generate
       a catalog of compact sources. This technique is based on a multi-scale filtering to
      remove extended emission from clouds to better determine the parameters corresponding to
       the embedded compact
       sources. In a second step we extract the sources by fitting 2D Gaussians with the \gc\ algorithm. 
   }
  % results heading (mandatory)
   {
     We have identified in total $10861$ compact sub-millimeter sources with fluxes above 5$\sigma$. Completeness tests show that this catalogue is 97\% complete above 5$\sigma$ and $>99$\% complete above 7$\sigma$. 
  Correlating this sample of clumps with mid-infrared point source catalogues (MSX at 21.3~{\mum} and WISE  at 22~{\mum}) we have determined a lower limit of {\totalmirmatchpercent}\% that are associated with embedded protostellar objects. We note that the proportion of clumps associated with mid-infrared sources increases with increasing flux density, achieving a rather constant fraction of $\sim$75\% of all clumps with fluxes over 5\,Jy/beam being associated with star-formation. Examining the source counts as a function of Galactic longitude we are able to identify the most prominent star forming regions in the Galaxy.
   }
  % conclusions heading (optional), leave it empty if necessary 
   {
   We present here the compact source catalog of the full \at\ survey and investigate their characteristic properties.
   From the fraction of the likely massive quiescent clumps ($\sim$25\%) we estimate a formation time-scale of $\sim7.5\pm2.5\times10^4$~yr for the   
   deeply
   embedded phase before the emergence of luminous YSOs. 
    Such a short duration for the formation of high-mass stars in massive clumps clearly proves
    that the earliest phases have to be dynamic with supersonic motions.
    }

   \maketitle
%
%________________________________________________________________

\section{Introduction}

\subsection{High-mass star-formation in the Galactic plane}

The dominant formation mechanism leading to the birth of high-mass stars
is still an enigma in modern astrophysics. Unlike for low-mass stars, there is
no clear, observationally constrained evolutionary sequence for individual protostars above 10~{\msol}.
This is due to the fact that they are rare and likely evolve on short time-scales, therefore 
it is challenging to identify and study them. {Yet, they are fundamental
building blocks of galaxies, they provide significant mechanical and radiative feedback to  the interstellar
medium and enrich it with heavy elements.} They are used as a tool to study
star-formation and the evolution of galaxies in various environments as a function of redshift \citep{Kennicutt1998},
it is therefore crucial to study them first locally, in our Galaxy in much greater detail.

Observations of high-mass stars are greatly hindered by the fact that they are still deeply 
embedded in their dust cocoons when reaching the 
main sequence. Since star-formation in general proceeds in the densest regions of molecular
clouds, dust is the best tracer to identify these locations 
(see \citealp{Evans99} for a review). In fact the first systematic
surveys for massive young stellar objects (MYSOs) have used the infrared emission of heated 
dust to pin down the most luminous sites in our Galaxy.

These studies used the IRAS survey \citep{Hughes1989, Wood1989, Bronfman1996, Molinari1996} to look for {\hii} regions and MYSOs \citep{Sridharan2002,Beuther2002} based on their infrared colors. These samples were then extended using data from the more sensitive {\sl Midcourse Space Experiment} (MSX; \citealp{Price2001}) point source catalogue (\citealp{Egan2003}). One notable example is the Red MSX Source (RMS; \citealp{Hoare2005}) Survey, which used a combination of near- and mid-infrared colors to identify a large sample of MYSO candidates (\citealt{Lumsden2002}). Their initial sample was, however, contaminated by asymptotic giant branch (AGB) stars and evolved stars whose mass loss has stopped and dust shells became detached (post-AGB stars and planetary nebulae). These contaminating sources have been identified through a set of multi-wavelength observations (e.g.\,\citealp{Urquhart2008}) and were removed from the final sample \citep{Lumsden2013}. However, since the infrared emission traces heated dust, this sample is strongly biased towards the more evolved MYSO and {\hii} region 
phases of high-mass star-formation.

The search for the earlier, thus colder stages began with the discovery 
of IR-dark clouds (IRDCs), which trace high dust column density
 \citep{Perault1996,Egan1998, Carey1998}. These clouds have been cataloged 
using MSX \citep{Simon2006, Rathborne2006} and the more sensitive
Spitzer Space Telescope \citep{Peretto2009,Butler2009,Peretto2010a, Rygl2010}. 
Their physical properties 
have also been extensively studied since then (e.g.\,\citealp{Pillai2006,Vasyunina2009,Ragan2012}).
Although IRDCs
have been considered for a long time to be the initial stages for high-mass star-formation, 
some of them have been shown to be simply holes in the  
interstellar medium \citep{Wilcock2012}, and 
in fact only a small fraction of them is likely to sustain massive star-formation
 \citep{Peretto2010b,Kauffmann2010}.
 As a consequence, 
 IRDCs also represent a biased and incomplete sample of high-mass star-forming
 sites.

To unambiguously trace high column densities, the optically thin emission from dust in the 
millimeter/sub-millimeter regime is the best tool~\citep{Andre2000}. Sensitive to both cold and warm dust
it is the least biased tracer of all embedded evolutionary stages of star-formation. 
Such studies have been first
performed on smaller areas of the sky in nearby low-mass star-forming regions until a few years ago 
(e.g.\,\citealp{M98, Johnstone2000,Johnstone2001, Motte2001b, M07, Enoch2007}), however 
larger area surveys are required to 
reveal statistically significant samples of high column density sites.
Thanks to the recent development of
large field of view bolometer cameras, surveys 
covering a substantial fraction of the Galactic plane became feasible. 
The ultimate goal of unbiased surveys is to identify (e.g.\,\citealp{M07,schuller2009,Molinari2010,Aguirre}) and then characterize large samples
of high-mass star-forming sites (e.g.\,\citealp{Dunham2011, Dunham2011b, Wienen2012, U2013mmb}).
Statistical studies are required to then constrain the evolutionary stages of high-mass
star-formation and reveal the corresponding time-scales for these stages.
Furthermore, such large-scale surveys allow the properties of
star-forming sites in various environments and conditions to be studied and therefore  
the dominant processes regulating high-mass star-formation can be pinned down.

Here we present results of one of these large-scale surveys, the APEX Telescope Large Area Survey of the Galaxy  
({\at}). 

\subsection{\ATLASGAL\  in the context of Galactic plane surveys}

The \at\ survey\footnote{
{Observing runs: 078.F-9040, 181.C-0885 (ESO); 
079.C-9501, 081.C-9501 (MPIfR); and Chilean data}}\citep{schuller2009}  imaged the Galactic plane between Galactic longitude,
$-60^\circ \le\ell \le +60^\circ$ and Galactic latitude $-1.5^\circ \le b \le +1.5^\circ$  at 870~$\umu$m with the 
LABOCA camera \citep{siringo2009} on the APEX Telescope \citep{gusten2006} in its first campaign.
In a following step, an extension towards Galactic longitude $-80^\circ \le \ell \le -60^\circ$
and Galactic latitude $-2^\circ \le b\le+1.0^\circ$ was added.
Altogether the total area of the survey covers
$\sim420^{\circ 2}$ of the Galactic plane at a 19$\rlap{.}{''}$2 spatial resolution.

The \ATLASGAL\ survey supersedes other ground-based surveys 
providing the most sensitive and 
complete view of the inner Galaxy at sub-millimeter wavelengths. 
So far only space based missions provide better sensitivity. 
In this context the \ATLASGAL\ survey is  outstanding because it provides a view of the
thermal dust emission at a comparable angular-resolution at 870~{\mum} as Herschel at 250~{\mum}, 
and a 2$\times$ better spatial resolution than Herschel at 500~{\mum}, where
the dust emission is optically thin. 
It is therefore well suited to study deeply embedded objects, 
the majority of which, are sites of on-going star-formation.

\at\ is well suited to complement other surveys of dust in the Galactic plane at various
 wavelengths, such as the mid-IR surveys with Spitzer (GLIMPSE at 3.6, 
4.8, 5.6 and 8~$\umu$m  \citealp{Benjamin2003} and MIPSGAL at 24~$\umu$m, \citealp
{Carey2009}) and WISE (3.6, 4.6, 11.8 and 22~$\umu$m,
\citealp{Wright2010}) probing the warm dust.
The Hi-Gal survey (Herschel Infrared Galactic plane survey; \citealp{Molinari2010}), 
uses the PACS and SPIRE instruments onboard Herschel to map the Galactic plane with 
unprecedented sensitivity from the far-IR to the sub-millimeter wavelength regime (at 70, 160, 250, 350 and 500~{\mum}). 
It provides complementary information
of the spectral energy distribution in the regime where the
thermal emission of cold and warm dust peak (between 10-500 K). 
\at\ also has a substantial overlap with the Bolocam Galactic Plane Survey (BGPS, \citealp{Aguirre})
at 1.1 mm and partially overlaps with the JCMT SCUBA-2 survey of the 
Galactic plane \citep{diFrancesco2008} at 850~$\umu$m.
There are common complexes also covered by the HOBYS program 
(Herschel imaging survey of OB young stellar objects; \citealp{M2010}).
The Coordinated Radio and Infrared Survey for High-Mass Star Formation at 5~GHz,
(CORNISH; \citealp{Hoare2012, Purcell2013}), and other surveys in the 
radio regime probe free-free emission of ionized gas 
surrounding OB type stars. 

The combination of these surveys therefore
provide a complete view of the spectral energy distribution (SED) from IR to
radio wavelengths, which is necessary to probe
the nature of dust clumps. 
This reveals purely dust sources, which can be starless or pre-stellar,
mid-IR bright protostars which just started to heat up their surroundings and
massive stars where the ionizing emission leads to the development
of {\uchii} regions which then expand becoming optically visible {\hii} regions. 
The dust emission at 870~$\umu$m measured in the \at\ survey 
is therefore sensitive to sources in all evolutionary stages with both cold and warm gas 
(\citealp{schuller2009}, \citealp{UCornish2013}).

The \at\ survey has been used to study various objects in specific environments in the Galaxy and
here we give an overview of them.
For a sample of Galactic bubbles, identified
from Spitzer-GLIMPSE data, \citet{Deharveng2010} use \at\ to study 
the dense and cold material at the borders of {\hii} regions.
\citet{Beuther2012} uses \at\ data to reveal the Galactic structure.
In a limited range of Galactic longitude ($10^\circ < \ell < 20^\circ$), \citet{Tackenberg2012} use \at\ to
identify a population of starless clumps. In a series of papers
the properties of dust clumps associated with various signposts of massive star-formation were determined.
\citet{U2013mmb} studied the physical properties of sources from the Methanol-MultiBeam survey~\citep{Green2009},
a sample of {\hii} regions from the CORNISH survey \citep{UCornish2013} and 
 MYSOs from the RMS survey (Urquhart et al.\,in prep).
The projected image of dust lacks, however, any information on the line-of-sight 
distribution of the material, for which spectroscopic observations are required.
Numerous molecular line follow-up observations have therefore 
been triggered by the \at\ survey. The MALT90 project maps over 2000~\at\ sources
in 16 lines~(\citealp{Foster2011,Jackson2013}). \citet{Wienen2012} presents an extensive
follow-up campaign of the $\sim$1000 brightest dust clumps in NH$_3$ in order
to determine kinematic distances and gas temperatures.

The first catalog of compact sources from the \at\ survey is
presented by \citet{Contreras2012} and focuses on a limited range in Galactic longitude ($ -30^{\circ}<\ell<21^{\circ}$). 
Here we aim to complement this work by covering the full area of the survey and
at the same time specifically addressing the population of embedded, smaller size-scale objects (see Sect.\,\ref{sec:comp}) 
compared to 
\citet{Contreras2012}.

The paper is organized as follows: Sect.\,\ref{sec:obs} describes the data processing, 
Sect.\,\ref{sec:se} presents the source extraction method and comparison with the previous 
catalog. Sect.\,\ref{sec:prop} presents the catalog and the properties of the extracted sources. 
In Sect.\,\ref{sec:ir} we compare the sources with mid-IR diagnostics to estimate their 
fraction associated with on-going star-formation and study their statistical properties.
Then we present their global properties, such as the Galactic distribution and typical fluxes
of quiescent and star-forming sources in Sect.\,\ref{sec:glon}.
Based on these statistics we estimate a Galactic star-formation rate and formation
time-scales in Sect.\,\ref{sec:timescales}.
We summarize the results in Sect.\,\ref{sec:sum}.
\afterpage{}

%
%__________________________________________________________________
\section{Observation and data reduction}\label{sec:obs}

This paper is based on all data taken for the survey between 2007 and 2010.
The observing strategy and the main steps of the data reduction procedure are 
described in detail by \citet{schuller2009} and \citet{Contreras2012}.
The data was reduced with the {\sl BoA} software 
\citep[]{Schuller2012}\footnote{http://www.eso.org/sci/activities/apexsv/labocasv.html}, and emission from larger scales was iteratively recovered until convergence was reached 
after $15$ iterations. 

The absolute position accuracy of the \at\ survey has been discussed in detail in 
\citet{Contreras2012}. The astrometry of the dataset 
and the derived source positions are estimated to be accurate to the pointing
accuracy of the telescope, which is $\sim2-3$\arcsec. 
The absolute flux uncertainty is estimated to be less 
than $\sim15\%$ \citep{schuller2009}. 
Variations in the sky emission (``sky-noise'') mimic emission from extended astronomical objects.
Ground based bolometer arrays are therefore not well suited to  
measure extended emission, since the emission from larger angular scales is
removed when subtracting the correlated noise from the maps. 
The final emission maps of the survey are sensitive to angular
scales up to $2\rlap{.}'5$, thus the current data reduction
is optimized to enhance compact sources. 
The final maps have been gridded on $3$ by $3$ degree
tiles with $\sim4\rlap{.}'5$ overlap between adjacent emission maps.
The pixel size is 6{\arcsec} which is $\sim$1/3 of the beam size. 
The data is publicly available at {\tt http://atlasgal.mpifr-bonn.mpg.de/}.

\begin{figure}[!htpb]
\centering
\includegraphics [width=6cm, angle=90]{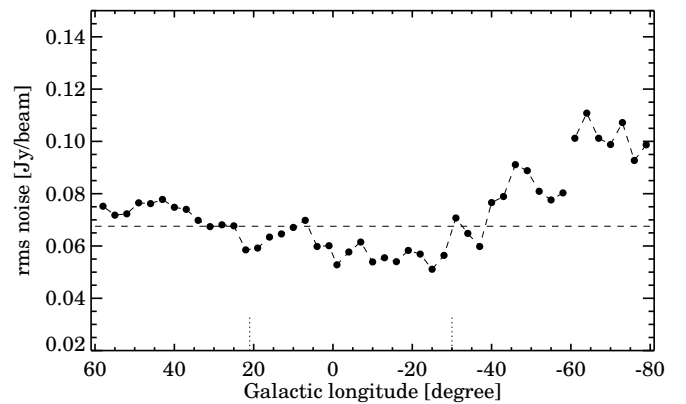}
\caption{The rms noise level determined on the individual tiles of maps using a Gaussian fitting to the distribution of the pixel values between b$\le1^{\circ}$. The region analyzed in \citet{Contreras2012} is indicated by dotted lines.
The horizontal dashed line shows the average 0.07 Jy/beam noise level for the main part of the survey ($|\ell|\le60^\circ$).
}\label{fig:noise}

\end{figure}

The average noise level is determined from the $b\le1^{\circ}$
portions of the maps where the vast majority of the emission originates from 
and because the noise increases rapidly towards the edges of the maps 
due to a lower number of overlapping coverages.
The noise was determined from a Gaussian fit to the distribution of the pixel 
values in the maps and is found to be on average $\sim$70~mJy/beam over the
whole survey region.
The average noise varies ($\sim$20\%) 
as a function of $\ell$ due to the 
non-homogenous coverage of the observed regions and varying observing conditions (see Fig.\,\ref{fig:noise}).
In the longitude ranges $\ell>40^{\circ}$ and $\ell<-40^{\circ}$ the 
average noise increases to $70-90$~mJy/beam, while  
in the central part of the plane the noise is $\sim50-60$~mJy/beam. 
In the extension, between $-80^\circ \le \ell \le -60^{\circ}$, the noise is higher 
($\sim110$~mJy/beam) than in 
the main part of the survey due to fewer coverages and shorter observing time.

In the following we extract the population of compact sources in the Galactic plane using
all \at\ data. However, due to the increased noise in the extension, we base our analysis only on the main part of
the survey within the $\ell \le \pm60^\circ$ in Galactic longitude range. 
The identified compact objects 
mainly correspond to starless, pre-cluster dense material, protostars, protoclusters,
compact {\hii} regions, as well as over-densities in the clumpy and filamentary cloud structures.

%		Map of the noise distribution
%______________________________________________ 
%______________________________________________ 
\afterpage{}

\section{Source extraction}\label{sec:se}
\afterpage

In the following we assume that the 870~\mum\ emission arises from thermal continuum emission from dust and that the 
contamination from free-free emission is negligible. 
The filter of LABOCA admits a 60~GHz frequency range centered on 345~GHz~\citep{siringo2009} 
thus covering the CO~(3-2) rotational transition at 
345.796~GHz and other lines. 
Here we do not account for any line contamination, nevertheless, as discussed in \citet{schuller2009}, the
contribution to the derived fluxes may be more than the usual 15\% flux calibration 
uncertainty, but only towards the most extreme sources, such as hot molecular cores, strong outflow sources, and bright photon-dominated regions and is not likely to be an issue for the vast majority of the \at\ sources.

\subsection{Overview of source extraction algorithms}\label{sec:overviewofalgorithms}

Systematic source identification algorithms have been developed 
to analyze ground-based millimeter/sub-millimeter maps, 
such as {\tt Clumpfind} \citep{Williams94}, and
Gaussian source fitting in different implementations,
originally developed by \citet{SG1990}. 
The arrival of fast mapping capabilities by the PACS and SPIRE instruments 
onboard Herschel boosted the development 
of various source-extraction algorithms. The space-based observations 
provide the possibility for efficient mapping, but also 
the imaging of a significant amount
of large-scale emission, which can not be recovered by ground-based surveys.
Therefore several new methods have been 
developed to deal with the extended emission and also to handle multi-wavelength data sets such as those of 
Herschel (e.g.\,{\sl MRE-GCL, Multi-resolution and \gc\ algorithm, }\citealp{M2010}, {\sl CuTeX:} \citealp{Molinari_cutex}, 
{\sl Getsources:} \citealp{M2012}). A detailed overview on these different
source-extraction algorithms, discussing their advantages and
disadvantages is given in \citet{M2012}. 

The majority of these algorithms assume a uniform noise distribution 
which is not realistic, especially for ground based observations that 
may be done at different meteorological 
conditions, different elevations, etc. It is therefore unavoidable that large area ground-based 
surveys show varying noise levels over the survey region. 
The {\tt SExtractor} algorithm \citep{BA1996} uses a $\sigma$-clipping method
to handle the varying noise level, and has been originally developed for optical and infrared images, 
identifying stars (point sources) and galaxies. 
However thermal emission from the spatially extended cloud structure 
usually shows a more complex morphology. For example, the population 
of compact sources is deeply embedded
in dense clouds.
\citet{Contreras2012} have successfully applied the {\tt SExtractor} algorithm to a substantial part of the 
\at\ survey ($ -30^{\circ}<\ell<21^{\circ}$). This method applied to dust emission maps 
pulls out properties of the whole clump as it is not optimized to separate compact 
emission from the more diffuse envelope.
Therefore it can be considered to work similarly as a contouring algorithm like 
{\sl Clumpfind} without any assumption on the source characteristics. 

Another, in many ways complementary approach is to assume a certain characteristic
or property for the embedded sources. The {\sl Gaussclumps} algorithm \citep{SG1990, Kramer98} 
was originally developed to 
identify coherent structures of 3-dimensional molecular line (position, position, velocity: ppv) data cubes, and assumes a 
Gaussian intensity distribution to identify structures. 
The interstellar medium exhibits a clumpy morphology
with the embedded sources on much smaller spatial scales
superimposed on this background emission. Since 
our main interest is to identify these more compact sources, i.e. cores and clumps, 
we base their identification on the assumption
that they exhibit a Gaussian intensity distribution.
The assumption of a Gaussian intensity profile is adequate for compact sources 
with sizes of a few times 
the beam (e.g.\,\citealp{M07}, see also Fig.\,\ref{fig:filtering_size_app}). 

Our source extraction method follows the scheme originally developed in \citet{M98} and presented
in \citet{M07}, 
and which is referred to as the {\sl MRE-GCL} method in the literature.
It has been used to identify compact sources of thermal dust emission by different groups
for complex regions, like W43~\citep{M03}, Cygnus-X~\citep{M07} and the NGC~6334-NGC~6357 complexes \citep{Russeil10}.
It has also been applied for smaller regions, 
such as RCW~106 \citep{Mookerjea2004}, NGC~2264 \citep{Peretto2006} and G327.3-0.6 \citep{Minier2009}.
The main steps are 
to first disentangle the compact sources from the more diffuse, extended emission of filamentary clouds. 
This is done by a multi-scale 
wavelet transformation that is used to filter out the larger scale structures (see examples in
Fig.\,\ref{fig:filtering_size}, Fig.\,\ref{fig:filtering_size_app})
and then use the \gc\ algorithm to extract sources. 
These steps of the method are described in more detail in the following two subsections, 
Sec. \ref{mr-decomp} and \ref{sec:extr-gc}. 
 
\subsection{Multi-scale decomposition}\label{mr-decomp}

Molecular clouds exhibit structures at various scales and part of their material is organized into large-scale
filaments (e.g.\,\citealp{Andre2010,Molinari2010b,Kainulainen2011}). 
Although ground based bolometers are insensitive 
to uniform diffuse emission, they recover a significant amount of extended structures 
of dust seen towards the Galactic plane. 
Since the {\at} maps contain emission from 19$\rlap{.}{''}$2 up to 2$\rlap{.}'$5 spatial scales,
it is desirable to remove the extended emission in order to extract the properties of embedded sources.

At distances up to 1~kpc the spatial resolution of the survey corresponds to $<$0.1~pc
 physical scales, thus rather individual cores. Placed at larger, beyond 10~kpc distance this translates
 to $>1$~pc size-scale objects, therefore corresponding to cloud structures.
{
As opposed to single complexes, 
the choice of a physical scale is not trivial since the 
 whole Galaxy is seen in projection along the line-of-sight. 
}

We therefore visually examined the structures at different scales
aiming to identify the best spherically symmetric, centrally
 condensed compact objects.
 {This way we choose to optimize the extraction method to be sensitive to structures 
 with angular scales from the 19$\rlap{.}{''}$2 resolution element to 
 $\sim$50{\arcsec}, which translate to a physical scale of $0.4-1$~pc at 4~kpc.}
 This is also a reasonable physical scale for compact structures assuming that a large fraction of the sources 
lie at the typical distance of~$\sim4$~kpc (e.g.\,
\citealt{Wienen2012}), while \citet{Peretto2010a} finds that 95\% of the IRDCs are at distances of $<6$~kpc 
with the mean distance between 3-5~kpc. 
 Here we aim therefore to extract the properties of compact
 sources with typical size-scales of 0.4~pc to 1~pc, commonly
 noted as clumps (e.g.\,\citealp{BT2007, MH09}). 

\begin{figure*}
   \centering
   {\rotatebox{0}{\includegraphics[width=6cm]{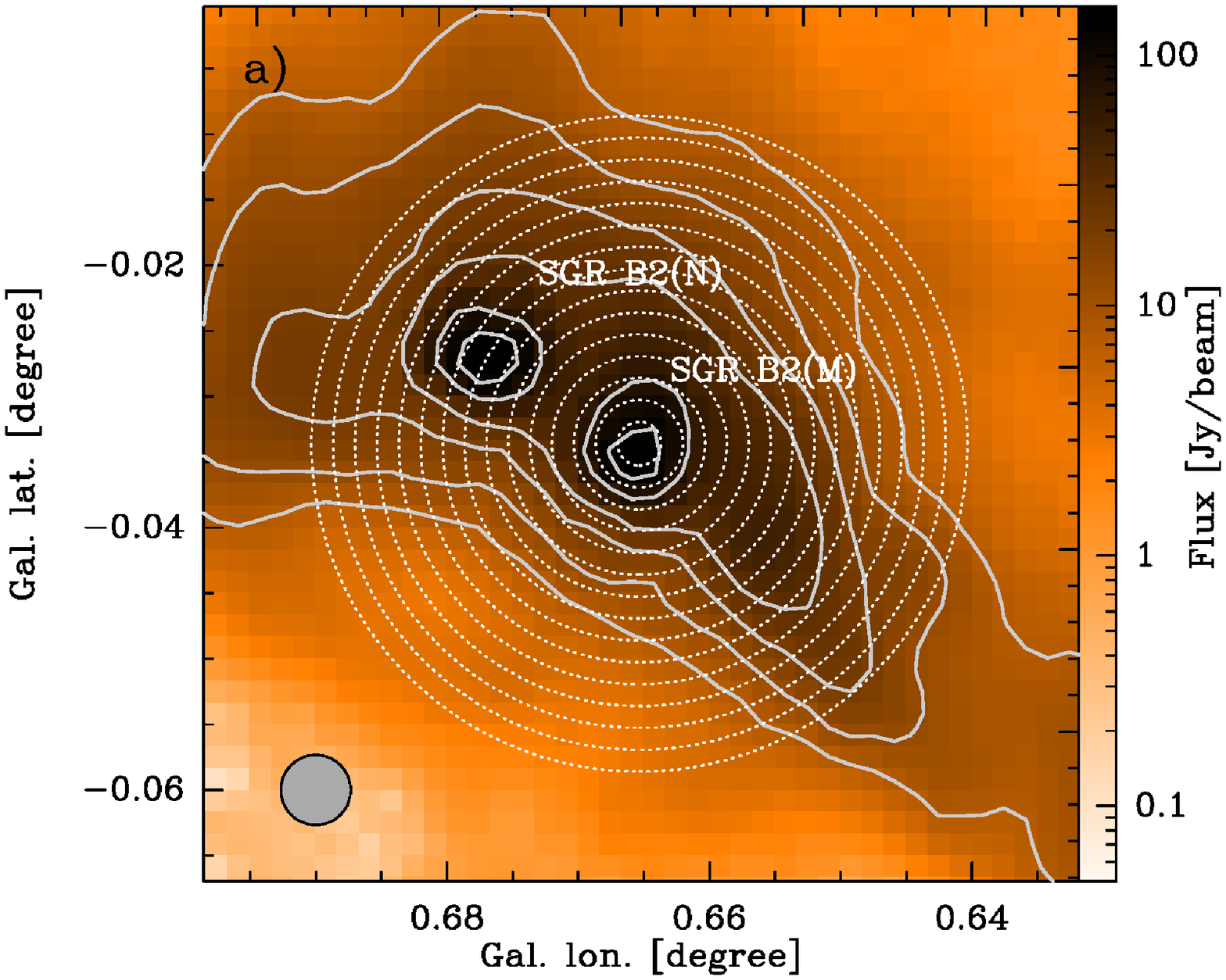}}}
   {\rotatebox{0}{\includegraphics[width=6cm]{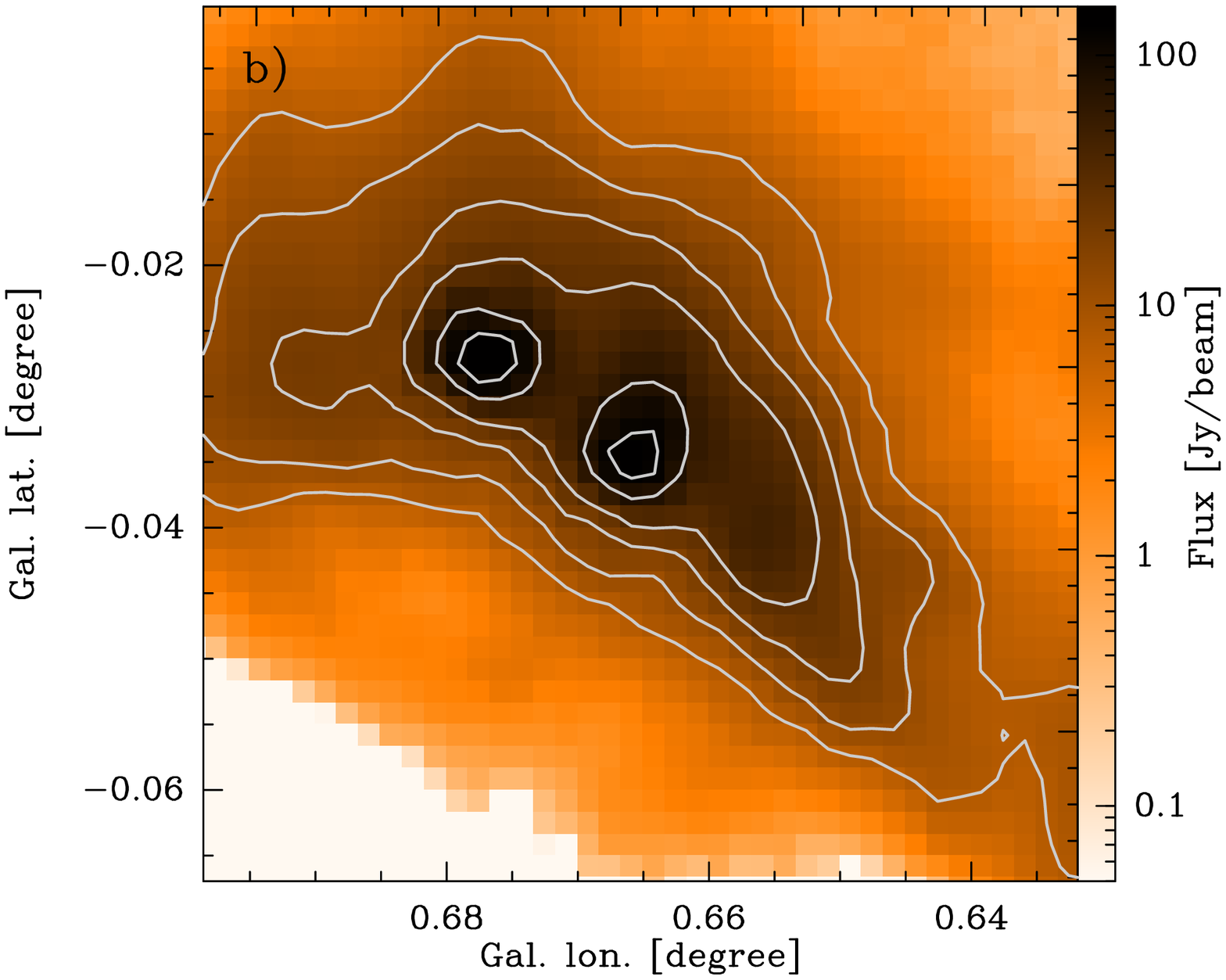}}}
   {\rotatebox{0}{\includegraphics[width=6cm]{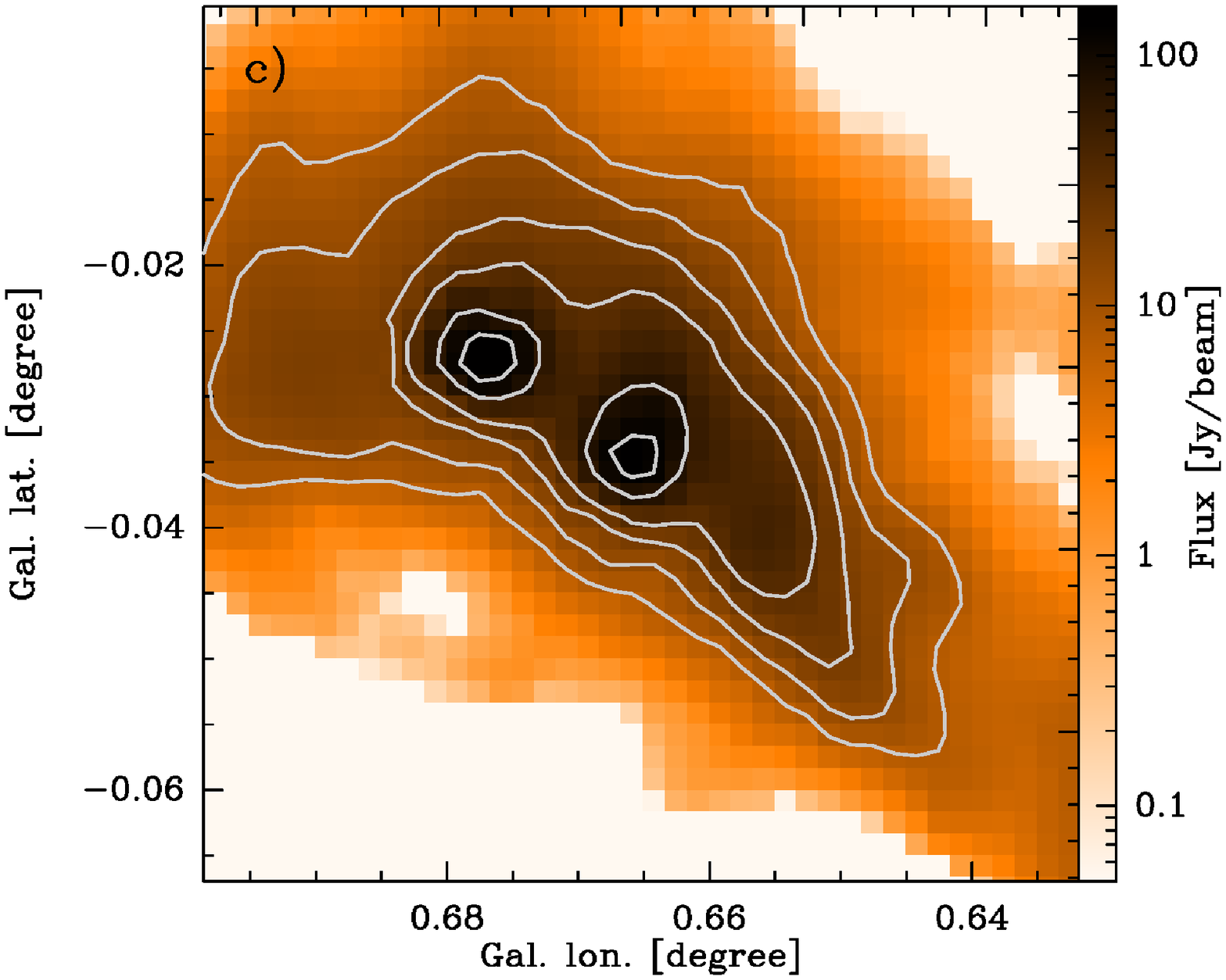}}}
   {\rotatebox{0}{\includegraphics[width=6cm]{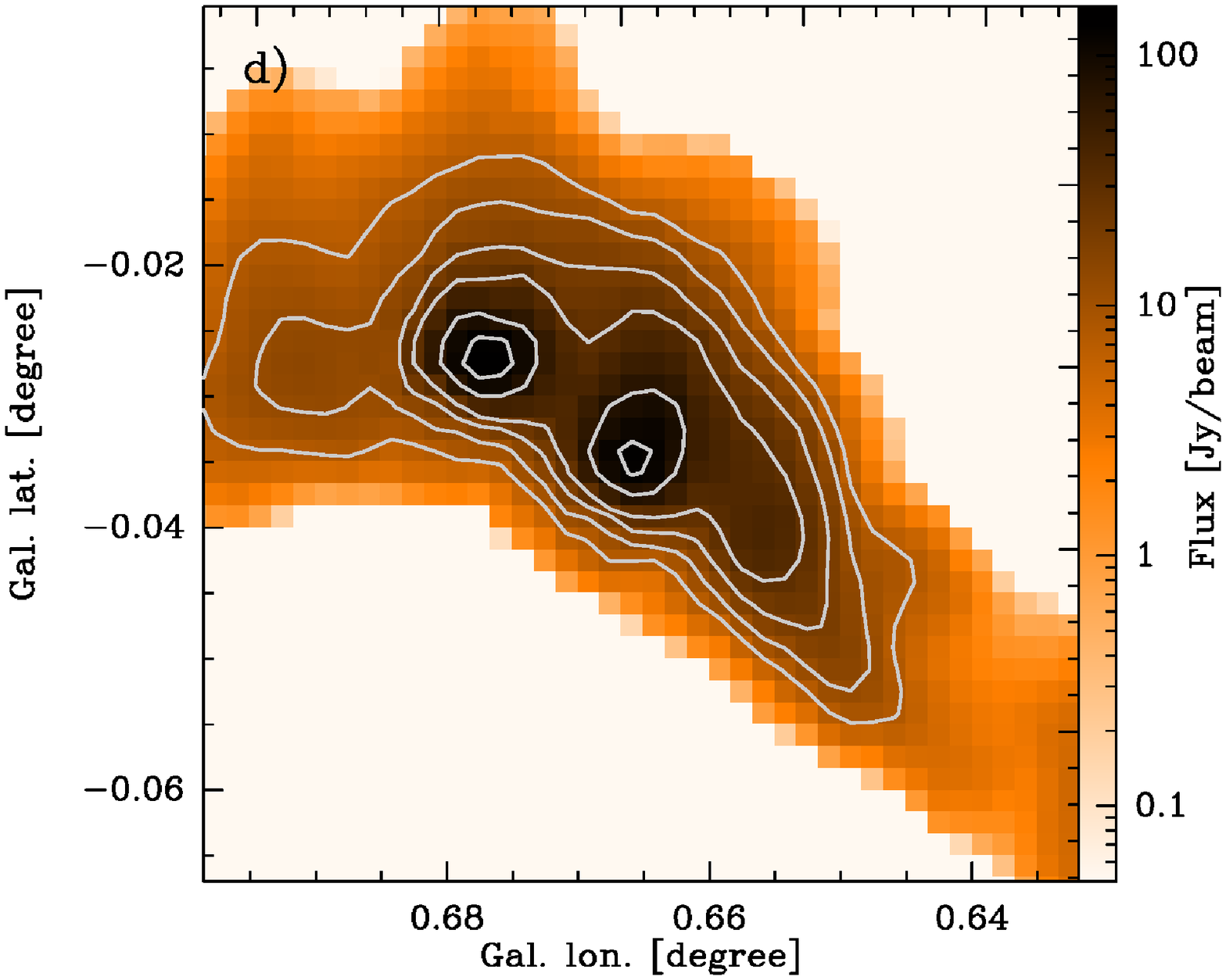}}}
   {\rotatebox{0}{\includegraphics[width=6cm]{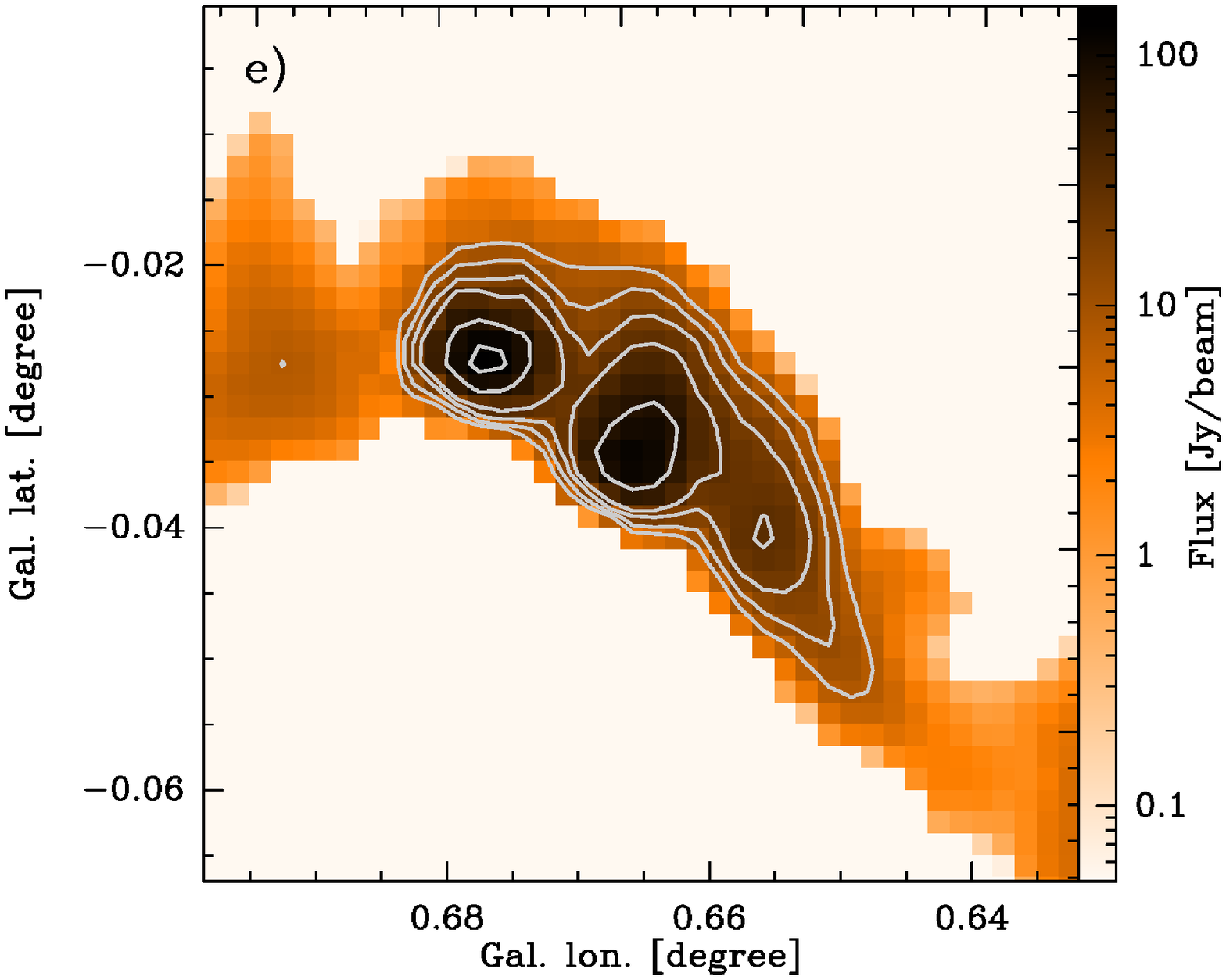}}}
   {\rotatebox{0}{\includegraphics[width=6cm]{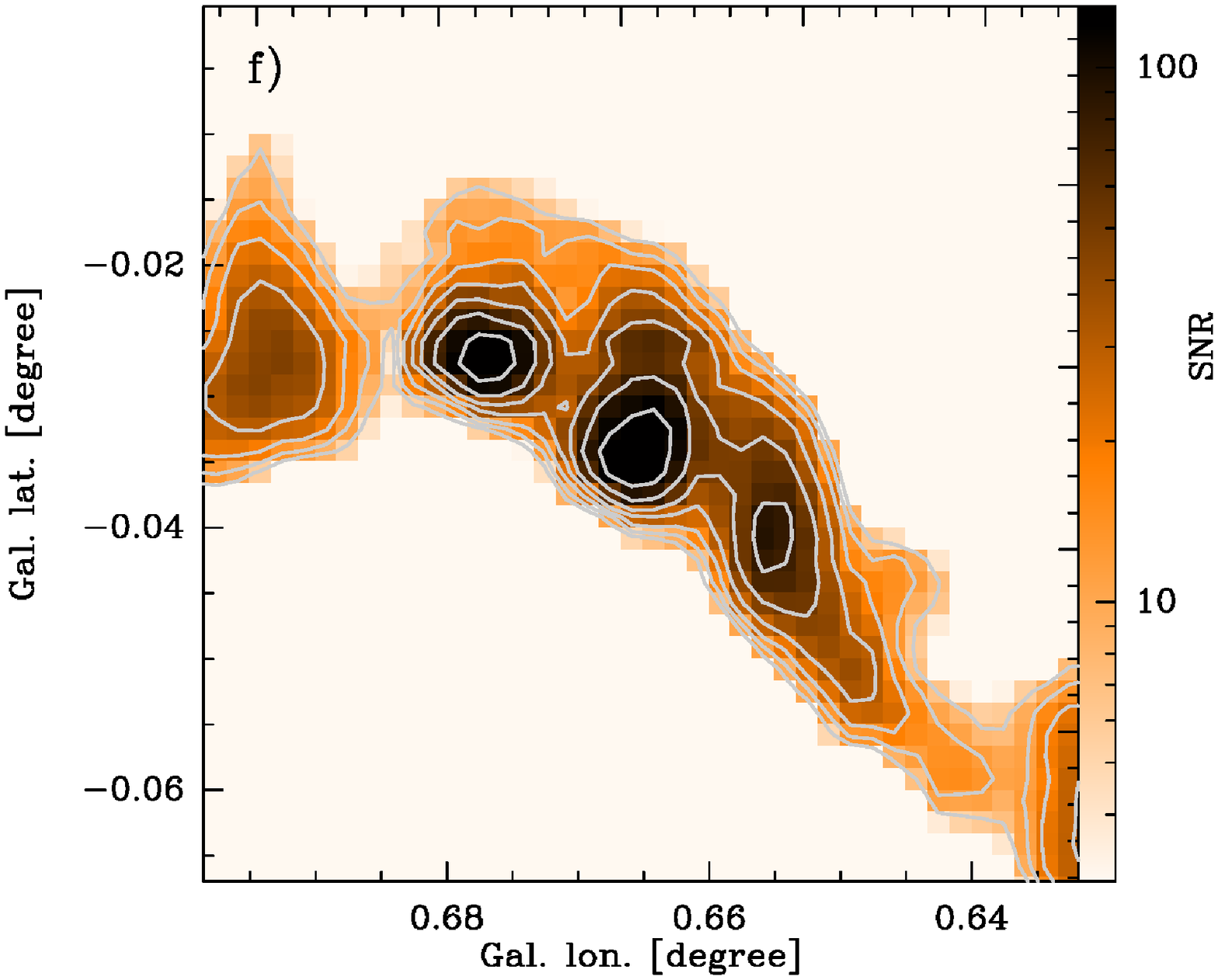}}}
    \caption{
                  We show an example of the multi-scale decomposition towards the
                  brightest region at 870~$\umu$m  in the whole survey which hosts SgrB2(N) and  SgrB2(M).
                  The impact of flux loss due to filtering is the most severe towards this region.
                  Contours start at 7.5~Jy/beam and increase on a logarithmic scale to 120~Jy/beam.
                  Dotted circles mark the regions where the radial averaging has been done (see Sect.\,\ref{sec:extr-gc}
                  and Fig.\,\ref{fig:radial_avg} for details). 
                  The beam size of 19.2\arcsec\ is shown in panel {\bf a)}.
                  Panels {\bf a)} to {\bf e)} show maps with different scales of 
                  filtering: from no filtering to maps where background emission is increasingly removed. 
                  The catalog is generated from a filtering level corresponding
                  to the {\textbf e)} panel, where emission until 2$\times$50\arcsec\ scales
                  is summed up.
                  The peak flux density of the object in this case decreases by 20\% from
                  the original images to the most compact one, while the size decreases
                  by only 10\%. 
                  The {\textbf f)} panel shows the filtered $snr$ map used as the input
                  for the source
                  extraction algorithm containing the same spatial scales
                  as the {\textbf e)} panel. The scaling is logarithmic between $3-120\sigma$, 
                  contours start at 7$\sigma$ increasing on a logarithmic scale.
                  }
   \label{fig:filtering_size}%
   \end{figure*}

Since our primary interest is in these embedded sources, 
we systematically remove emission from cloud structures at larger scales that we consider as 
extended emission
originating from the embedding cloud. 
To do this we decomposed the emission into different spatial scales using a wavelet transformation 
with the method of \citet{Starck06}.
Emission 
from smaller than twice the maximum required 50\arcsec\ scale was then summed up.
Due to the wavelet decomposition negative artifacts appear
around bright sources and we have set these negative artifacts to zero %. 
{in order to help the convergence of the algorithm (see 
\citealp{Kramer98,M07}).}
The emission with different scales is illustrated
in Fig.\,\ref{fig:filtering_size} (for an isolated source, see Fig.\,\ref{fig:filtering_size_app}). %, 

%______________________________________________ 

%%%%%%%%%%%%%%%%%%%%%%%%%%%%%%
% JUSTIFICATION OF THE EXTRACTION METHOD
\afterpage{}

\subsection{Extraction of compact sources -- Gaussclumps}\label{sec:extr-gc}

Although \gc\ does not deal with varying noise levels, it is very important 
to consider the non-uniform noise distribution of the maps to avoid detection of 
spurious sources and missing genuine sources in a region with lower noise.
To overcome this, we used the weight maps and calculated 
{from pixel to pixel}
signal-to-noise maps
using the formula: $snr=flux\, \times \sqrt{weight}$, where $flux$ corresponds to the flux density in the
emission maps and the weight is computed when combining signals of all bolometers into a map. 
The weight is related to the noise as $1/{rms^2}$ where the $rms$ is the standard deviation of the 
signals of the bolometers (see \citealp{schuller2009} for more details).
The intensity of the weight maps exhibits variations of $\sim$10\% on a 60\arcsec\ scale,
 so the spatial variations are smaller than the expected size of compact sources.
Therefore we use the signal-to-noise maps as the input files for the source extraction method,
and use \gc\ with a 
threshold of $\sim$ 5, which corresponds to 5$\times$ the local $rms$
\footnote{
The input parameters for \gc\ were an initial guess for the size of 
1.5$\times$ the beam FWHM,
and the stiffness parameters ($s_0, s_a, s_c$) have been set to 1.
These control the weighting
of the Gaussian function in order to fit a peak intensity and position close to the
observed value. For
a detailed description of these parameters
we refer to \citet{SG1990} and \citet{Kramer98}.
}.
The choice of this threshold is justified in Sect.\,\ref{sec:completeness}.

The position and size of the sources were derived from the output
of \gc\, while the peak flux values were computed using the flux values from the $snr$ maps 
multiplied by the noise level measured on the weight maps.

Since we apply this procedure for each tile and there is a $\sim$4.5\arcmin\ overlap between 
the individual \at\ tiles, sources 
falling on these edges
are found twice by the algorithm. We eliminated these duplicates by 
systematically keeping those detections
that are closer to the center of the map, in which they are identified.
After rejecting duplicate sources, we also rejected sources with sizes smaller than the beam
as these are likely to correspond to noise peaks in the maps 
(see e.g.\,\citealp{Kramer98}).

As a sanity check we have visually inspected all sources with outlying (i.e. the highest and 
lowest) fluxes and aspect ratios and found them to be genuine sources. 
{
We discuss the impact of the filtering on the extracted source parameters in 
App.\,\ref{app:radial-fluxprofile} and \ref{app:souparams}. 
}

All of the survey region was inspected for missed  sources 
and we found that  visible but undetected structures
fall below the $snr$ threshold used for the extraction.

\subsection{Completeness limit}\label{sec:completeness}

We estimate the completeness limit of our source extraction method by 
adding artificial sources into the input maps of {\gc}.
Using the same input parameters as for the
catalog we measure the number of  
extracted artificial sources as a function of peak flux, following a similar procedure described by 
 \citet{Contreras2012}. For these tests we have used a simulated 
field of $20\times9^{\circ 2}$ size, which corresponds to 20 \at\ tiles
{and injected} 9989 {artificial} sources. 
This is similar to the {total} number {of sources}
we find for the full {survey}. 
To mimic a uniform background noise
the first set of artificial sources were
injected on maps with a normal distribution of noise values with a standard deviation of 
 $\sigma=60$ mJy/beam.
The injected sources 
have peak flux densities following a uniform distribution up to 1.5~Jy/beam, with 
30\arcsec\ FWHM size, an aspect ratio of 1 and a Gaussian flux distribution in order to imitate spherically structured cores.

The second test was made with the same characteristics for the artificial sources, but 
 injected on a background map with a varying noise level. For this we
used a map with real observations, but smoothed with a Gaussian kernel of 31 by 31 pixel array
 which has a FWHM of 5~by~5 pixels in order to imitate the extended emission
from molecular clouds. A third test was done using the maps with varying background noise 
and the same filtering level as for the full catalog. We run our extraction algorithm 
down to a fixed threshold of $\sim$ 3$\sigma$ and 
therefore extracted sources down to 0.2 Jy/beam. This threshold was chosen to be lower 
than that used for the source extraction on the real data in order to explore the lower limits of our completeness.

\begin{figure}\centering
   {\rotatebox{90}{\includegraphics[width=7cm]{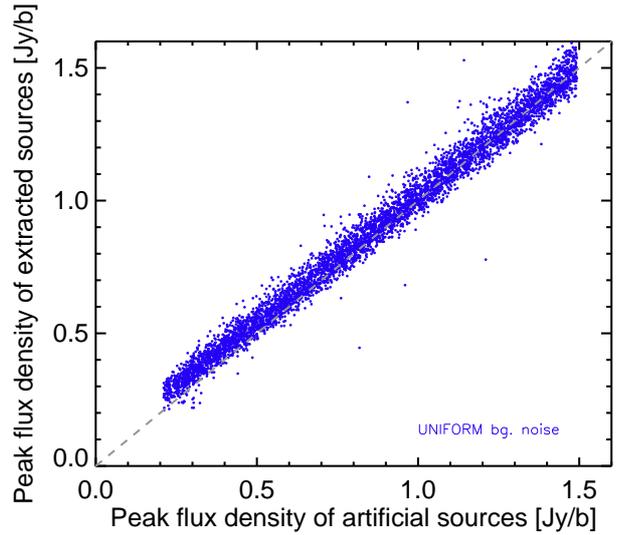}}}
    \caption{
                                For clarity we only show a comparison between the 
                                peak flux densities of the extracted  
                                and the artificial sources for
                                the uniform noise simulations;  
                                the result obtained from the other two tests are similar.
                                }
   \label{fig:completeness-pfl}%
   \end{figure}

As a consistency check we have compared the fluxes of the injected and recovered sources and find good agreement between these values for all three simulations. Fig.\,\ref{fig:completeness-pfl} shows the comparison between the injected and extracted peak flux densities for the case of normal noise distribution. In all of the simulations we find that the measured source properties lie close to unity compared to the injected parameters for high peak flux densities. Close to the detection threshold we see a deviation from unity with the measured fluxes being offset to slightly higher than the injected values. This offset is at most 1$\sigma$ (60 mJy) in the peak flux densities introducing $<$20\% error in the peak fluxes 
for the weakest sources and is a result of flux boosting due to the constructive interference with the noise which determines the fitting of the peak flux. 

\begin{figure}
   {\rotatebox{90}{\includegraphics[width=7cm]{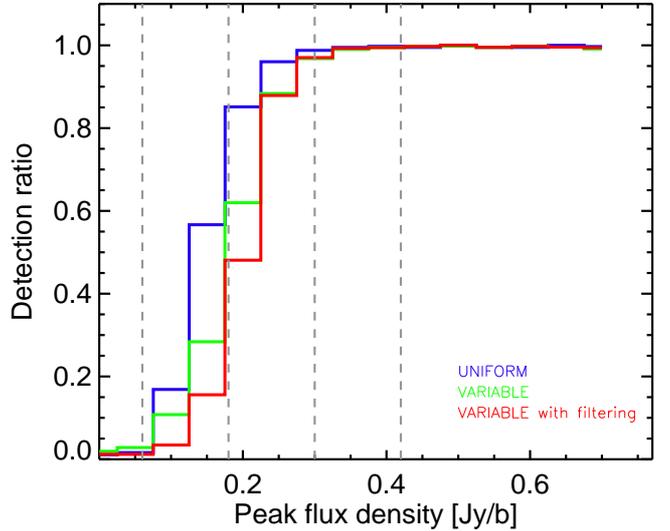}}}
    \caption{
                                Detection ratio of recovered versus injected sources
                                         shown as a function of peak flux.
                  			Blue, green and red line show the uniform and the varying noise
					extraction without and with filtering, respectively. The filtering
					corresponds to the 100\arcsec\ filtering level we used to produce the catalog.
					Grey dashed lines
					correspond to 1$\sigma$, 3$\sigma$, 5$\sigma$, 7$\sigma$
					noise levels. 
                 }
   \label{fig:completeness}%
   \end{figure}

Fig.\,\ref{fig:completeness} shows the detection ratio of the sources as a function of peak flux
density, {where we allowed a positional offset of 10\arcsec\ between
the artificial and the extracted sources.}
We find that our catalog is complete to 99.4\% above
{\compvar}$\sigma$ and 97\% of the injected sources are recovered at 5${\sigma}$. 
This was therefore chosen as a detection threshold
{for the extraction for the full survey}. 
{A lower value would result in a less
reliable catalog and the completeness level at lower flux densities
would also be limited, which is not desirable for statistical studies.}
We stress that below a {\compvar}$\sigma$ noise level, measured on the input maps,
our catalog is not complete, and genuine sources may be missed. Our completeness
level is comparable to that of other source extraction methods, such as the one used by \citet{Contreras2012}. 

%______________________________________________ 
%______________________________________________ 
%______________________________________________ 
%______________________________________________ 

\subsection{Physical parameters of the compact sources}\label{sec:phys}

\begin{figure}[!h]
   \centering
   {\rotatebox{90}{\includegraphics[width=0.9\linewidth]{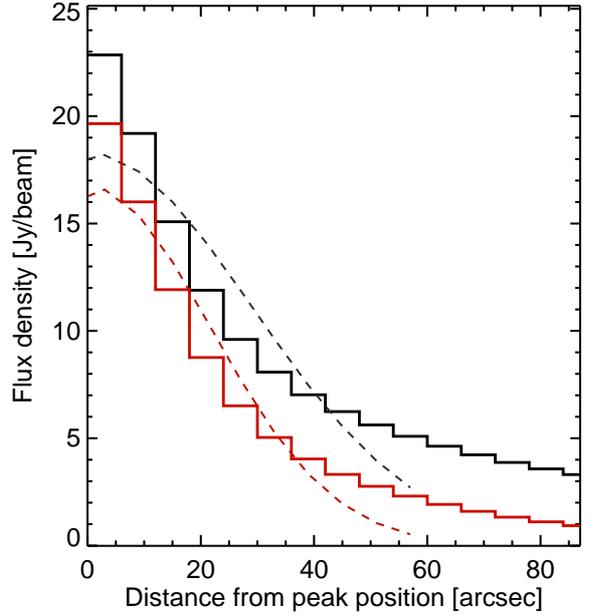}}}
    \caption{
                  Azimuthally averaged emission profile averaged over the 100 brightest sources.
                  Black line corresponds to the original images, while red line
                  shows the images filtered above 100\arcsec\ scale. Dashed lines show
                  the Gaussian fits to these profiles.
                  }
   \label{fig:radial_avg}%
   \end{figure}

Since the sources
are identified and extracted from maps where the extended emission was removed, 
the corresponding physical parameters, such as the size and the 
peak flux are different compared to the original emission maps. These parameters are more likely to
reflect the real properties of the embedded compact sources as illustrated 
in Fig.\,\ref{fig:filtering_size} 
(another example is shown in Fig.\,\ref{fig:filtering_size_app} for an isolated source)
and in Fig.\,\ref{fig:radial_avg}, where we show the azimuthally averaged
intensity profile from the original and the filtered maps used for the source extraction. 
It illustrates that by removing
the extended emission, the measurement of the Gaussian parameters
of the embedded source improves.
We have tested the impact of the multi-scale decomposition on the measured fluxes and sizes 
by extracting the sources using different
scales. We found that the impact on the measured size is in most cases 
very small, because the majority of the sources are only marginally 
resolved, and their morphology is dominated by a compact emission. 
For the peak flux values we find that the difference is less than 20\% for isolated sources,
however in complex regions, especially for bright, massive clouds the extended
emission contributes more to the total flux. This can be as high as 
$50$\% of the total flux in such regions (see also App.\,\ref{sec:souparams} for more details). 

We measure the physical parameters of the sources using the formulae from \citet{schuller2009}
and assuming optically thin emission of dust at 870~\mum. The column density is derived from:
\begin{equation}
N({H_2})=\frac{F_\nu\,R}{B_\nu(T_d)\,\Omega\,\kappa_{\nu}\,\mu_{H2}\,m_H}
\end{equation}
where {$F_\nu$ is the peak flux density}, $\Omega$ is the beam solid angle, $\mu_{H2}$ is the mean molecular weight of the interstellar medium
with respect to hydrogen molecules, which is equal to 2.8 \citep{Kauffmann2008}\footnote{The mean molecular weight is commonly defined with respect to free
particles in the gas, in which case $\mu=2.33$ and the derived column density reflects the number density of free particles. 
Using $\mu_{H2}=2.8$ gives the column density of $H_2$ molecules, and one has to multiply our $N({H_2})$
values by a factor of $\sim$1.2 to be consistent with other studies {using $\mu=2.33$}.
}, 
and $m_H$ is the mass of an hydrogen atom. We adopt here the same assumptions as \citet{schuller2009}, a 
gas-to-dust mass ratio (R) of 100, and  $\kappa_{\nu}=1.85$ cm$^2$~g$^{-1}$, which is interpolated to 870~$\umu$m from Table 1, 
Col.\,5 of \citet{OH1994}. At our completeness level of 7$\sigma$ we are thus sensitive to column densities 
between $\sim7\times10^{21}-4\times10^{22}$~cm$^{-2}$ for warm ($T_d=30$~K) and cold ($T_d=10$~K) gas, respectively.
We estimate the mass as: 
\begin{equation}
M=\frac{S_\nu\,R\,d^2}{B_\nu(T_d)\,\kappa_{\nu}}
\end{equation}
where {$S_\nu$ is the integrated flux} and $d$ corresponds to the distance of the sources.
For a typical dust temperature
of $T_d=15$~K, 
our completeness limit corresponds  
to $\sim$270~\msol\ at 8~kpc and 1~\msol\ at 0.5~kpc.
The \at\ survey therefore probes all massive clumps in the inner Galaxy and in the $<1$~kpc nearby regions
intermediate-mass cores as well.

Since with \at\ we probe the whole inner Galaxy, it is crucial to assess what 
fraction of the extracted sources may sustain high-mass star-formation.
\citet{M07} adopted a limit of 40~\msol\ for defining Massive Dense Cores (MDCs)
based on an unbiased study of the high-mass star-forming complex, Cygnus-X.
The sources were found to have an average size of 0.13~pc and the mass was calculated assuming a density profile of $\rho(r) \propto r^{-2}$. 
Using high angular-resolution observations these MDCs have been confirmed
 to host individual high-mass
protostars \citep{Bontemps2010, Csengeri2011b}. 
We extrapolate this definition to our average spatial resolution of 0.4~pc (see Sect.\,\ref{mr-decomp}), 
and find that the same $\rho(r) \propto r^{-2}$ density profile corresponds to minimum mass of $\sim$125~{\msol}. 
However, as observed in Cygnus-X, it is very likely that
MDCs are embedded in massive clumps on the 0.4~pc scale of our resolution. 
For example, the massive clump of DR21(OH) with 0.6~pc size is fragmented into
 3~MDCs on a 0.13~pc scale. Each of these MDCs were shown
to host individual high-mass protostars~\citep{Csengeri2011b}.
If the \at\ clumps fragment on smaller than 0.4~pc scale, 
they should exhibit a shallower density profile.
Therefore we estimate a minimum mass-range between the $\rho(r) \propto r^{-2}$ density profile 
(extreme lower limit, $\sim$125~{\msol}) and a uniform density profile (extreme upper limit, $\sim$1150~{\msol}).
We adopt here
an average value of these two limiting cases of 650~{\msol} as 
an approximate threshold for \at\
clumps likely to host MDCs and high-mass protostars.
{
This is consistent with other studies similarly suggesting
that clumps below 1000~\msol\
can as well potentially form massive stars as long as they are compact (e.g.\,\citealp{Lada2003,U2013mmb}) 
}\footnote{
{
Our criterion for clumps to form massive stars is more selective compared to \citet{Tackenberg2012}, who adopt a mass limit of 
1000~\msol. They select on average larger structures with larger integrated fluxes. 
Our analysis yields on average 6$\times$ lower masses due to the
different source identification and a different $\kappa_{\nu}$. Consequently, their mass limit
corresponds to 1000~\msol$/6\simeq$170~\msol in our analysis. Our criterion for clumps to form massive stars is therefore four times more conservative.
}
}.

%______________________________________________ 
%______________________________________________ 
\afterpage{}

\subsection{Comparison to previous work}\label{sec:comp}
The source identification method described in  \citet{Contreras2012} has been 
extended to the full
\at\ survey\footnote{The full catalog using {\tt SExtractor} is available on this web-page: {\tt{http://atlasgal.mpifr-bonn.mpg.de}}.}
and here we give a comparison of the two 
methods. 

As a first step we checked what fraction of our sources coincide 
with the source masks from \citet{Contreras2012} to
compare the efficiency of identifying sources between the two methods. 
We find altogether 26 sources that
are identified in the current catalog, but were not found by {\tt SExtractor} and there are further
326 sources which were identified by {\tt SExtractor}, but later rejected due to  
having a smaller number of pixels than the area of the beam.
We {visually} inspected all of these sources and found them to be genuine sources.

We find 2391 sources identified by {\tt SExtractor}, but not identified with the current method.
This is partly due to the fact that \citet{Contreras2012} lists sources to a lower, 3$\sigma$ threshold, and
therefore the majority of these sources are primarily weak and fall below our 5$\sigma$ detection limit. 
A fraction of them belong to structures which are removed by the filtering we apply here.
Sources with peak fluxes above our detection threshold were individually inspected, 
and were found to be part of cloud structures that are decomposed in a different way by our method.

As pointed out in Sect.\,\ref{sec:overviewofalgorithms}, the {\tt SExtractor} software has originally been developed
for optical/infrared images to find stars and galaxies, which show lower level of extended background emission
 than dust
continuum from molecular clouds. 
Differences between the two catalogs come from the fact that the \textsl{MRE-GCL} method is more efficient 
identifying small and compact sources, while {\tt SExtractor} is better suited to
find larger size-scale sources.  The two catalogs are therefore 
complementary in terms of source characteristics.

%______________________________________________ 
   \begin{figure*}[!htpb]
   {\rotatebox{0}{\includegraphics[width=0.9\linewidth]{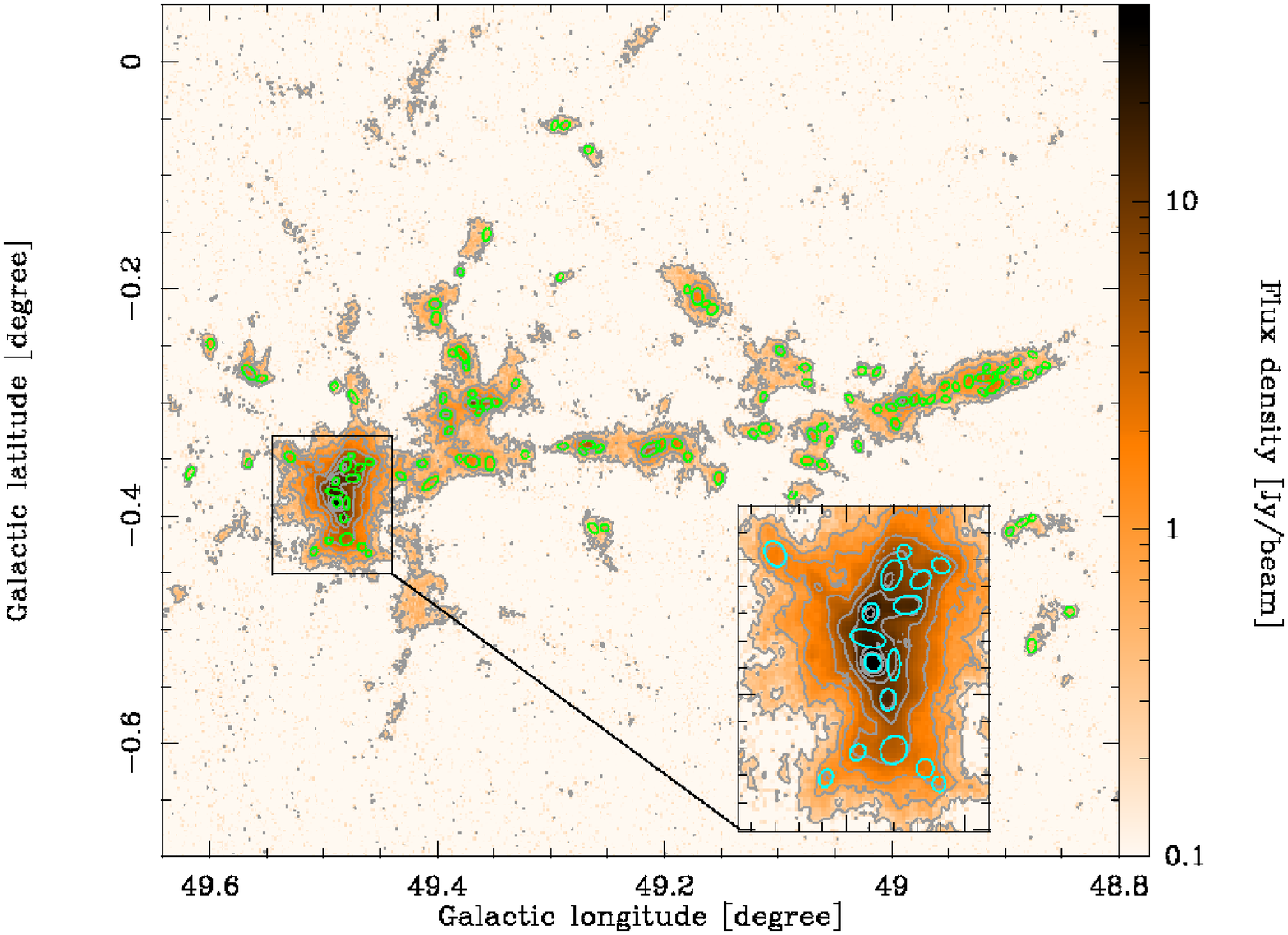}}}
   {\rotatebox{0}{\includegraphics[width=0.9\linewidth]{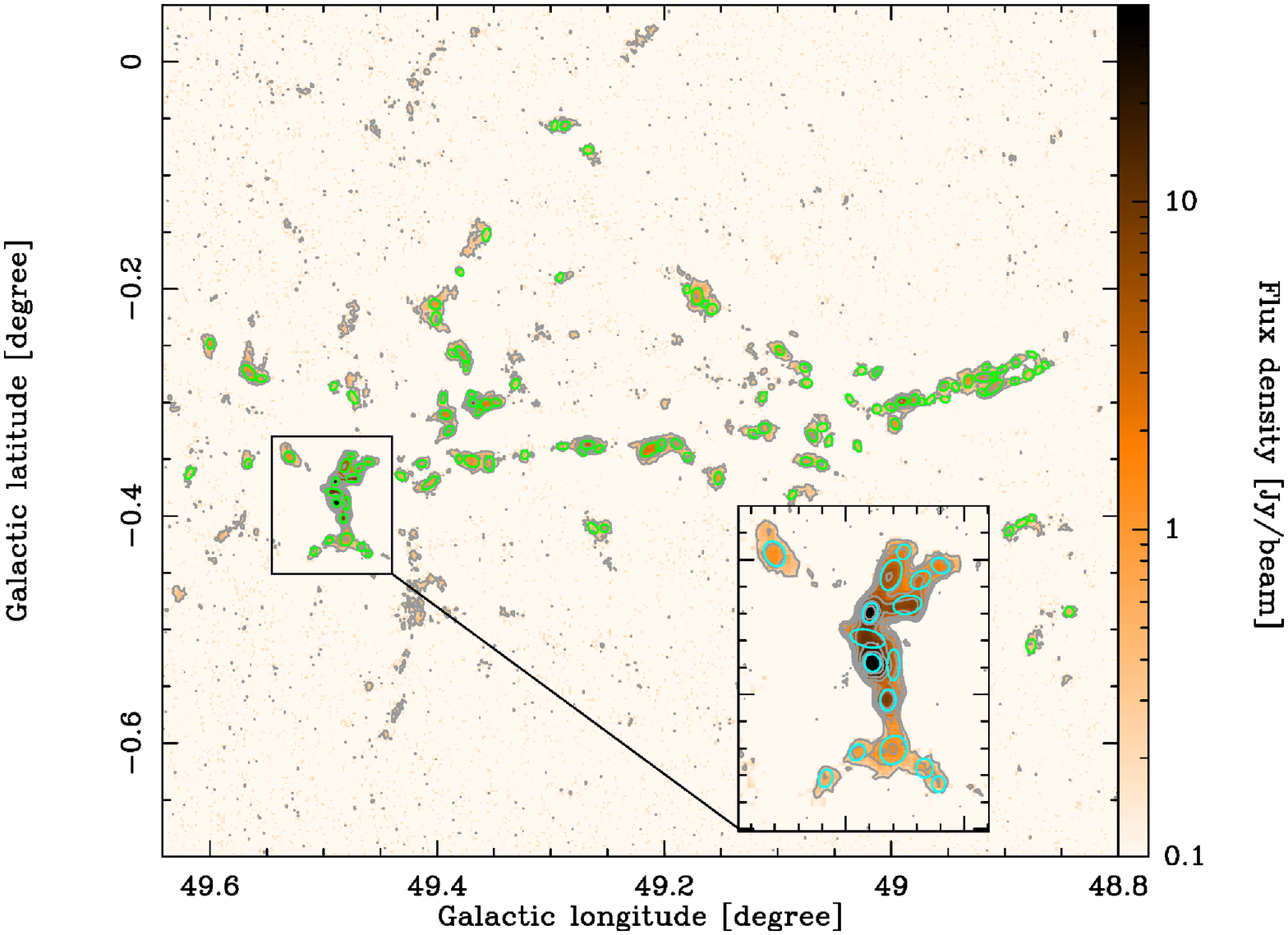}}}
    \caption{
                 {\bf Left:} 	Slice of the {\at} survey showing the active star-forming region, W51,
                                       which contains both 
                  			diffuse emission and compact sources. Color scale
					starts at a 0.1 Jy/beam and goes to 40 Jy/beam.
					Dark gray contours start from $3\sigma$ level (0.21 Jy/beam)
					and continue show $10\sigma$, $20\sigma$, $40\sigma$, $80\sigma$, 
					$160\sigma$, $320\sigma$ and $500\sigma$. 
					Green circles show the position of the extracted 
					sources. A zoom towards the brightest
					continuum sources is shown in the inset, also known
					as W51 Main. 
                 {\bf Right:} The same region as on the left panel but with from
                 			emission larger than 100\arcsec\ scale removed.
                 }
   \label{fig:filtering_comp}%
   \end{figure*}
%______________________________________________ 
\afterpage{}

%--------------------------
%    RESULTS: THE CATALOG
\section{Compact sources in the Galactic plane}\label{sec:prop}

Here we present  the result of the source extraction method described above. In total 
we have found \nsou\ sources in the $-60^\circ \le \ell \le +60^\circ$ longitude and 
$-1.5^\circ\le b\le+1.5^\circ$ latitude range, covering the I$^{\rm{st}}$ and IV$^{\rm{th}}$
Galactic quadrants. Table\,\ref{tab:table} presents a sample of the compact source 
catalog, the full catalog is available in electronic form\footnote{CDS reference, {\tt{http://atlasgal.mpifr-bonn.mpg.de}}}. 
The table structure is as follows: column (1) corresponds to the source id, then we give the source name (2), 
and the position in 
galactic coordinates (3, 4) and in J2000 equatorial coordinates (5, 6) (in hexagesimal format).
The physical parameters of the sources follow: the beam convolved major ($\Theta_{\rm maj}$) and minor 
axis ($\Theta_{\rm min}$) in arc seconds (7, 8), the position angle of the fitted Gaussian measured from north to east (9),
the average FWHM source size (beam convolved)\footnote{We determine the geometric average source size
from the geometrical mean of the Gaussian axes: $FWHM=\sqrt{\Theta_{\rm maj} \times \Theta_{\rm min}}$} (10) 
the peak flux (11) and the integrated flux (12) calculated assuming a 2D-gaussian shape for the sources:
S$_{\nu}$=F$_{\nu}\times (FWHM/FWHM_{\rm bm})^2$, where F$_{\nu}$ is the peak flux, 
$FWHM$ is the geometric size of the source,
$FWHM_{\rm bm}$ is $19\rlap{.}{''}2$, the $snr$ determined from the weight maps (13). %, 

As an example we show the identified sources in one of the most complex regions in the Galaxy in 
Fig.\,\ref{fig:filtering_comp}.
The left panel shows the large-scale emission of the W51 complex with a zoom on the most active site of star-formation 
associated with W51 Main (e.g.\,\citealp{Gaume1993}). The right panel shows
the filtered image which was used for the source extraction. 
This illustrates how the algorithm performs in complex environments: the filtering clearly removes large scale emission,
while all bright sources are recovered in the filtered maps.

We identified \nsouext\ sources in
in the extension between 280$^\circ \le \ell \le 300^\circ$ longitude and 
$-2.0^\circ \le b \le 1.0^\circ$ latitude range towards the Carina-arm.

\subsection{Flux Distribution}

%%%%%% PEAK FLUX

\begin{figure}[!h]
   \centering
   {\rotatebox{90}{\includegraphics[width=6cm]{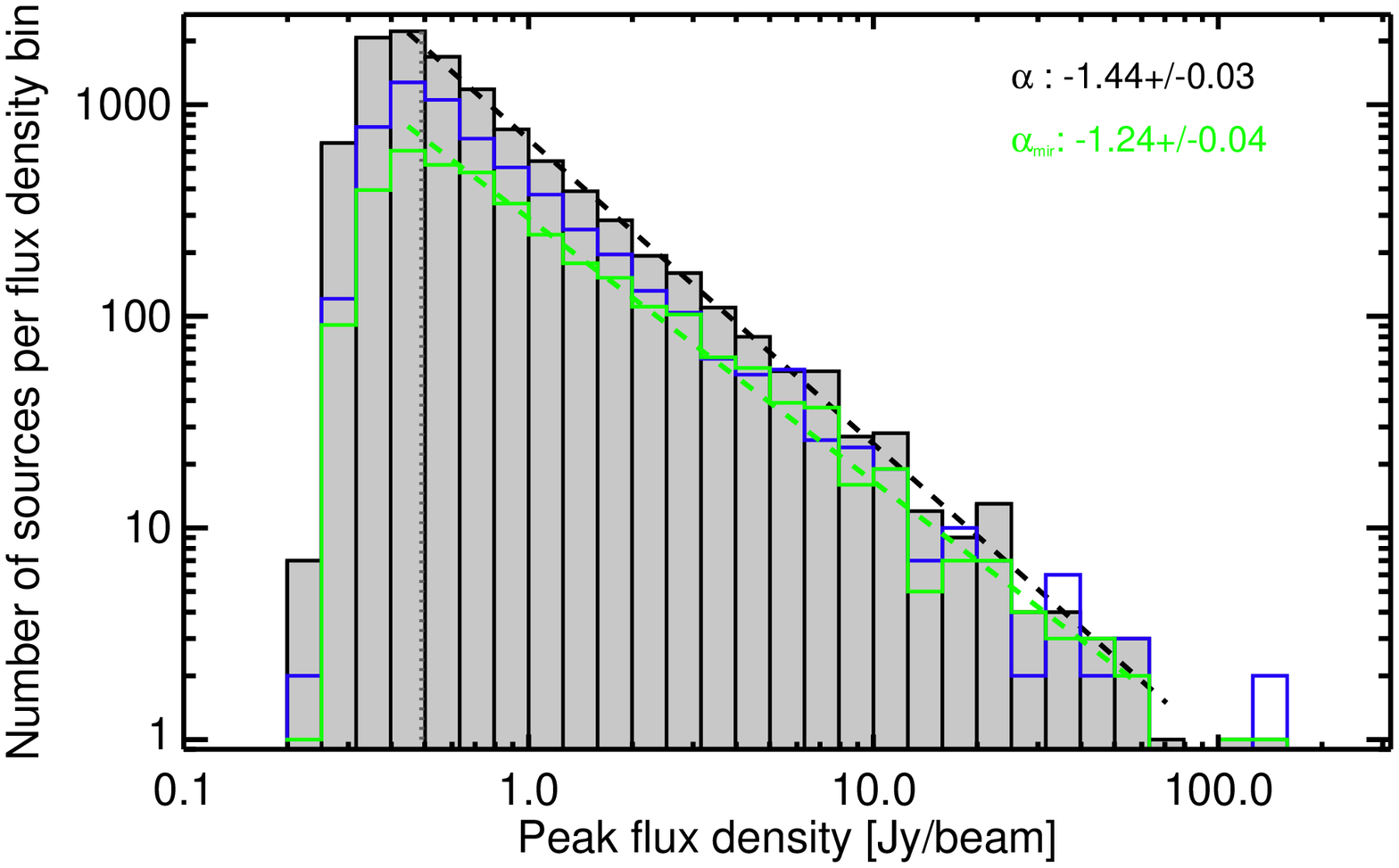}}}
    \caption{
                  Distribution of the peak flux density of the detected sources. The \at\
                  sources from the current paper are shown in black histogram with grey
                  color. Blue line shows the sources from \citet{Contreras2012}.
                  Black dashed line shows the measured slope of the distribution,
                  and green line indicates the star-forming \at\ clumps (see Sect.\,\ref{sec:ir} for details). A
                  black dotted line corresponds to our 7$\sigma$ completeness limit when using an average
                  of 70 mJy/beam noise level. Fit to the slope of the distributions are
                  shown in dashed colored lines. 
                  We note that here we
		 show the source count per bin, while in \citet{R2010} and \citet{Contreras2012}
		 the differential source count,  
		 $\Delta N/\Delta F_\nu$ is shown.
                  }
   \label{fig:histo_flux}%
   \end{figure}

We show the distribution of the peak flux densities in Fig.\,\ref{fig:histo_flux} 
with a linear least-square
fit to the flux bins above the completeness limit and determine a slope of 
$N/\Delta F_\nu \sim F_\nu^{\alpha}$, where $\alpha = -1.44\pm0.03$. 
As a comparison, 
we plot the
sources of \citet{Contreras2012} in blue, which shows an identical slope of $\alpha_{Cont} = -1.47\pm0.04$. 
Despite the differences of the
extraction methods, the two distributions 
compare well and we find the derived slopes to be consistent. 
 
\begin{figure}[!h]
   \centering
   {\rotatebox{90}{\includegraphics[width=5.3cm]{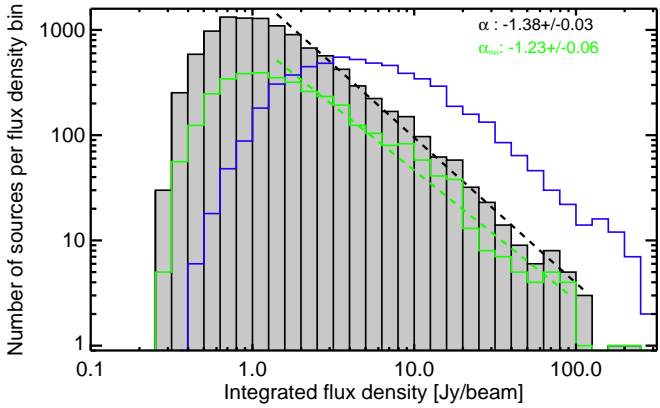}}}
    \caption{
                  Distribution of the integrated flux density of the detected sources.
                  Black dashed line shows the measured slope of the distribution,
                  green line shows the same distribution for the star-forming \at\ clumps (see Sect.\,\ref{sec:ir} for details).
                  The blue histogram shows the sources from \citet{Contreras2012}.
                  }
   \label{fig:histo_intflux}%
   \end{figure}

%%%%%% INTEGRATED FLUX
The distribution of the integrated intensity is shown in Fig.\,\ref{fig:histo_intflux}. Here we find larger 
differences
between the results obtained by \citet{Contreras2012} and our method.
We derive systematically lower integrated intensities for the individual sources
compared to \citet{Contreras2012}, and also we find fewer sources with large integrated flux. 
This illustrates the differences in the concept of the two catalogs, also discussed in
Sect.\,\ref{sec:comp}. Our method is better at identifying centrally condensed, more compact objects while
\citet{Contreras2012} extracts more irregular and diffuse structures of clumps and small clouds.
The observed shift in the distribution of integrated flux density is simply due to a combined effect of smaller
peak fluxes derived for the individual sources (due to a more efficient removal of the background emission)
and smaller size-scales (see also Sect.\,\ref{sec:size-distr}) compared to \citet{Contreras2012}.

{
We calculated the differential flux density distribution ($\Delta$N/$\Delta F_\nu$) which has a slope of $-2.35\pm0.05$. This compares well with the value of $-2.4$ obtained from the BGPS catalog \citep{R2010}, and $-2.2$ for the slope of the clump mass function measured by \citet{Tackenberg2012} for a more limited sample of ATLASGAL sources extracted with the Clumpfind algorithm. 
 }
%______________________________________________ 
%______________________________________________ 
\afterpage{}

%______________________________________________ 
%______________________________________________ 
%______________________________________________ 
   \begin{figure}[!htpb]
   \centering
   {\rotatebox{90}{\includegraphics[width=7cm]{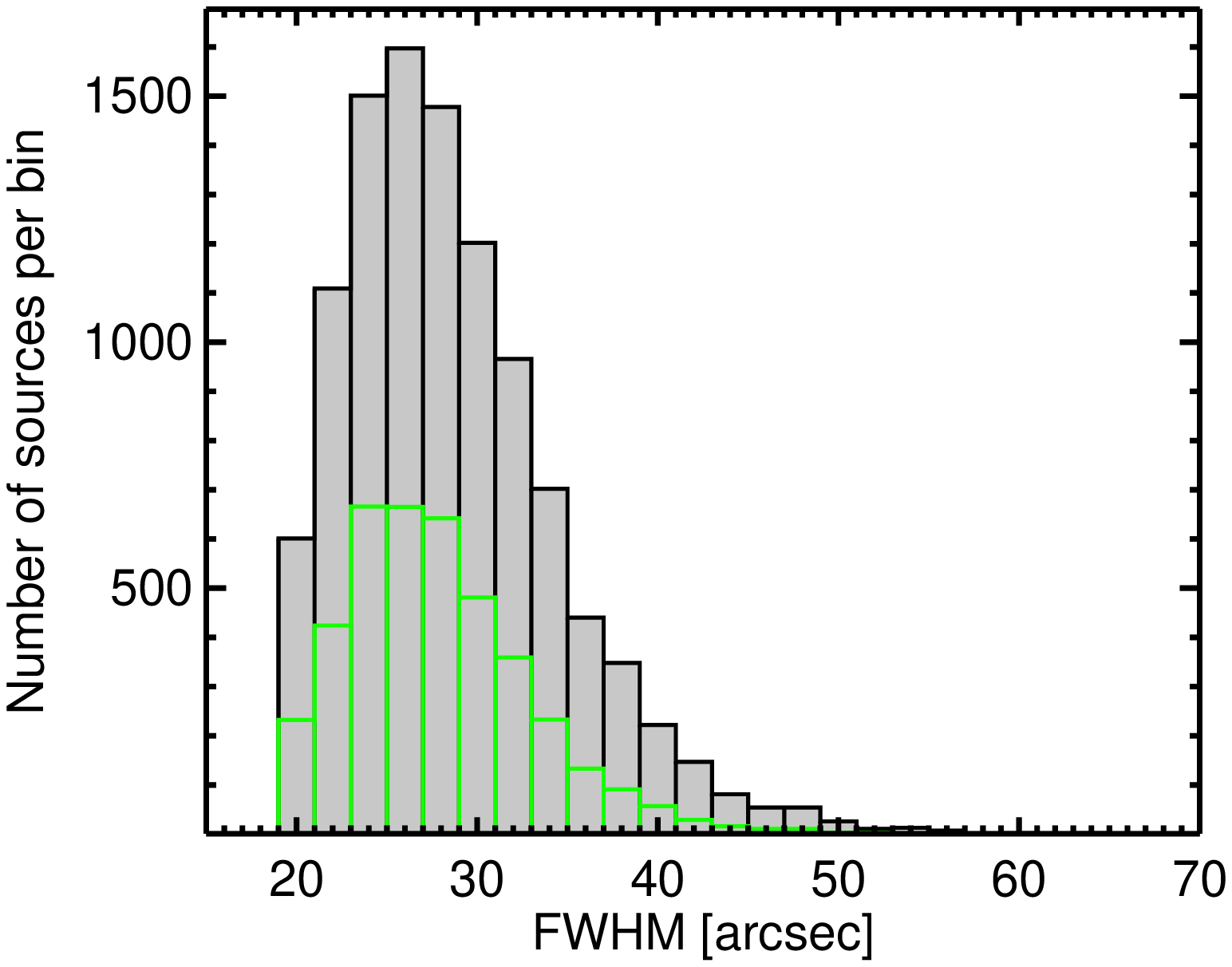}}}
   {\rotatebox{90}{\includegraphics[width=7cm]{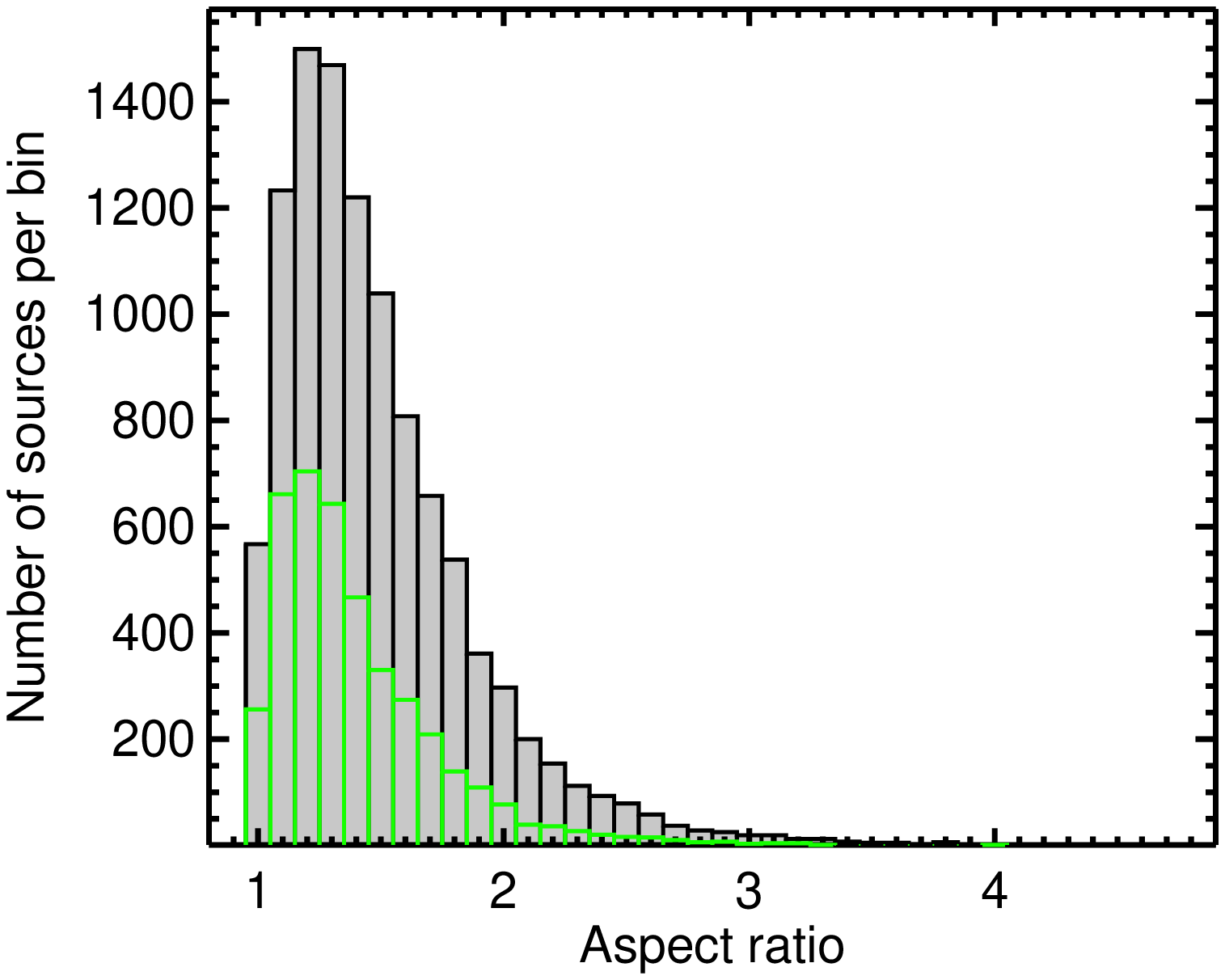}}}
    \caption{
                 {\bf Top:} 	
                  Distribution of the beam convolved FWHM sizes of the detected sources.
                  Green line shows the \at\ sources with mid-IR associations (see Sect.\,\ref{sec:ir} for details).
                 {\bf Bottom:} 	
                  Same as top panel for the aspect ratio of the detected sources. 
                  }
   \label{fig:histo_fwhm}%
   \end{figure}
%______________________________________________ 

\afterpage{}

\subsection{Sizes and aspect ratios}\label{sec:size-distr}

The distribution of the FWHM of the geometric sizes peak at $\sim$25\arcsec\ as shown 
in Fig.\,\ref{fig:histo_fwhm} (top-panel), suggesting that most of the sources are resolved by our $19\rlap{.}''2$ beam.
Although we measure source sizes up to 55\arcsec\, the majority of the sources 
exhibit compact characteristics with sizes of $<35${\arcsec}, and with a median of 27{\arcsec}. 

In the lower panel of Fig.\,\ref{fig:histo_fwhm} we present a plot showing the aspect ratio distribution of the sample. This has a mean of 1.5 with a median value of 1.4
and decreases with increasing peak flux.
For example considering sources above 5~Jy/beam peak flux density, the mean aspect ratio drops to 1.3 with 
a median of 1.2, while the \citet{Contreras2012} sources have a median of 1.5 for these brightest sources.
Again, our extraction method was optimized for compact dust condensations.  
{The brightest sources generally exhibit a rather spherical morphology.}
Elongated sources are
more likely to be inhomogeneities in the cloud structure mimicking embedded sources 
or simply the blend of several 
sources. 

%_____________________________________________________________
%                                 A rotated Table in landscape  
%  In the preamble, use:   \usepackage{lscape}
%-------------------------------------------------------------
 \begin{table*}
 \caption{Dust condensations identified in the {\sl ATLASGAL} survey.}\label{tab:table}
\centering
\begin{tabular}{cccccccccccc} 
\hline
Id & Name & Ra & Dec & $\Theta_{\rm maj}$ & $\Theta_{\rm min}$ & PA & FWHM & $F_\nu$ & $S_\nu$ &  snr \\
& & [J2000] & [J2000] & [arcsec] & [arcsec] & [$^\circ$] & [arcsec] & [Jy/beam] & [Jy] & & \\
\hline \hline
         1   &      G300.1627$-$0.0899   &  12:27:08.6   &  $-$62:49:51.6   &        33   &        23   &        97   &        28   &      1.18   &      2.51   &     12.44   \\
         2   &      G300.2169$-$0.1106   &  12:27:35.9   &  $-$62:51:24.2   &        31   &        26   &        21   &        29   &      1.04   &      2.40   &     11.14   \\
         3   &      G300.5038$-$0.1763   &  12:30:03.5   &  $-$62:56:50.4   &        25   &        24   &        88   &        24   &      3.31   &      5.56   &     32.88   \\
         4   &      G300.8250+1.1517   &  12:33:40.9   &  $-$61:38:51.0   &        32   &        25   &        42   &        28   &      1.52   &      3.43   &     12.26   \\
         5   &      G300.9097+0.8811   &  12:34:14.5   &  $-$61:55:23.4   &        39   &        29   &        28   &        34   &      1.36   &      4.32   &     11.63   \\
         6   &      G300.9688+1.1456   &  12:34:53.2   &  $-$61:39:47.3   &        32   &        28   &       125   &        30   &      6.54   &     16.16   &     53.74   \\
         7   &      G301.0140+1.1137   &  12:35:15.0   &  $-$61:41:52.2   &        32   &        25   &        59   &        29   &      1.09   &      2.51   &     10.47   \\
         8   &      G301.1164+0.9596   &  12:36:02.1   &  $-$61:51:28.6   &        51   &        34   &       $-$21   &        41   &      1.61   &      7.71   &     13.46   \\
         9   &      G301.1169+0.9771   &  12:36:02.8   &  $-$61:50:25.9   &        40   &        33   &        80   &        36   &      1.61   &      5.89   &     14.69   \\
        10   &      G301.1365$-$0.2256   &  12:35:35.2   &  $-$63:02:31.9   &        26   &        22   &       102   &        24   &     21.43   &     34.55   &    120.43   \\
        11   &      G301.1385+1.0092   &  12:36:14.7   &  $-$61:48:35.0   &        48   &        33   &       127   &        40   &      1.06   &      4.68   &     10.01   \\
        12   &      G301.6798+0.2456   &  12:40:33.1   &  $-$62:35:59.1   &        28   &        24   &        81   &        26   &      1.41   &      2.59   &     11.77   \\
        13   &      G301.7313+1.1038   &  12:41:17.6   &  $-$61:44:39.6   &        39   &        31   &       157   &        35   &      2.07   &      6.89   &     18.53   \\
        14   &      G301.7414+1.1013   &  12:41:22.6   &  $-$61:44:50.4   &        38   &        28   &       128   &        33   &      1.29   &      3.90   &     12.53   \\
        15   &      G301.8138+0.7811   &  12:41:53.3   &  $-$62:04:12.1   &        27   &        26   &        44   &        26   &      1.46   &      2.80   &     11.38   \\
        16   &      G302.0208+0.2517   &  12:43:31.0   &  $-$62:36:22.0   &        35   &        25   &        15   &        30   &      2.06   &      5.14   &     17.54   \\
        17   &      G302.0318$-$0.0607   &  12:43:31.7   &  $-$62:55:07.4   &        30   &        27   &        18   &        28   &      2.62   &      5.95   &     24.47   \\
        18   &      G302.0327+0.6254   &  12:43:43.0   &  $-$62:13:58.6   &        27   &        21   &        65   &        24   &      1.49   &      2.39   &     14.40   \\
        19   &      G302.3912+0.2804   &  12:46:44.4   &  $-$62:35:11.3   &        37   &        27   &       $-$28   &        32   &      2.12   &      6.08   &     22.71   \\
        20   &      G302.4861$-$0.0310   &  12:47:31.4   &  $-$62:53:58.0   &        30   &        25   &        74   &        27   &      1.89   &      3.96   &     17.65   \\
\hline
\end{tabular}
 \tablefoot{The full table is available only in electronic form at the CDS via anonymous ftp to cdsarc.u-strasbg.fr (130.79.125.5) or via http://cdsweb.u-strasbg.fr/cgi-bin/qcat?J/A\&A/.}
\end{table*}

\afterpage{}

%--------------------------
%    ANALYSIS OF CATALOG

\section{Mid-IR diagnostics to characterize star-formation activity}\label{sec:ir}

Mid-IR emission traces warm dust, generally heated by embedded 
protostars, (M)YSOs, OB stars or evolved stars undergoing mass-loss. 
{Dust surrounding evolved stars
is less frequently detected than from protostars.}
Only a handful of such objects up to a distance of 5~kpc are detected at
870~\mum\ with LABOCA within the sensitivity limit (on average $\sigma$\,$\sim$\,$70$~mJy/beam) 
of \at\ \citep{Ladjal2010}.
{Since}
the majority of the evolved stars are expected to emit weakly in the sub-millimeter 
the \at\ sources {are expected to} be dominated by young, star-forming
objects. As a consequence, their association with mid-IR sources can generally be used 
as a proxy to trace on-going star-formation. 

MSX surveyed the Galactic plane at mid-IR wavelengths with an angular resolution (18{$\rlap{.}''$}3 at 8-21~\mum)
similar to that of the \at\ survey. 
With a higher spatial resolution and sensitivity, surveys with the Spitzer 
and WISE space telescopes uncover a larger population of mid-IR sources in the Galactic plane.
The multi-color Spitzer GLIMPSE and MIPSGAL surveys probe fainter, thus even younger or lower mass
protostars and they represent 
to date the highest spatial resolution and brightness sensitivity at mid-IR wavelengths.
However, due to source confusion it is not straightforward to unambiguously associate
Spitzer sources with dust peaks and the line of sight contamination from chance alignments with field stars is high. 
MIPSGAL at 24~\mum\ has  
only a factor of 2 higher sensitivity and an angular resolution of 6\arcsec\ \citep{Carey2009} compared to WISE at 22~{\mum} with 12{\arcsec} resolution. 
Since here we only study the statistical properties 
of the mid-IR associations with \at\ sources, the best combination of surveys are
the MSX and WISE datasets with available point source catalogs. 
They cover a continuous brightness sensitivity 
from the brightest sources in the Galactic plane to weaker, deeply embedded young protostars
and have a comparable angular resolution as {\at} at 870~{\mum}.
 
\citet{Contreras2012} used data only from MSX at 21.3~\mum\ to give a crude estimate on the
proportion of star-forming clumps of {\at}. 
We base the characterization of the 
mid-IR content of dust clumps using the 21.3~$\umu$m filters on MSX (E-band), and the
22~$\umu$m filter on WISE (band 4). The former provides a sensitivity
down to $\sim2-6$ Jy \citep{Egan99}, while the latter has 
a 5$\sigma$ point source sensitivity in unconfused regions of 6 mJy \citep{Wright2010}. 
{MSX suffers from saturation only for the brightest, 
few hundred Jy bright sources, 
and although the WISE 22~$\umu$m band starts to saturate at $\sim$10~Jy reliable photometry can be extracted up to 330~Jy by fitting the wings of the PSF \citep{Cutri2012}.
} 
Therefore the two catalogs have a substantial overlap in sensitivity (see also Sect.\,\ref{sec:wise-msx})
and combining these gives a continuous 21-22~\mum\ sensitivity from 6~mJy to several hundred Jy, covering 5 orders of magnitudes.

{
In Sect.\,\ref{sec:msx} we compare source counts using the 21~\mum\ as well as  
the more sensitive 8~{\mum} (A-Band) information from MSX. This dataset is relevant
for the brightest sources and is directly comparable to the mid-IR analysis of 
\citet{Contreras2012}.
In Sect.\,\ref{sec:wise} we use the more sensitive WISE point source catalog to complement this 
sample and analyze the color properties of the {\at}-mid-IR associations. }
We merge these two 
samples to derive the final number of star-forming \at\ sources in Sect.\,\ref{sec:wise-msx} and
study their characteristic colors in Sect.\,\ref{sec:cc}.

\afterpage{}

\subsection{Association with MSX}\label{sec:msx}
We have searched for MSX counterparts of \at\ sources within a 60\arcsec\ search radius using the 
MSX Point Source Catalog (PSC) v2.3 \citep{Egan2003}. In the next step we determined
 a more appropriate matching radius by fitting a 
Gaussian to the
normalized distribution of the angular separation between the MSX sources and \at\ peak positions and then 
take 3$\sigma$ as a match radius corresponding to $\sim${\distlimitmsx}{\arcsec} 
(Fig.\,\ref{fig:histo_dist}).

We find a total number of {\totalmsxmatch} sources to have a corresponding MSX source
within this angular separation and with detectable flux at 21.3~{\mum}. This corresponds to
{\totalmsxmatchpercent}\% of the total \at\ sources (Table~\ref{tab:table2}) and is a more conservative
estimate than the 34.8\% found by \citet{Contreras2012}. 
Using a similar 30{\arcsec} search radius as \citet{Contreras2012},
we find the same fraction of the sources with an MSX counterpart. 

We compare the color properties of these {\totalmsxmatchpercent}\% matches with the
color selection criteria used in the literature to search for embedded massive objects.
We find that 97\% of the sources fulfill the minimum criteria of
$\rm {F_E > F_D}$, which are the flux measurements at 14.65~{\mum} (D-band) and 21.3~{\mum} (E-Band)
and corresponds to a rising SED towards longer wavelengths.
\citet{Schuller2006} use a selection criteria of
$\rm {F_E/F_D} \ge 2$, and from our matches we find that 1165
fulfill this, which is 77\% of the sources with reliable flux measurements in both of these bands. 
\citet{Lumsden2002} adopts a less conservative requirement by using $\rm {F_E} \ge 2\times \rm{F_A}$ 
and we find that 86\% of our matches fulfill this, where $\rm{F_A}$ corresponds to the flux in the
A-band of MSX (8~{\mum}). We note that extending
the match radius to 30\arcsec\ leads to a lower fraction of sources fulfilling these color 
properties, suggesting that the larger match radius would contain more sources with chance line-of-sight alignment
or merging of nearby sources with different evolutionary status.
Our position restricted estimate of the mid-IR content of \at\ sources therefore corresponds 
to mostly embedded objects, while it is likely that the \citet{Contreras2012} 
matches include more significant contamination of chance alignment sources.

Since MSX has the highest sensitivity in the 8~\mum\ band (0.1-0.2 Jy, \citealp{Egan2003}), we
derive source counts of \at\ matches in this band as well following the same method 
as described above. Similarly as above, here we derive
a {\distlimitmsxeight}\arcsec\ distance-limit and find \totalmsxmatcheight\ sources matched to dust condensations. 
This number represents {\totalmsxmatchpercenteight}\% of the \at\ sources. Out of these,
 \totalmsxmatchtwoband\ fulfill our selection criteria of the 21~\mum\ matches, the rest are new associations. 
 From these sources (only detected at 8~\mum\ and not at 21~{\mum}) we 
 find 130 sources which are only detected at the shortest wavelengths (i.e. not detected in the C and D bands), 
 suggesting that some of these new
 associations may be contaminated with evolved stars or PDR emission features. 
 Therefore, despite of the higher sensitivity, 
 we do not significantly increase the fraction of star-forming clumps
using the 8~\mum\ of MSX instead of the 21~{\mum} band.
 
We note that between 8~\mum\ and 22~\mum\ 
not only the heated dust from the inner envelope emits, but there is diffuse emission from 
polycyclic aromatic hydrocarbon (PAHs) 
in the surrounding nebulosity. For the most complex region in the Galactic center \citet{Schuller2006}
estimates 10\% of the sources to be actually diffuse PDR emission rather then protostellar sources.

% ABOUT THE WISE CATALOG & PHOTOMETRY:
% http://wise2.ipac.caltech.edu/docs/release/allsky/expsup/sec5_3.html#snr
\subsection{Association with WISE}\label{sec:wise}

Since WISE is  more sensitive than MSX in all bands 
(by a factor of 500 at 21~$\umu$m), 
there is a higher probability of chance alignment between WISE and \at\ sources. 
These can be field stars, low-mass YSOs and brown dwarfs from the foreground, 
and evolved stars. In the following we use the WISE point source catalog \citep{Wright2010}, and 
with a 30\arcsec\ search radius we find 
a 96\% match between \at\ and WISE sources.  
{
We investigated the color properties of these first associations based only on
positional coincidence 
in order to get a first hint on their nature. 
In fact, as expected, we found that such associations are largely contaminated
by chance alignment with field stars.
}

\begin{figure}[!htpb]
   \centering
   {\rotatebox{90}{\includegraphics[width=7cm]{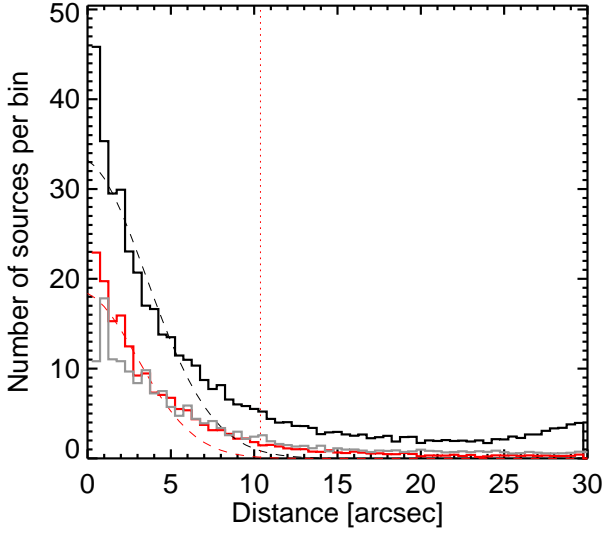}}}
    \caption{
                  Distribution of the normalized distance between the dust continuum peaks and the mid-IR sources within 30{\arcsec}.
                  Black line shows all sources from the WISE catalog, while the red line corresponds to the distribution of 
                  only the red sources (see text for details). As a comparison the same distribution for all MSX sources is shown 
                  in gray.
                  Dashed lines show the Gaussian fits to these distributions, the maximum angular distance between
                  mid-IR sources and \at\ sources was determined as 3$\sigma$ from this and is 
                  shown in dashed line.}
   \label{fig:histo_dist}%
   \end{figure}

It is necessary therefore to determine a more reliable estimate of the number
of embedded mid-IR sources associated with \at\ clumps by 
eliminating chance alignments as far as possible. For this we first limited 
the search radius to find the best potential matches (Fig.\,\ref{fig:histo_dist}) using a similar strategy 
 as was discussed above for MSX (see Sect.\,\ref{sec:msx}). 
However, given the large number of WISE matches,
{we study the distribution of the angular offset between mid-IR and
dust peaks only using  
embedded sources 
}
with characteristic red colors  (F$_{4.5{\mu}m}$/F$_{3.6{\mu}m} > 0.75$ and 
F$_{4.5{\mu}m}$/F$_{12{\mu}m} < 1$\footnote{This corresponds to $[3.6]-[4.5] > 0.24$ and
$[4.5]-[12] < 0.0$.}, S. Lumsden priv.\,comm). 
We arrive to a similar search radius of {\distlimit}{\arcsec}, as for MSX. 
{
For the eventual matching between the \at\ and WISE positions we do not use any color requirements, but the above derived limiting angular offset.
We then put a further} constrain on the matched sources by requiring a S/N~$>$~10 
for the 22~\mum\ band WISE measurements in order to ensure their nature of
embedded sources instead of a chance alignment with a field star or evolved stars, which are not detected 
at 22~{\mum} (see e.g.\,\citealp{Schuller2006}). 
We discuss in more detail the 
justification and the color properties of these matches in Sect.\,\ref{sec:cc}.

With these selection criteria we find \totalwisematch\ \at\ sources 
to have a WISE association. This corresponds to {\totalwisematchpercent}\%
of all the \at\ sources. 

From these matches we find 
that only a negligible fraction have S/N lower than 2 or a bad quality flag\footnote{
Objects flagged as "D", "H", "O", or "P" are
likely to be spurious sources according to the WISE catalog description.} 
at 12~{\mum} (band 3). Therefore the majority (89\%) of our 
matches are detected at least in two WISE bands. Interestingly, 
we find that 97\% of these matches have a measured flux at 4.5~{\mum} (band 2) with a S/N
greater than 2 (this value drops to 93\% when restricting to S/N $>$ 10). In the
 4.5~\mum\ and the 12~\mum\ bands we find that 86\% of the 
sources are detected with S/N $>$ 2. 
At the shortest wavelengths (3.6~{\mum}, band 1) 89\% of the matched sources have fluxes
above 2 S/N. Altogether 82\% of our matches (2559 sources) have fluxes measured in all four bands above 
S/N $>$ 2. 

From these 2559 sources, 2481 (97\%) fulfill the $[3.6]>[4.5]>[12]>[24]$ criteria 
corresponding to rising SEDs in the mid-IR wavelength range. We find that
96\% of them also fulfill the red object selection criteria used above 
(F$_{4.5{\mu}m}$/F$_{3.6{\mu}m} > 0.75$ and 
F$_{4.5{\mu}m}$/F$_{12{\mu}m} < 1$), while 99\% of them fulfill the
 F$_{22{\mu}m}>2\times$F$_{12{\mu}m}$ criteria, which is similar to
 that of \citet{Lumsden2002}.
 The majority of these
associations are therefore reddened, most likely embedded star-forming objects. We study in more
detail the color properties of this sample in
Sect.\,\ref{sec:cc}. 

{
In the following we refer to \at\ sources
with associated mid-IR emission as star-forming clumps, while
to the rest of the sources
as quiescent. 
They represent a lower limit to the true proportion of star-forming sources
since there is likely to be a number of clumps associated with embedded sources that fall 
below the 10$\sigma$ detection threshold we used for the WISE catalog.
}

\subsection{Correlation between 21~\mum\ and 870~\mum\ fluxes of star-forming \at\ sources}\label{sec:wise-msx}

\begin{figure}[!htpb]
   \centering
   {\rotatebox{90}{\includegraphics[width=7cm]{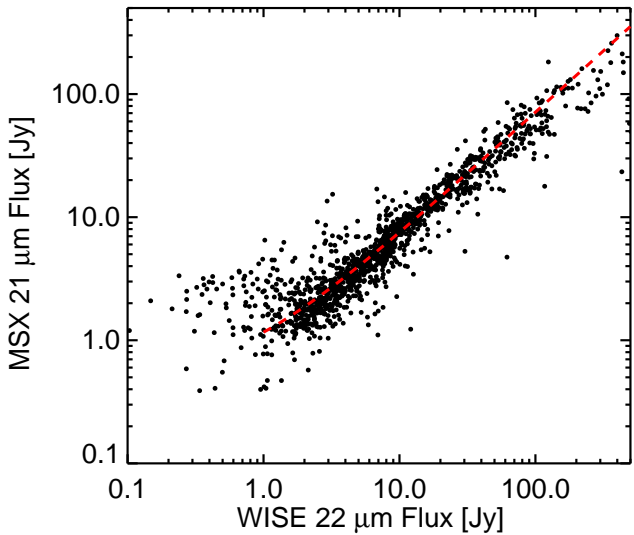}}}
    \caption{
                  Comparison of the WISE 22~\mum\ and MSX 21~\mum\ fluxes for the sources
                  that are found in both catalogs. Red dashed line shows a robust linear fit with a slope of $0.71\pm0.01$.
                  }
   \label{fig:msx_wise_fluxcorr_check}%
   \end{figure}

Comparing the $21-22$~\mum\ flux densities between the MSX and WISE sources, we find that \overlap\
sources are found in both catalogs. To check our independently
matched \at\ sources with MSX and WISE, we show in Fig.\,\ref{fig:msx_wise_fluxcorr_check} the
comparison of the 21~\mum\ flux density from the two catalogs\footnote{We convert the magnitudes given in the WISE catalog using the zero-level
fluxes from \citet{Wright2010}, a color correction factor of 1.0 and a correction factor of 0.9 from \citet{Cutri2012}.}. 
These independently derived flux densities from both surveys show a good
correlation 
{
(with a Pearson correlation coefficient of 0.88 for sources between $10-80$~Jy fluxes)}, 
however it is also clear that below $\sim$10~Jy the flux measurements of MSX show an increased
scatter. This is fully consistent with the varying noise level in the MSX data at different positions
in the Galactic plane leading to less reliable 
measurements at the lower flux limit. For the 
brightest sources the WISE photometry seems to give systematically higher fluxes.

Taking into account this overlap between the two catalogs we arrive at a total number 
of {\totalmirmatch} associations at 21-22~\mum\  with \at\ sources.
This is {\totalmirmatchpercent}\% of the total number of dust condensations
from the \at\ catalog. This result compares well
with \citet{Contreras2012}, who derived an upper limit on the fraction of
star-forming clumps of $\sim$50\% 
using a less robust test for mid-IR emission with MSX. Here
our matching criteria is more conservative to associate 
embedded sources with the \at\ clumps (see also Sect.\,\ref{sec:cc}) resulting in a more robust sample.
Therefore our derived fraction is a strict lower limit of the fraction of
star-forming dust clumps in the inner Galaxy. 

\begin{figure}[!htpb]
   \centering
   {\rotatebox{90}{\includegraphics[width=7cm]{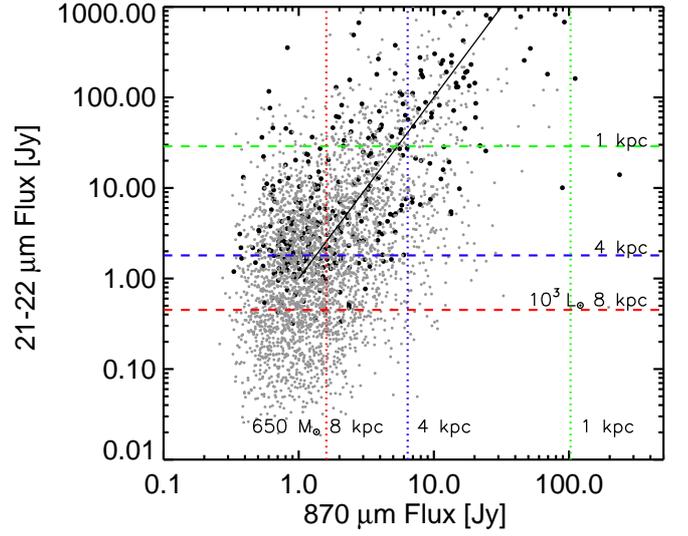}}}
    \caption{
                  Integrated flux density from \at\ versus the flux density at 22~$\umu$m. Gray dots
                  show the WISE, black dots show the flux measurement from MSX associations.
                  For sources found in both the WISE and MSX catalogs, only the WISE flux is shown.
                  A trend is seen, the brightest sources having higher 22~$\umu$m flux (see text for more details).
                  Dashed lines show the flux-limit at 22~\mum\ corresponding to a B3 protostar at
                  1, 4 and 8~kpc. The mass limit for high-mass star-forming clumps 
                  at the same distances
                  is shown in dotted lines.
                  Black solid line shows a relation of $S_{22{\mu}m}\sim S_{870{\mu}m}^2$.
                  }
   \label{fig:msx_wise_fluxcorr}%
   \end{figure}

Fig.\,\ref{fig:msx_wise_fluxcorr} shows the measured 22~\mum\
flux densities from WISE and MSX versus the 870~\mum\ flux density from {\at}. 
The mid-IR flux density seems to increase with increasing sub-millimeter flux density,
in particular the brightest dust clumps exhibit brighter 22~\mum\ fluxes. 
We determine the Pearson coefficient to check for correlation between these two parameters using the WISE sources with fluxes between $10-300$~Jy, where the photometry is
the most reliable. 
We obtain a correlation coefficient of 0.29 with a significance value of $<0.001$ which
suggests a weak correlation between these parameters. As a comparison
we plot a relation of $S_{22{\mu}m}\sim S_{870{\mu}m}^2$ to highlight the trend seen in the data.
This may reflect that
the more massive and luminous YSOs form within more massive dust clumps producing
the distinct tail of the distribution, however 
distance and temperature effects may also play a role.

We also show in Fig.\,\ref{fig:msx_wise_fluxcorr} the 22~\mum\ flux density limit corresponding to a
ZAMS B3 type star with a bolometric luminosity of $10^3$ L$_{\odot}$ at 
1, 4, and 8~kpc distance based on the color properties defined by \citet{Wood1989}
and adopted here following \citet{M07}.
 We also show the mass-limit required to host 
MDCs, precursors to high-mass stars, corresponding to 650~{\msol} (see Sect.\,\ref{sec:phys})
at these distances. We adopt here a dust temperature of 18~K,
because at larger scales the dust
temperature is dominated by the interstellar radiation field at approximately this value~\citep{Bernard2010}.
Similar values have been used  
by \citet{M07}, who adopt T$=20$~K for MDCs and \citet{Wienen2012} as well, who determine
a gas kinetic temperature of 17~K for a sample of \at\ clumps. 

For 1, 4 and 8~kpc this mass-limit translates to a 
sub-millimeter flux density limit of $102.6, 6.4, 1.6$~Jy, respectively in order 
to potentially form high-mass stars.
{
To eventually determine the fraction of massive clumps
distance information is essential \citep{Wienen2012}.
}

\begin{figure}[!htpb]
   \centering
   {\rotatebox{90}{\includegraphics[width=7cm]{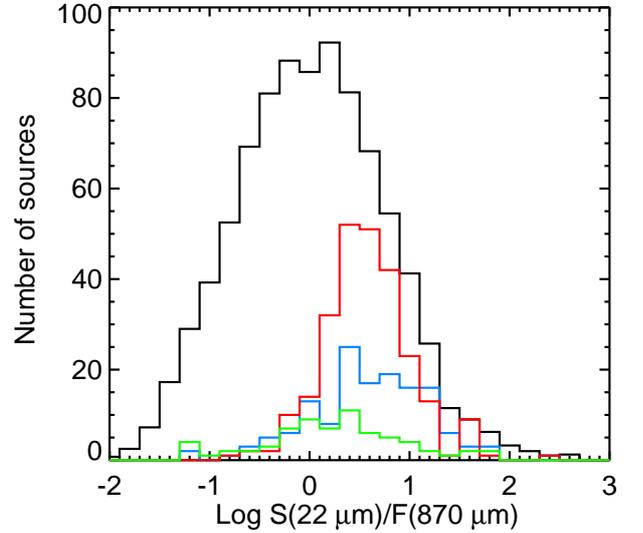}}}
    \caption{
                  Histogram of the ratio of 22~$\umu$m and 870~$\umu$m peak fluxes. Black line
                  shows the whole sample (scaled down by a factor of 4), red line shows the RMS associations, blue corresponds to
                  the CORNISH sources, while green indicates the MMB sample.
                  }
   \label{fig:fluxratio}%
   \end{figure}

Fig.\,\ref{fig:fluxratio} shows the histogram of the ratio of the 22~\mum\ and 870~\mum\
fluxes. We find a broad range for this flux ratio with a peak around 1. 
We plot also the {\at}-MMB, the {\at}-CORNISH and the {\at}-RMS 
associations from \citet{U2013mmb} and \citet{UCornish2013} respectively.
The MMB sources tend to show a smaller flux ratio, than the CORNISH and RMS samples. 
We note, however, that the CORNISH sample can be considered as a subsample of
the RMS sources as they consist of MYSOs and {\hii} regions. The peak of the
distribution for all star-forming \at\ sources is shifted towards smaller flux ratios
than for the MMB and more evolved samples suggesting an evolutionary trend
with lower flux ratios corresponding to colder sources.

%______________________________________________ 
%______________________________________________ 
%______________________________________________ 
%______________________________________________ 

%______________________________________________ 
%______________________________________________ 

%______________________________________________ 
%______________________________________________ 

%______________________________________________ 
   \begin{figure}[!htpb]
   \centering
   {\rotatebox{90}{\includegraphics[width=0.9\linewidth]{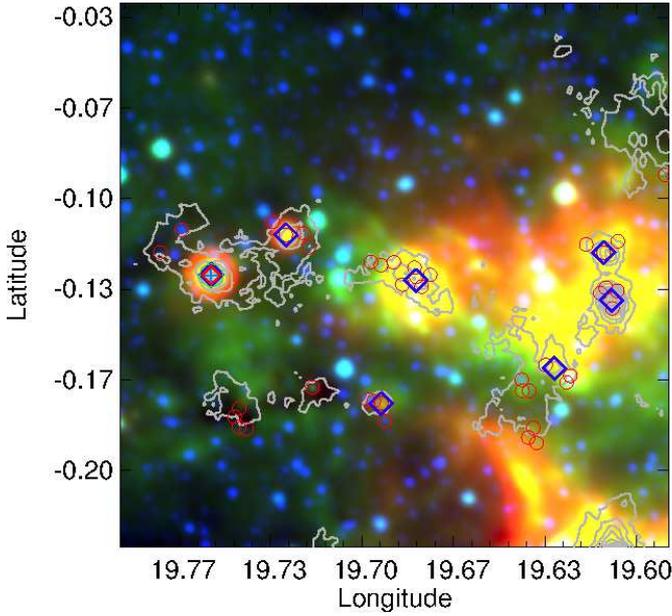}}}
    \caption{
                  {
                  An example showing the distribution of \at\ and WISE point sources 
                  on a color composite image from WISE (blue: 3.4~$\umu$m,
                  green: 12~$\umu$m, red: 22~$\umu$m) with the 870~$\umu$m contours overlaid from {\at}.
                  Red circles show all the WISE sources in within 30\arcsec\ of the \at\
                  sources and dark blue diamonds indicate the associated \at\ and WISE positions
                  likely hosting embedded sources.
                  }
                  Light-blue crosses mark the position of CORNISH sources \citep{UCornish2013}, while
                  green circles show the positions of all MMB sources of \citet{U2013mmb}.
                  Thick red circles show the RMS sources (Urquhart et al\., in prep.). Note that
                  the source G19.7713$-$0.1150 likely hosts already an {\hii} region as it is
                  part of the CORNISH sources of \citet{UCornish2013}, but it is also a known
                  RMS source and has associated methanol maser emission from the MMB-survey.
                  }
   \label{fig:wise_ex}%
   \end{figure}
%______________________________________________ 

\subsection{Infrared colors of ATLASGAL sources}\label{sec:cc}

We use the WISE matched \at\ sources which have measured fluxes above 2 S/N in 
all bands (2559 objects, $\sim$25~\% of the identified \at\ sources) 
to study their characteristic colors. 
To demonstrate the results of these matches we
show a 3-color composite image in Fig.\,\ref{fig:wise_ex} with the star-forming \at\ sources
indicated together with all the WISE sources found within a 30\arcsec\ search radius.
We also use the list of \at\ matched
MMB associations \citep{U2013mmb}, CORNISH sources \citep{UCornish2013} and
RMS sources (Urquhart et al., {\sl in prep}) to illustrate the location of these sources 
with a known evolutionary stage on these plots. In general there are several WISE sources
found within the 30\arcsec\ search radius, and a substantial fraction of them are field stars
appearing in blue on the figure. The requirement of emission at 22~{\mum}, however, eliminates
a major fraction of these sources. 

\begin{figure}
   \centering
   {\rotatebox{90}{\includegraphics[width=0.7\linewidth]{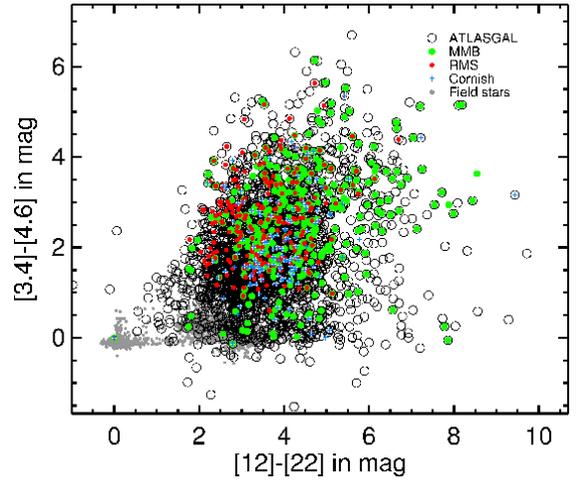}}}
    \caption{
                  Color-color plots of WISE matched {\at} sources
                  with measurable WISE fluxes in all bands.
                  Field stars from a randomly chosen $\sim0.5^{\circ}\times0.5^{\circ}$ area
on the sky around the central coordinates of $\ell$=280$^\circ$, $b=15^\circ$
are shown in gray.
                }
   \label{fig:wise_cc}%
   \end{figure}

As shown in Fig.\,\ref{fig:wise_cc} (as well as Figs.\,\ref{fig:cc1}-\ref{fig:cc4})
the majority of these matches separate well from field stars in the color-color space, suggesting that the contamination
of chance alignments is small. Clearly, the mid-IR sources associated with \at\ clumps
are deeply embedded showing characteristic reddened colors. Their average colors are
summarized in Table\,\ref{tab:table-col}. The outlying points on the
diagram likely correspond to mismatched sources, where our method erroneously associated
a nearby field star to the \at\ source. This may happen in complex regions affected by saturation, where
the 22~\mum\ flux measurement in the WISE catalog is flagged, but there is a nearby red source which
can be a chance alignment.

Comparing
the positions of embedded {\hii} regions, RMS  and MMB sources in Fig.\,\ref{fig:wise_cc}
we see that especially the former two occupy a well defined region with
colors between 1.5 $<$ [3.4]-[4.6] $<$ 6 and 2 $<$ [12]-[22] $<$ 8. These sources
correspond to more evolved stages of high-mass star-formation, {\hii} regions and MYSOs. 
The {\hii} regions seem to occupy a well-determined region of the plot, while the 
RMS sources show more dispersion, likely because this sample is a 
combination of MYSOs and {\hii} regions. The largest scatter is seen for the MMB sources, which 
includes many of the more extreme reddened objects than the two other samples.  
This is consistent with the general view on the occurrence
of Class II methanol masers, which predominantly trace high-mass star-formation,
and are also frequently found around {\uchii} regions (e.g.\,\citealp{U2013mmb}). 

A large fraction of the unclassified \at\ sources are spread in the color-space, similarly to 
the MMB sources 
suggesting that they may host the early 
stage of intermediate- and high-mass star-formation.
There are several sources with moderately reddened colors and they are likely 
low- to intermediate-mass nearby star-forming cores.
Using this plot we identify a number of highly reddened extreme sources which
have not been revealed by the other surveys. These may be of potential interest to study
in greater detail.

%______________________________________________ 
%______________________________________________ 

\section{Galactic distribution of star-forming and quiescent \at\ sources}\label{sec:glon}
\subsection{Galactic longitude}

\begin{figure*}[!htpb]
   \centering
   {\rotatebox{90}{\includegraphics[width=0.7\linewidth]{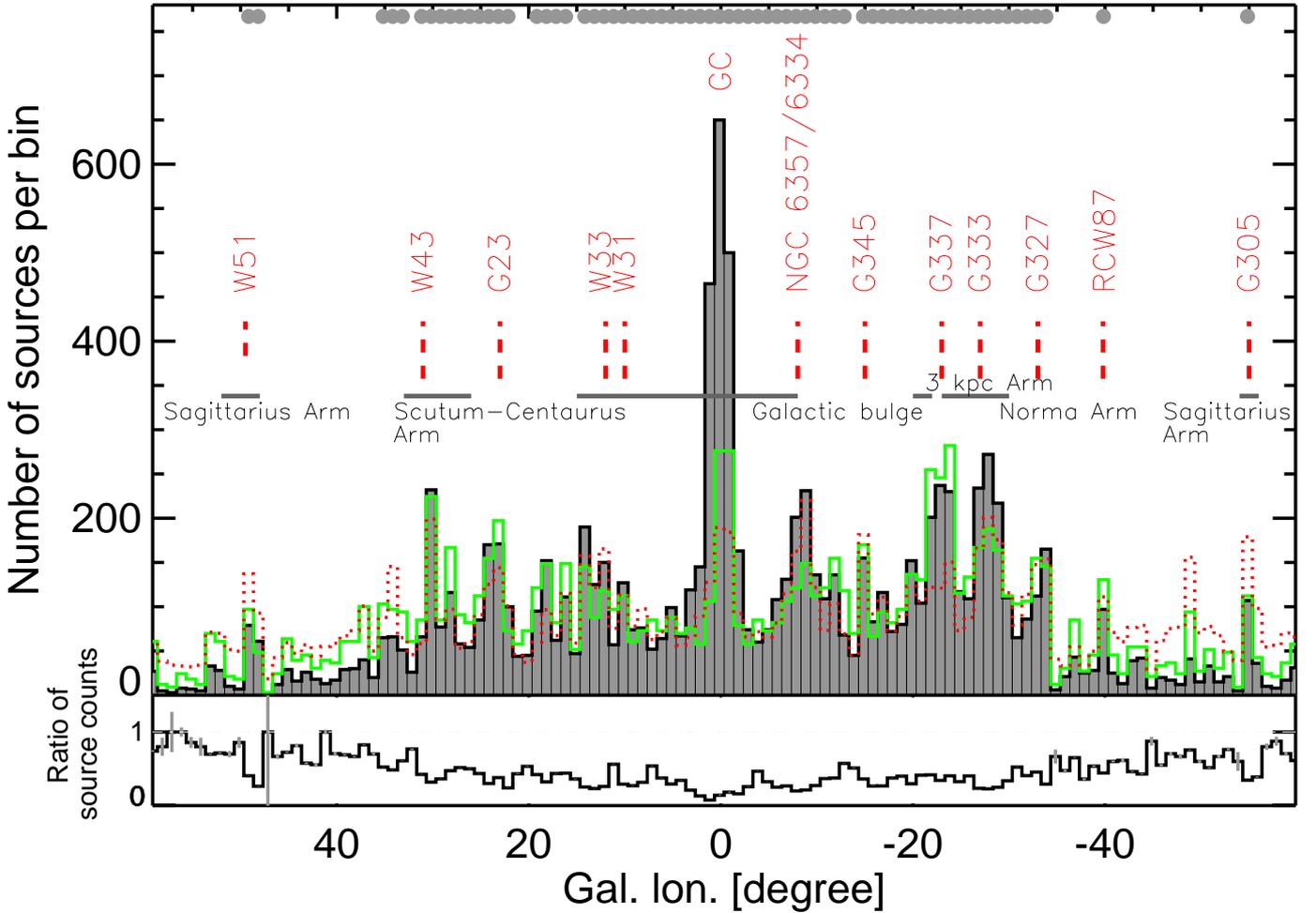}}}
    \caption{
                 {\bf Top:} 	
                  Distribution of the \at\ sources in Galactic longitude binned in $1^{\circ}$. 
                  Gray lines indicate the spiral arms and the dust complexes
                  are labeled in red. Gray dots in the upper line of the panel 
                  show the bins with $>7\sigma$ peaks.
                  Green line shows the distribution of star-forming \at\ sources
                  and have been scaled to the dust source counts.
                  Red dotted line shows the WISE selected red sources
                  according to a color criteria (see text for details).
                  {\bf Bottom:}
                  The distribution of the ratio of star-forming to quiescent \at\ sources
                  is shown in black.
                  Error bars in gray indicate the propagated Poisson-error from
                  the two distributions.
                  }
   \label{fig:histo_glon}%
   \end{figure*}

The distribution of quiescent and star-forming \at\ sources as a function of Galactic longitude is shown in Fig.\,\ref{fig:histo_glon}.
With over 10 000 sources we are able to trace various aspects of the Galactic structure. The strongest peak 
corresponds to the Galactic center region, with the highest number density of \at\ sources. We see additional
over-densities toward the 
direction of the Scutum-arm, the Norma-arm and Sagittarius-arm. 
In addition there are several known star-forming complexes appearing as 
statistically significant ($>7\sigma$) peaks. The G305 complex stands out from the average source density, but
we find other peaks indicating massive dust complexes, such as those around $\ell=327^{\circ}$, $\ell=333^{\circ}$ towards the Norma-tangent and $\ell=337^{\circ}$, which is in the direction of the 3-kpc arm.
These are spatially very extended complexes  rich in embedded sources. 
The $\ell=333^\circ$ complex contains sources mostly at about 3.5~kpc \citep{Simpson2012} 
in the Crux arm, between the Sagittarius- and Norma-arms along the line of sight \citep{Bronfman2000}. 
The source counts in Fig.\,\ref{fig:histo_glon} are likely to be lower beyond $\ell=327^\circ$, 
because there is only one arm on the line of sight with the Sagittarius-arm being quite out of the plane.   
Likewise, the $\ell=337^\circ$ region is in the tangent region of the 3-kpc arm, but there is strong molecular emission
both from the Norma and the Crux spiral arms. Therefore the peaks in Fig.\,\ref{fig:histo_glon} are related to 
the superposition of spiral arms on the line of sight. 
We also find a narrow peak
toward $\ell=345^{\circ}$. In this region there are several clouds spread over relatively large 
Galactic latitudes between  $b=-0.9^{\circ}$ to $b=1.3^{\circ}$. These features are all found  within a very narrow
Galactic longitude range of $ 345.2^{\circ} < \ell < 345.5^{\circ}$, and as a consequence appear as a very narrow peak
on Fig.\,\ref{fig:histo_glon}. This complex has been studied in more detail in \citet{Lopez2011}.

The relatively nearby complexes of NGC 6334 and NGC 6357 \citep{Russeil10} also appear. 
The peaks at positive Galactic longitudes at $\ell=10-12^{\circ}$ correspond to the complexes associated with 
W31 and W33, and
there is a clear excess of sources around $\ell=15-17^{\circ}$ dominated by the 
known star-forming regions M16 and M17. 
There are other peaks associated with mini star-burst regions like W43 and W51. Interestingly, the active
star-forming region, W49, does not appear as a prominent peak. This can be explained due to its larger 
distance of 12~kpc \citep{Gwinn1992}, compared to the other regions, such as W43 and W51, which are located 
at 6 and 5.4~kpc, respectively \citep{Quang2011,Sato2010}. As a consequence towards W49 the physical resolution is not sufficient to
resolve individual clumps, and only a few massive cloud fragments are identified.
Alternatively another difference could arise from the different stage of evolution between molecular complexes. %,

The Galactic structure traced by the population of compact sources 
shown in Fig.\,\ref{fig:histo_glon} is in general very similar to that reported by \citet{Beuther2012}, who use the 
same data but  a different source identification method. 
In Fig.\,\ref{fig:histo_glon} we also show the normalized Galactic distribution of the star-forming \at\ sources.
These correlate well with the positions of the peaks associated with molecular complexes as discussed above. 
The fraction of star-forming versus quiescent sources is lower between $-35^{\circ} < \ell < 35^{\circ}$, 
with the lowest ratio found towards the Galactic center region. This ratio suggests
 a lower star-formation rate (SFR) toward the Galactic center region, although this
strikingly low ratio of star-forming versus quiescent sources can also 
be partially due to a much higher extinction (see also \citealp{Beuther2012}). 
Using various datasets other authors 
arrive at the same conclusion, that the Galactic center
region has an exceptionally low SFR \citep{Immer2012, U2013mmb, Longmore2013}. The mechanism responsible for the
lack of star-formation in this region has been investigated by \citet{Kruijssen2013}, however
no single physical mechanism (such as turbulence, tidal forces, feedback effects, cosmic
ray heating, etc.) was identified to be sufficient to to suppress the SFR to such low levels. 
In contrast to the Galactic center region, the $\ell > |35|^{\circ}$ range, the SFR seems to be
rather uniform, on average the fraction of \at\ sources exhibiting star-formation activity is 67$\pm$17\%.

As a comparison, we selected sources from the WISE catalog covering the same area as \at\ and 
put a color-cut as a selection criteria of (F$_{4.5{\mu}m}$/F$_{3.6{\mu}m} > 0.75$ and 
F$_{4.5{\mu}m}$/F$_{12{\mu}m} < 1$) (S. Lumsden priv.\,comm), to
select red objects. 
Here we find that {the embedded sources have a comparable distribution to the dust peaks.
}
Compared to the \citet{Robitaille2008} GLIMPSE YSOs used in \citet{Beuther2012}, 
{we find that many of the YSO peaks coincide with the dust peaks,
likely because our sample of red, embedded objects 
is more complete, than the other studies.
}
In the $\ell>|35|^{\circ}$ range, however, there is an increase in the number of
embedded red objects with respect to the dust source counts, which is 
very likely due to a larger contamination
of YSOs as well as evolved stars. 
In this sample this excess is clearly seen in the $\ell > |35|^{\circ}$ range,
where more nearby sections of the Galactic arms fall and hence the infrared source sample starts to be dominated by 
nearby low mass embedded objects. 

\subsection{Galactic latitude}

\begin{figure}
   \centering
   {\rotatebox{90}{\includegraphics[width=0.7\linewidth]{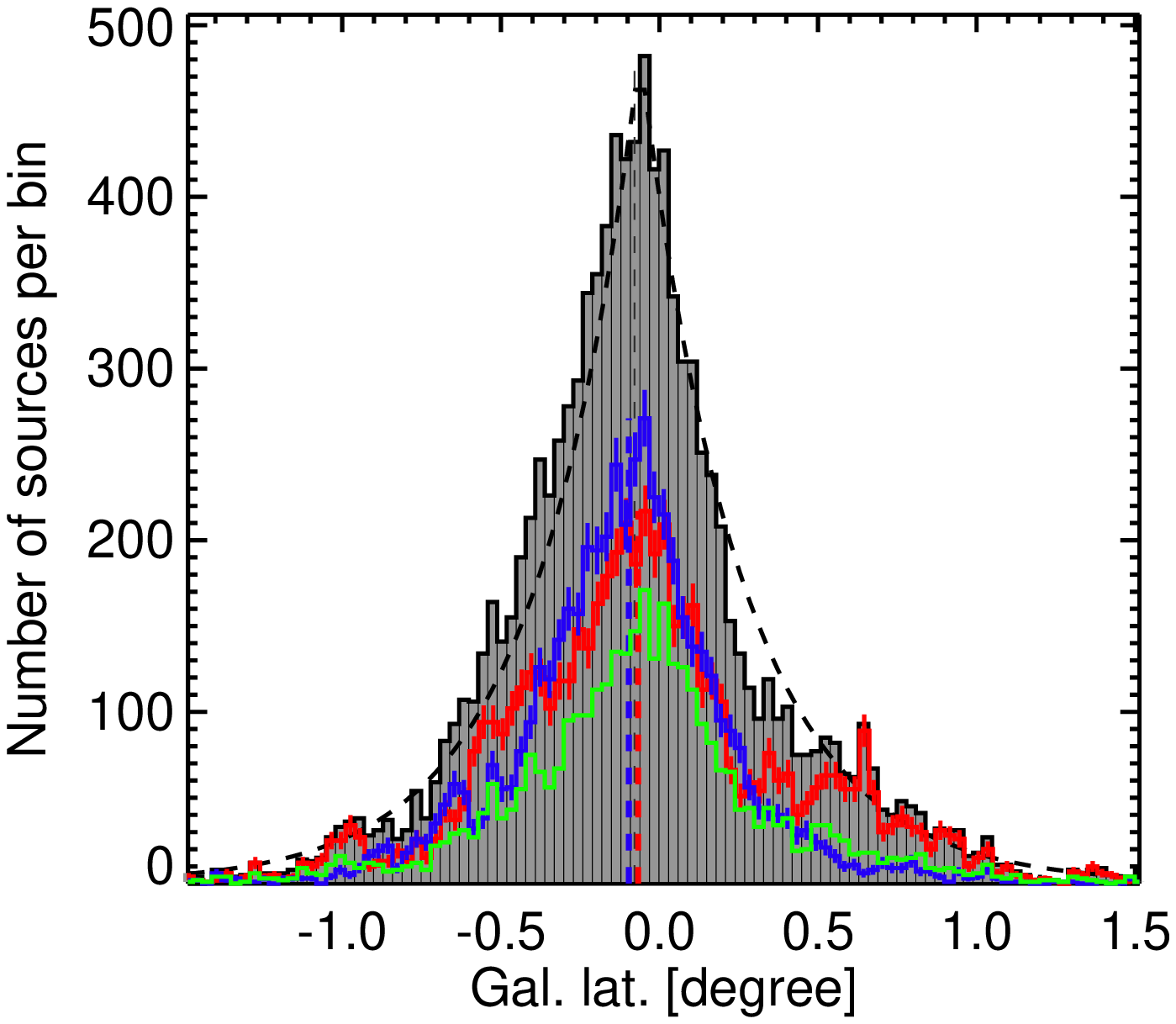}}}
    \caption{
                  Distribution of the detected sources binned in Galactic latitude. 
                  Gray histogram corresponds to all detected sources, while blue line shows those
                  lying in the I$^{\rm st}$ Galactic quadrant,
                  while red line shows sources from the IV$^{\rm th}$ quadrant. Dashed line
                  shows the function $400e^{-|b|/0.32^\circ}$ suggesting
                  a 0.32$^\circ$ scale height for the \at\ sources. Dashed colored lines 
                  correspond to the peak position of the I$^{\rm st}$ and IV$^{\rm th}$ quadrant source
                  distributions.
                  Green line shows the normalized distribution of star-forming \at\ 
                  to the total number of sources.
                  }
   \label{fig:histo_glat}%
   \end{figure}

Despite a similar number of sources being found below and above the Galactic
plane, the Galactic latitude distribution of \at\ sources is asymmetric 
and shows a characteristic shift towards 
negative latitudes (Fig.\,\ref{fig:histo_glat}).
We determine an offset 
of $-0.076^\circ\pm0.006^\circ$ to the peak of the 
distribution, which is very similar to the value of $-0.07^\circ$ determined by \citet{Beuther2012}. 
Other studies using similar datasets have also reported this shift
(e.g.\,\citealp{schuller2009, R2010}), like studies using various other tracers of (massive) star-formation, i.e. UC-{\hii} regions, 
\citep{Bronfman2000}, GLIMPSE YSOs \citep{Robitaille2008} and molecular gas 
\citep{Cohen1977,Bronfman1988}.
This shift is larger
when using only IV$^{\rm th}$ quadrant sources. 
However, this is not seen for samples of \at\ sources associated with UC-{\hii}
regions and MMB sources. 
It is therefore likely that in \at\
a large number of local clumps are seen tracing the Sun's position off the Galactic plane.

We estimate the Galactic scale height for dust
condensations using an exponential function and
find that a value of $0.32^\circ$ gives a good fit to the data. This means 
that the majority of dust sources are confined to a very narrow region around 
the Galactic midplane. As a comparison, the scale height of OB stars from IRAS is 
$0.6^\circ$ derived by \citet{Wood1989}, which is similar to the $0.8^\circ$ 
using MSX data to select OB type stars by \citet{Lumsden2002}. Our estimate
for the embedded stages is therefore much lower than based only on mid-IR diagnostics.
This can be partly explained with contamination of the previous samples with evolved stars,
which has been claimed for the \citet{Wood1989} sample by \citet{Becker1994}, who estimate
$0.4^\circ$ scale height using UC-{\hii} regions. Recent studies
suggest a lower scale height closer to the value suggested from {\at}, 
\citet{Walsh2011} estimates $0.4^\circ$ using
H$_2$O masers, similarly to \citet{Urquhart2011}.
Our estimate of a $0.32^\circ$ scale-height corresponds to 47~pc at 8.4~kpc~\citep{Reid2009},
which is the same as derived by \citet{Beuther2012}.

We also report here inhomogeneities in the distribution of IV$^{\rm th}$
and I$^{\rm st}$ quadrant sources. The most prominent is an excess in sources at
negative latitudes around $\sim-1.0^\circ$. 
This bump seems to be associated to sources from the IV$^{\rm th}$ quadrant.

As a comparison, the normalized distribution the star-forming \at\ sources is shown 
in green in Fig.\,\ref{fig:histo_glat}. The star-forming \at\ sources are well fitted
with the same scale height as for the total distribution. 

%______________________________________________ 
%______________________________________________ 
%______________________________________________ 
%______________________________________________ 

\section{Quiescent and star-forming dust clumps: clues to star-formation processes and time-scales}\label{sec:timescales}

The \at\ survey provides an unbiased view of the embedded stages of massive star-formation 
from the onset of collapse to the emergence of {\hii} regions. Therefore it provides
unprecedented statistics to study the properties of dust condensations in various stages:
quiescent phase prior to the onset of collapse, as well as
actively star-forming cores. Here we use these statistics to investigate the physical properties 
of the quiescent versus star-forming cores (Sect.\,\ref{sec:fluxdistr}),
derive a global star-formation rate for the Galaxy (Sect.\,\ref{sec:sfr}) and provide 
a good estimate of time-scales for these phases (Sect.\,\ref{sec:sf-timescale}).

\subsection{Properties of quiescent and star-forming clumps}\label{sec:fluxdistr}

In Fig.\,\ref{fig:histo_flux} we show the flux density distribution of  \at\ sources with mid-IR counterpart, 
i.e. embedded sources with ongoing star-formation, corresponding to {\totalmirmatchpercent}\% of the \at\ 
sources (i.e.\,both MSX and
WISE associations).
Their peak flux distribution is found to be very similar to that of all
\at\ sources and we derive
a slope of $\alpha=-1.25\pm0.04$.  This value suggests a
shallower distribution of the peak fluxes compared to that of all \at\ sources, although this 
is clearly due to the lower number of star-forming \at\ sources
at the lower peak flux-density range.
We note that the distribution of the peak flux density exhibits this uniform scaling 
over more than two orders of magnitude for both star-forming and all 
\at\ clumps.

\begin{figure}
   \centering
   {\rotatebox{90}{\includegraphics[width=0.5\linewidth]{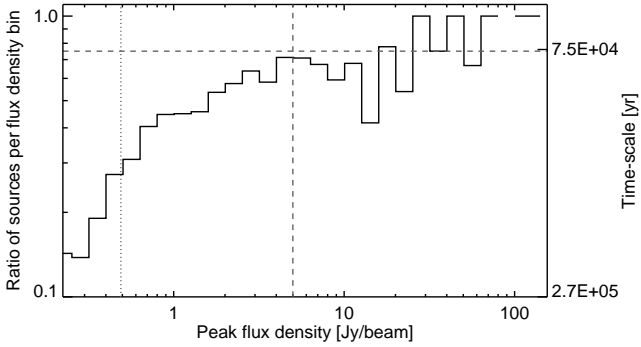}}}
    \caption{
                  Ratio of star-forming versus all \at\ sources as
                  a function of peak flux. The 75\% is indicated by a dashed line.
                  This fraction is fulfilled above 5~Jy beam averaged flux density.
                  Dotted line shows the 7$\sigma$ completeness level of
                  the catalog.
                  }
   \label{fig:histo_intflux_ratio}%
   \end{figure}

To further investigate the origin of this shallower distribution, 
in Fig.\,\ref{fig:histo_intflux_ratio} we show the beam averaged flux density versus the fraction of star-forming
\at\ sources compared to all sources. Clearly, the fraction of star-forming clumps increase with increasing 
peak flux density suggesting that the vast majority of the brightest clumps are actively forming stars.  
This fraction seems to be a constant of $\sim$0.75 within the errors for all sources 
above a beam averaged flux density of 5~Jy,
{
corresponding to $0.7-4.2\times10^{23}$~cm$^{-2}$ column density for warm (T$_d$=30~K)
and cold (T$_d=10$~K) gas.}
Adopting an average distance of 4.5~kpc and a temperature of 18~K,  
this threshold gives a similar value to the 650~{\msol} that
we obtained from the extrapolation of the MDCs in Cygnus-X from \citet{M07} 
to the \at\ survey (see Sect.\,\ref{sec:phys}) suggesting that
these sources could potentially sustain high-mass star-formation.
{ 
This increase in the fraction of star-forming sources 
may also indicate that there is a change in the star-formation process
in massive clumps}, although this transition is likely smeared to a continuous trend
due to distance effects in Fig.\,\ref{fig:histo_intflux_ratio}.

We point out that there is only a minor fraction of massive \at\ clumps above 
this threshold that do not
seem to harbor mid-IR embedded sources. 
These massive, quiescent looking sources are interesting because they may be 
either pristine massive clumps 
(i.e.\, the high-mass analogs of low-mass pre-stellar cores) at the onset of collapse processes. 
Alternatively, they can harbor
low- to intermediate mass embedded protostars (massive proto-stellar clumps, prior to the MYSO phase),
which are too faint to be detected with WISE at 22~$\umu$m at larger distances.

{
Investigating the appearance of these sources,  
we find a mean aspect ratio of 1.38 with a 
standard error of $0.005$ for star-forming \at\ sources, while  
all sources have a mean of $1.49$.
A KS-test further rejects the null hypothesis that they are drawn from
the same distribution with a significance level of $<0.0001$.
This shows that sources hosting embedded
protostars exhibit a rather spherical, less elongated morphology.
}

%______________________________________________ 
%______________________________________________ 

\subsection{The Galactic star-formation rate}\label{sec:sfr}

Adopting the color estimates of \citet{Wood1989}, similarly to
\citet{M07} and \citet{Russeil10}, we
extrapolate a flux density of $\sim$30~Jy at 21~{\mum} 
at a distance of 1~kpc for a B3 type star with $10^3~L_{\odot}$ 
(see also Sect.\,\ref{sec:wise-msx}) which corresponds to
$\sim$75~mJy at the far side of the Galaxy at 20~kpc.   
This is above the 
WISE sensitivity limit, therefore we are sensitive
to all B3 and earlier type massive stars in the Galaxy.

Based on the assumption that with WISE and MSX we cover all high-mass star-forming sites
within 20~kpc, we derive a crude estimate here for the 
current rate of star-formation for these deeply 
embedded objects. 

The time-scale for the protostellar evolutionary stage for high-mass stars is debated. If the accretion rate is constant
and is at the order of the observed $\sim10^{-4}\,${\msol}~yr$^{-1}$\citep{Klaassen2007}, at least $10^5$ years or longer
are needed to build up a 10~\msol\ star. 
On the other hand, there are a few studies suggesting that the time-scale for high-mass protostars 
can be shorter (e.g.\citealp{M07}), or similarly long \citep{Duarte2013} as for low-mass proto-stars, where
a life-time at the order of $\sim10^5$~yr has been derived \citep{Evans2009}. Theoretical
models predict a star-formation time-scale at the order of $10^5$~yr (e.g.\,\citealp{MT03,Stella2011}).
We adopt therefore 
{
the so far best observationally derived value of $\sim3\pm1\times10^5$~yr by \citet{Duarte2013}
}
for the highly embedded stage 
and account for a multiplicity of $\sim2$, which has been seen towards MDCs in Cygnus-X 
at high angular-resolution \citep{Bontemps2010}. 
We estimate a 
$dN/dt = \frac{{2\times\totalmirmatch}}{3\pm1\times10^5\,yr} = 0.023\pm0.007$~stars yr$^{-1}$ for the current
formation rate for high-mass stars. Assuming that they form on 
average $8-10$~{\msol} stars we estimate a star-formation rate of $0.21\pm0.07$~\msol\ yr$^{-1}$ for massive stars.

Taking the IMF of \citet{Kroupa1993} and neglecting the brown-dwarf population (objects with 
$\rm M <0.08$\,{\msol}),
8.6\% of the total mass is in OB-type stars with masses between 10-120\,{\msol}.
Based on this we can extrapolate the above derived star-formation rate 
to a Galactic star-formation rate of 
$\frac{0.21\pm0.07}{0.086}=2.44\pm0.81$~\msol\ yr$^{-1}$.
This estimate is based on the assumption that all the star-forming
\at\ sources form massive stars, which number is certainly
contaminated with intermediate-mass nearby objects, and is at the same time 
also incomplete for the brightest sources saturated in
both the MSX and WISE catalogs.
Nevertheless, it provides an independent and comparable value
for the global star-formation rate in our Galaxy
derived by other methods. Using the YSO population revealed by
Spitzer, \citet{RW2010} suggest
a value between 0.68 to 1.45~\msol\ yr$^{-1}$, while
\citet{Diehl2006} estimate 4~\msol\ yr$^{-1}$ based on radioactive $^{26}$Al measurements.
Comparing
these different tracers in a homogenous way 
\citet{Chomiuk2011} arrive to an estimate of 2~\msol\ yr$^{-1}$.  
Our estimate is consistent with these studies 
considering the large uncertainties in the exact number
of clumps forming massive stars, the
protostellar 
life-time estimates
and the poorly constrained factor of multiplicity.

\subsection{Formation time-scales}\label{sec:sf-timescale}

As discussed in Sect.\,\ref{sec:msx}, we expect our mid-IR characterization
to be more robust against chance alignments compared to \citet{Contreras2012}.
Other studies, such as the BGPS, base the mid-IR diagnostics of their sources
on the GLIMPSE and MSX 
point source catalogs. \citet{Dunham2011} finds that 44\% of the BOLOCAM sources have a mid-IR 
counterpart, however they consider 
a 50\% chance alignment and arrive to a conservative estimate of 20\% of the BGPS 
sources to have an
embedded mid-IR source.

Our analysis (Sect.\,\ref{sec:msx}, \ref{sec:wise}) is based on 
requiring a detected source
at 22~\mum\ benefitting from a continuous sensitivity between WISE and MSX with a similar 
spatial resolution as the \at\ survey, and 
thereby restricting the number of chance alignments with field stars and nearby Class II/III YSOs which
have no or only weak mid-IR emission. Nevertheless our sample can suffer from a minor contamination 
by AGB stars and other red 
sources that are not distinguishable from our sample based on their colors.  
Our estimate to be more robust against chance alignments  
is also supported by the color-color plots suggesting that indeed 
most of our matches are deeply embedded sources. As a consequence, 
our estimate of $\sim${\totalmirmatchpercent}\% 
is  better constrained than previous studies. 

Taking the 650~\msol\ limit for dense clumps to form massive stars, this translates
to a 5~Jy limit at a distance limit of 4.5~kpc.
Assuming that all quiescent \at\ sources above this limit are prone to collapse and will form massive 
stars, we estimate the time-scale for the pre-collapse phase for the most massive clumps in the Galaxy.
Adopting the estimate for a protostellar life-time of 
 $3\pm1\times10^5$~yr (see also Sect.\,\ref{sec:sfr}) 
and considering the fact that only $\sim$25\% of the sources are quiescent, this phase
can not last longer than 7.5$\pm2.5\times10^4$~yr. This estimate is consistent with other
studies (e.g.\,\citealp{M07,Russeil10, M2010, Tackenberg2012}) and is one order of magnitude lower than that of
low-mass cores (e.g.\,\citealp{Enoch2007, HFuller2008}), suggesting 
that massive cores, (precursors of high-mass stars) form in a fast, dynamic process.

In the dynamic scenario the formation time-scale is dictated by the crossing-time.
Considering the above derived 7.5$\pm2.5\times10^4$~yr as the crossing time for the pre-stellar and 
Class~0
like embedded phase, and the average physical size of 0.4~pc, a 
velocity dispersion of $\sim5.4\pm1.7$~km~s$^{-1}$
is required. This is in good agreement with kinematic studies of
high-mass star-forming sites, where flows of dense gas have been observed 
at similar relative velocities (e.g.\,\citealp{Schneider2010,Csengeri2011,Csengeri2011b,Quang2013}). This provides a
coherent view of dynamical processes to be at the origin of massive
star-formation in our Galaxy.

%_____________________________________________________________
%                                 Source counts  
%-------------------------------------------------------------
 \begin{table}
 \caption{Summary of cross-matches of dust condensations identified in the {\sl ATLASGAL} survey.}\label{tab:table2}
\centering
\begin{tabular}{rrrrrrrrrrrrrrr} 
\hline
 Catalogue & Matched sources & Fraction [\%] \\
\hline\hline 
MSX-\at\  &  \totalmsxmatch      & \totalmsxmatchpercent \\
WISE-\at\ &  \totalwisematch     & \totalwisematchpercent \\
MSX-WISE-\at\ & \totalmirmatch & \totalmirmatchpercent \\
\hline
\end{tabular}
\end{table}

%_____________________________________________________________
%                                 COLORS 
%-------------------------------------------------------------
 \begin{table}
 \caption{Mean and standard deviation of WISE colors of dust condensations.}\label{tab:table-col}
\centering
\begin{tabular}{rrrrr} 
\hline
 Sample & [3.6]-[4.6] & [4.6]-[12] & [12]-[22] & [4.6]-[22] \\
\hline\hline 
AG-WISE  &  1.76$\pm$1.02 &   3.48$\pm$1.56 &            3.45$\pm$0.93 & 6.93$\pm$1.74\\
CORNISH &  1.56$\pm$1.28&   3.34$\pm$2.35 &           2.89$\pm$1.94 &  6.23$\pm$3.94\\
RMS         &  0.86$\pm$1.36  &    1.04$\pm$1.62 &          1.18$\pm$1.77 &2.21$\pm$3.30\\
MMB         & 1.40$\pm$1.57 &    1.51$\pm$1.86 &           2.31$\pm$2.37 & 3.82$\pm$3.78\\
\hline
\end{tabular}
\end{table}

%--------------------------
%    SUMMARY
\section{Summary}\label{sec:sum}
We have produced a catalog of embedded sources in the \at\ survey using
a multi-scale decomposition tool to remove extended emission and then
used a Gaussian source fitting algorithm (\textsl{MRE-GCL}, \citealp{M2010}). 
This method is optimized to identify the population of centrally condensed,
compact structures. Here we summarize our main results :

\begin{enumerate}
\item
We identify \nsou\ \at\ sources in the main part of the survey, between $|\ell|<60^\circ$ and
$|b|<1.5^\circ$. Galactic plane. In the extension region, between $-80^\circ<\ell<-60^\circ$ and
$-2^\circ<b<1^\circ$, with a higher average
noise level we extract \nsouext\ sources. Our catalog is complete
to $>99$\% above $7\sigma$. 

\item
We find good correlation in the distribution of peak flux density
compared to the values from a different method by \citet{Contreras2012} which 
was optimized for larger size-scale sources corresponding to
clumps and cloud structures. We derive a slope of $\alpha\sim1.44\pm0.03$
for the distribution of the peak flux density, which is found
to be consistent with other surveys, such as the BGPS.

\item
We use the \textsl{MSX} and \textsl{WISE} point source catalogs to assess the
star-formation activity of the \at\ sources and find that {\totalmirmatchpercent}\% 
of them to harbor embedded mid-IR sources. Color-color plots
of the \textsl{WISE}-\at\ matches demonstrate the characteristic reddened
colors of these sources. 

\item
The Galactic distribution of \at\ sources shows peaks toward the
most prominent star-forming
complexes in our Galaxy. We find that the star-forming sources exhibit similar distribution
and similarly peak at the position of rich complexes. Considering all WISE sources with
characteristic red color we find a good correlation between the \at\ clumps
and red objects, suggesting that star-formation mainly takes place in large
complexes. 

\item
We determine the Galactic scale-height for the dust sources of $\sim0.32^\circ$, which is smaller
than previous estimates using mid-IR surveys, however it is close to the value determined
by surveys of young massive star-forming regions, e.g. using H$_2$O or methanol masers \citep{Walsh2011,U2013mmb}.

\item
From the fraction of star-forming sources we estimate a Galactic star-formation rate
of $\sim$$2.44\pm0.81$~\msol\ yr$^{-1}$. Although this value is subject to large uncertainties due to
the unknown factor of multiplicity, as well as the time-scale for the protostellar evolutionary phase,
we still find a good agreement with other estimates based on various datasets.

\item
Comparing the fraction of star-forming versus quiescent \at\ sources, 
we show that the fraction of embedded sources
exhibiting star-formation activity increases with the beam-averaged flux density. This ratio is found to be a
rather constant value of 75\% for 5~Jy beam averaged flux density, suggesting that the lifetime 
for the IR-dark evolutionary phase is short. 
We estimate an upper limit for the IR-dark phase of 7.5$\pm2.5\times$10$^4$ years.

\end{enumerate}

%    Acknowledgement
\acknowledgement
{
We thank the referee for their comments which helped to improve the manuscript.
}
%People:
This work was partially funded by the ERC Advanced Investigator Grant GLOSTAR (247078) and was partially carried out within the Collaborative Research Council 956, sub-project A6, funded by the Deutsche Forschungsgemeinschaft (DFG). LB acknowledges support from CONICYT project PFB-06.
%APEX:
This paper is based on data acquired with the Atacama Pathfinder EXperiment (APEX). APEX
is a collaboration between the Max Planck Institute for Radioastronomy, the European
Southern Observatory, and the Onsala Space Observatory.
%MSX:
This research made use of data products from the Midcourse Space Experiment. Processing of the data was funded by the Ballistic Missile Defense Organization with additional support from NASA Office of Space Science. This research has also made use of the NASA/ IPAC Infrared Science Archive, which is operated by the Jet Propulsion Laboratory, California Institute of Technology, under contract with the National Aeronautics and Space Administration. 
%WISE:
This publication makes use of data products from the Wide-field Infrared Survey Explorer, which is a joint project of the University of California, Los Angeles, and the Jet Propulsion Laboratory/California Institute of Technology, funded by the National Aeronautics and Space Administration.

%--------------------------
%    BIBLIO
\bibliography{ag-compactsou_v0}
\bibliographystyle{aa}

\begin{appendix} %Second online appendix
\section{Source parameters from {\sl Gaussclumps}}\label{sec:souparams}
{
Gaussian decomposition procedures, such as \gc\ 
have been widely used in the literature, both
on 3D cubes of molecular line observations, as well as 2D maps of dust emission.
The reliability of the algorithm has been tested and discussed
partly in \citet{SG1990} and more in detail in \citet{Kramer98}.
We refer the reader to these papers for a detailed discussion
of the algorithm.

However, since this is the largest survey where \gc\ 
has been applied, we performed several test to assess the reliability 
of the detections. In Sect.\,\ref{app:radial-fluxprofile} we discuss
the radial profiles of the sources and the impact of the filtering
on the measured sizes, while in Sect.\,\ref{app:souparams} we
compare the extracted peak fluxes to the original values in the 
filtered and non-filtered maps.
}

\subsection{Sizes and radial flux density profiles}\label{app:radial-fluxprofile}

\begin{figure*}
   \centering
   {\rotatebox{0}{\includegraphics[width=6cm]{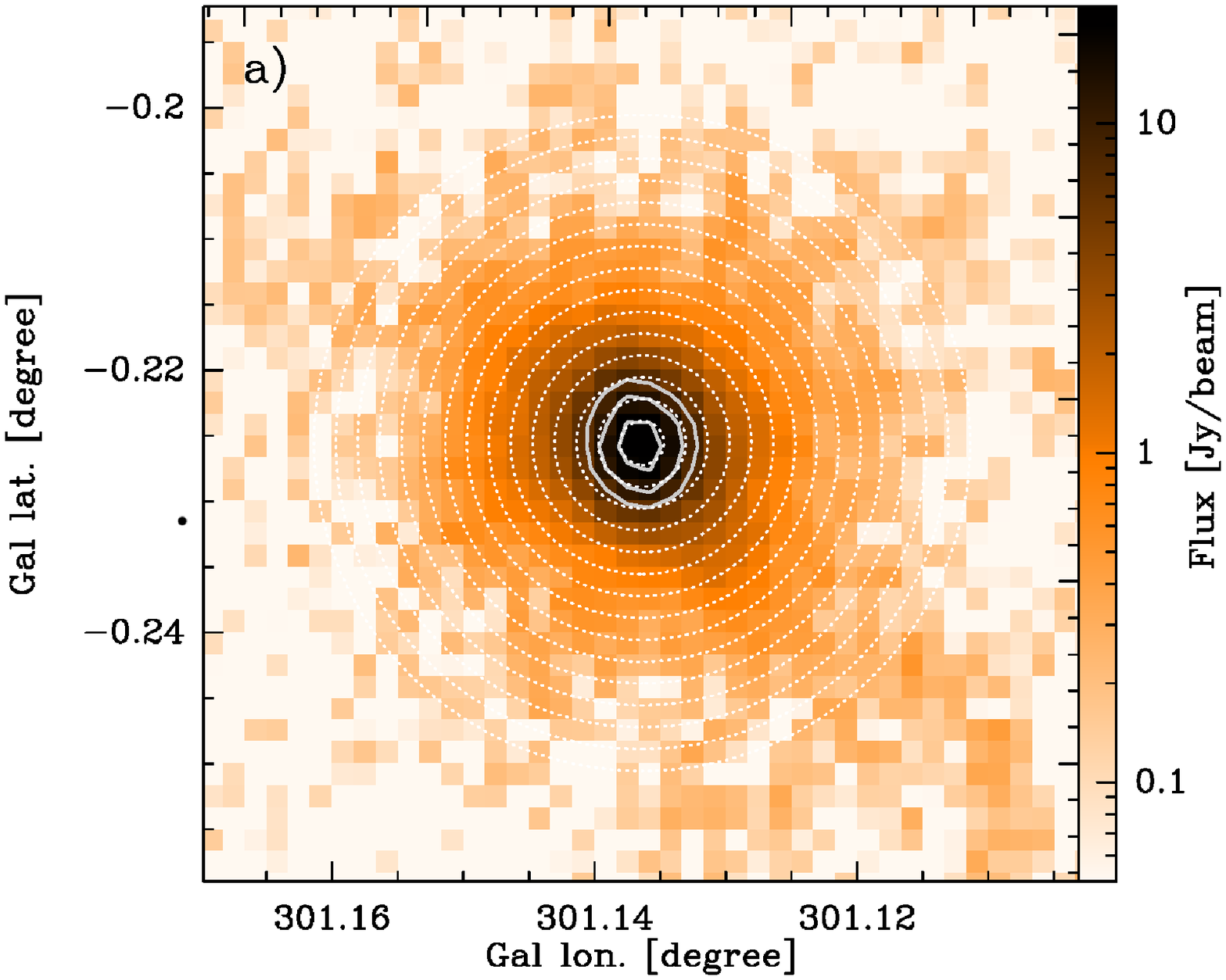}}}
   {\rotatebox{0}{\includegraphics[width=6cm]{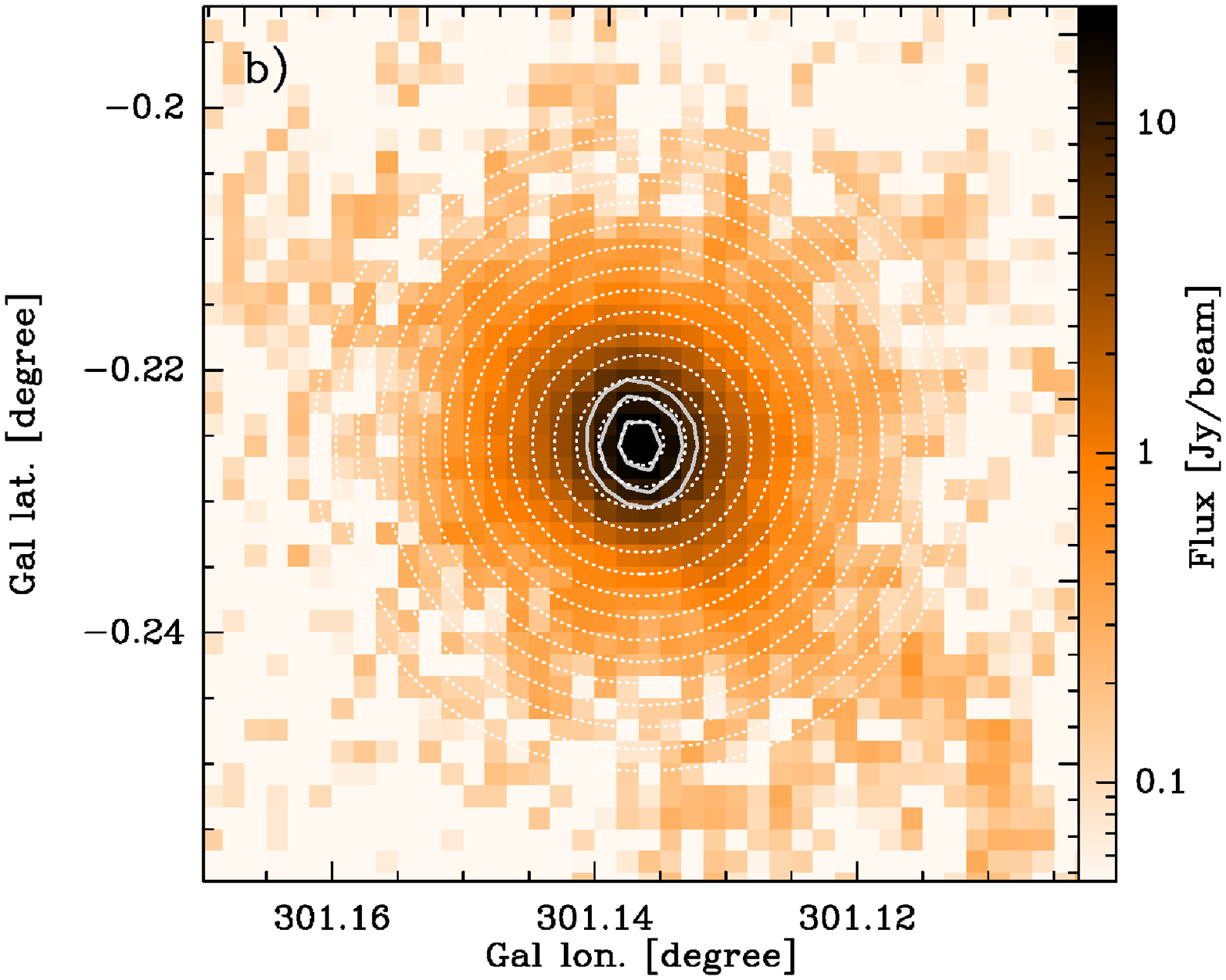}}}
   {\rotatebox{0}{\includegraphics[width=6cm]{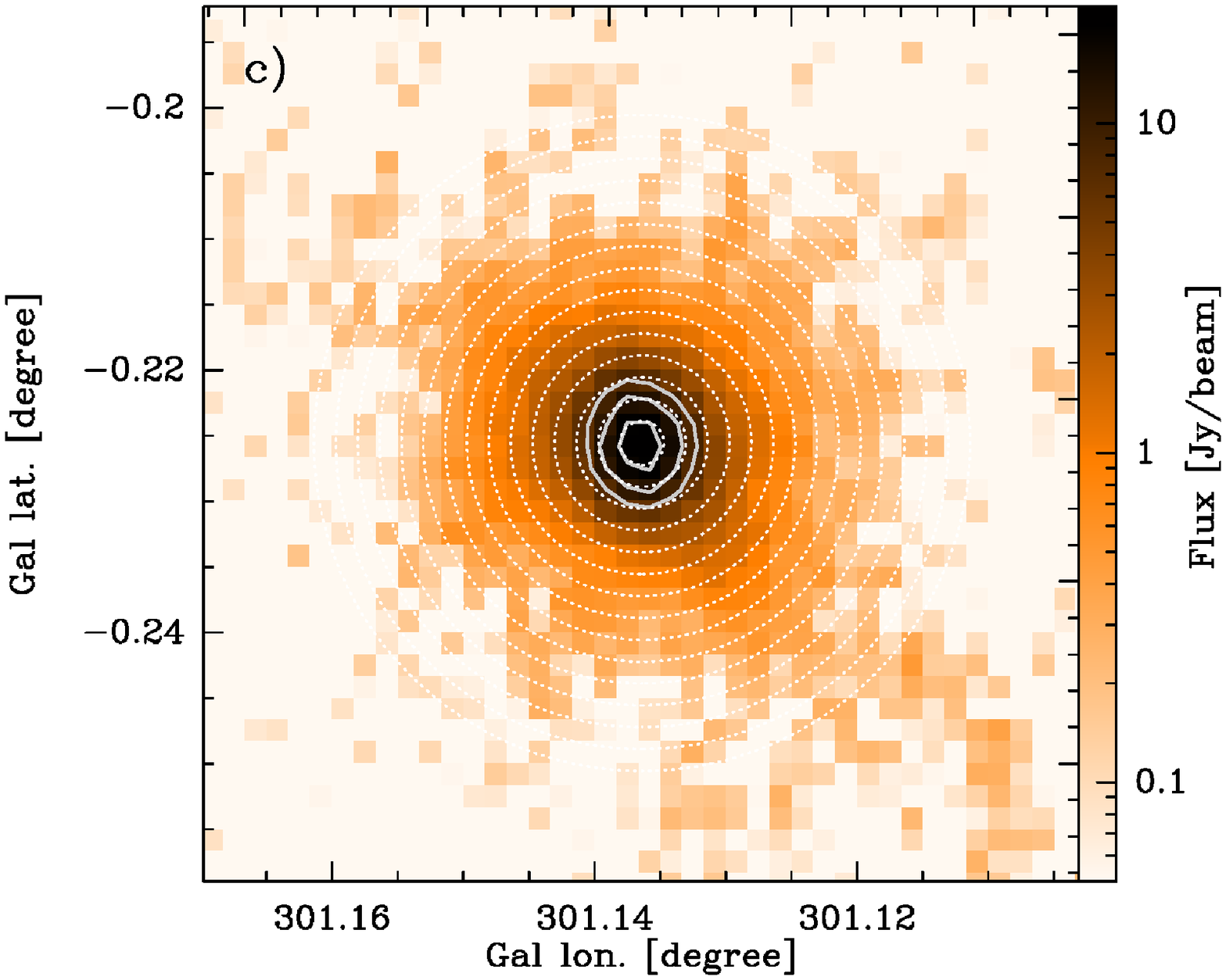}}}
\hspace*{-1cm}   {\rotatebox{0}{\includegraphics[width=6cm]{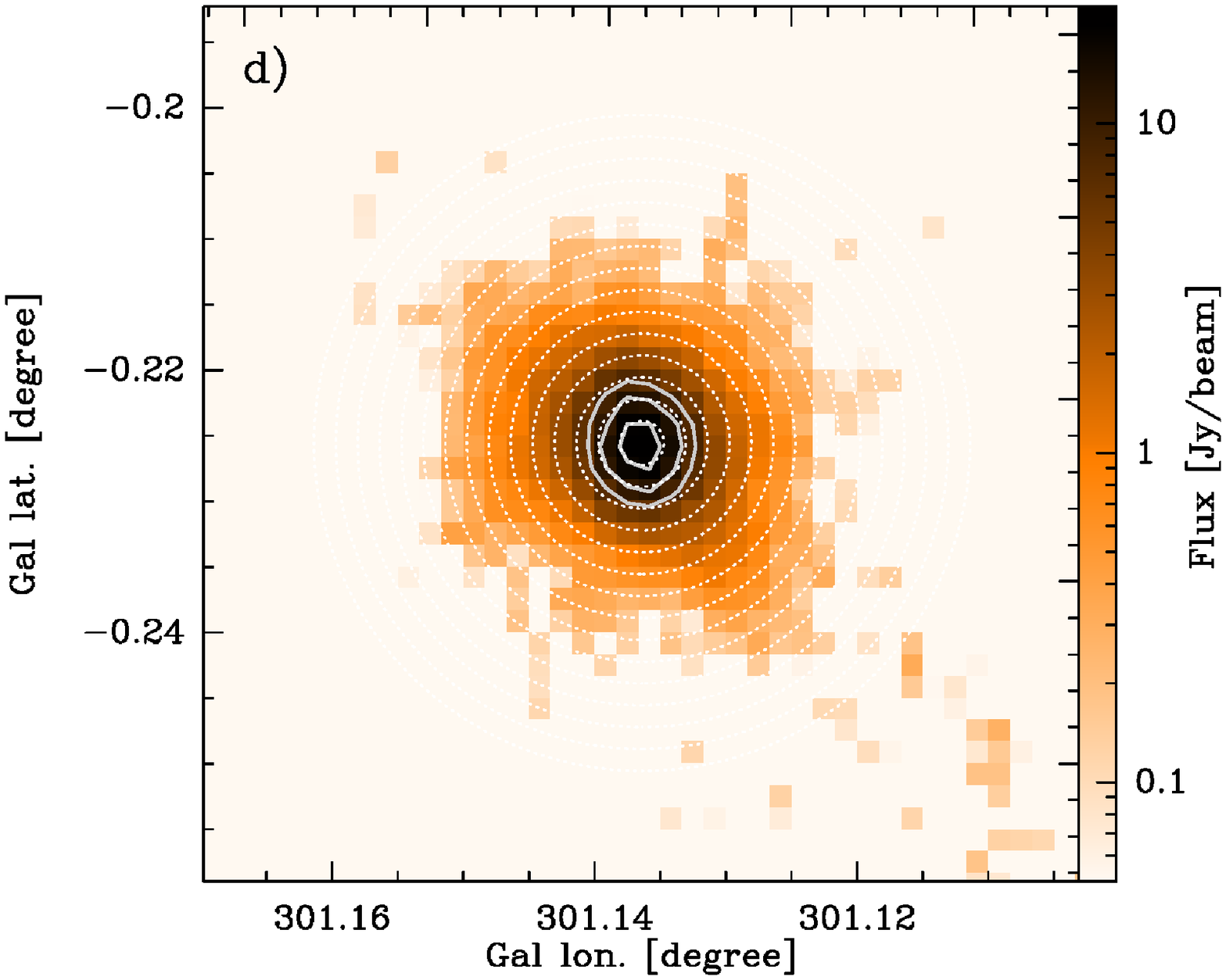}}}
   {\rotatebox{0}{\includegraphics[width=6cm]{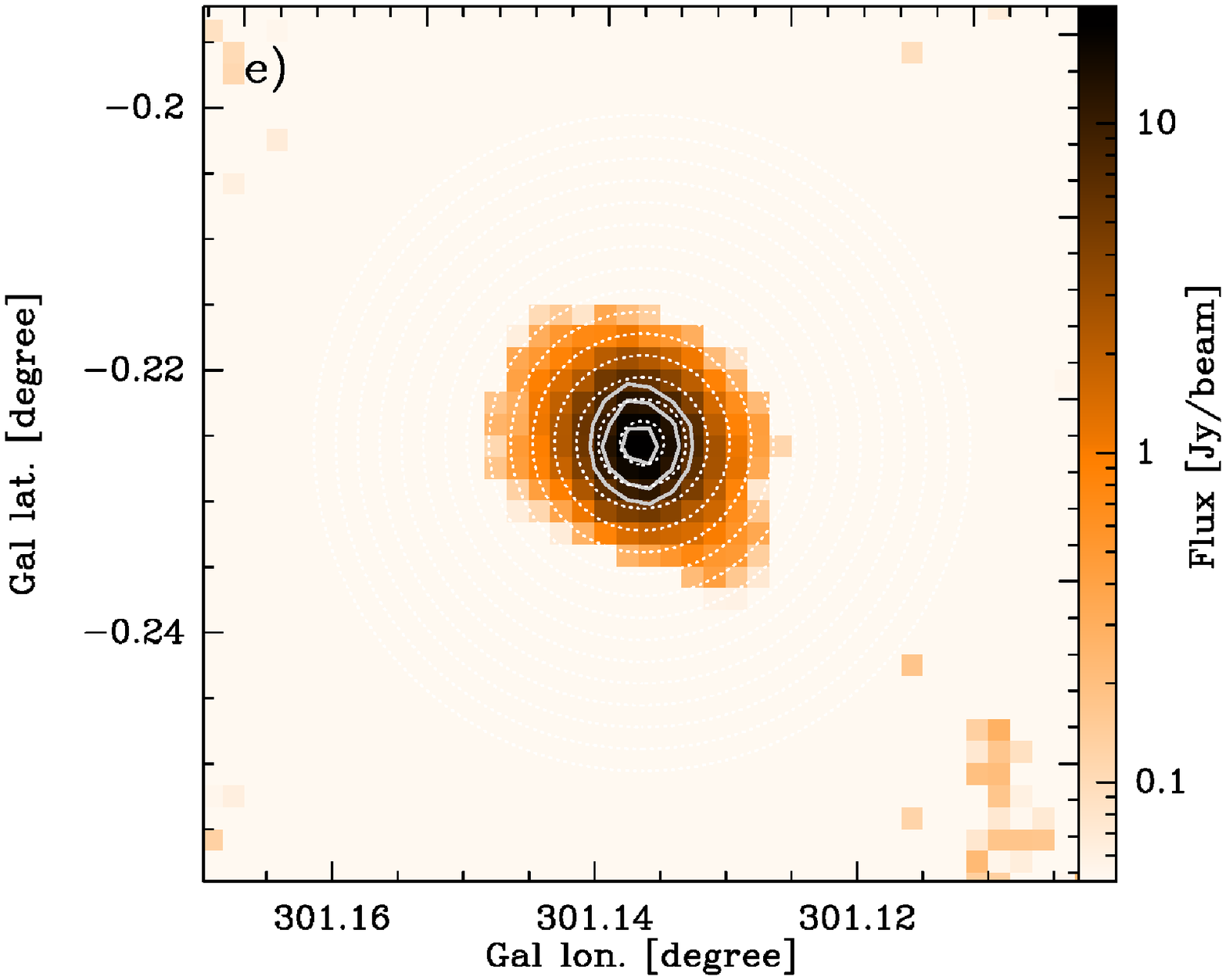}}}
\hspace*{+0.4cm}    
{\rotatebox{90}{\includegraphics[width=5.0cm]{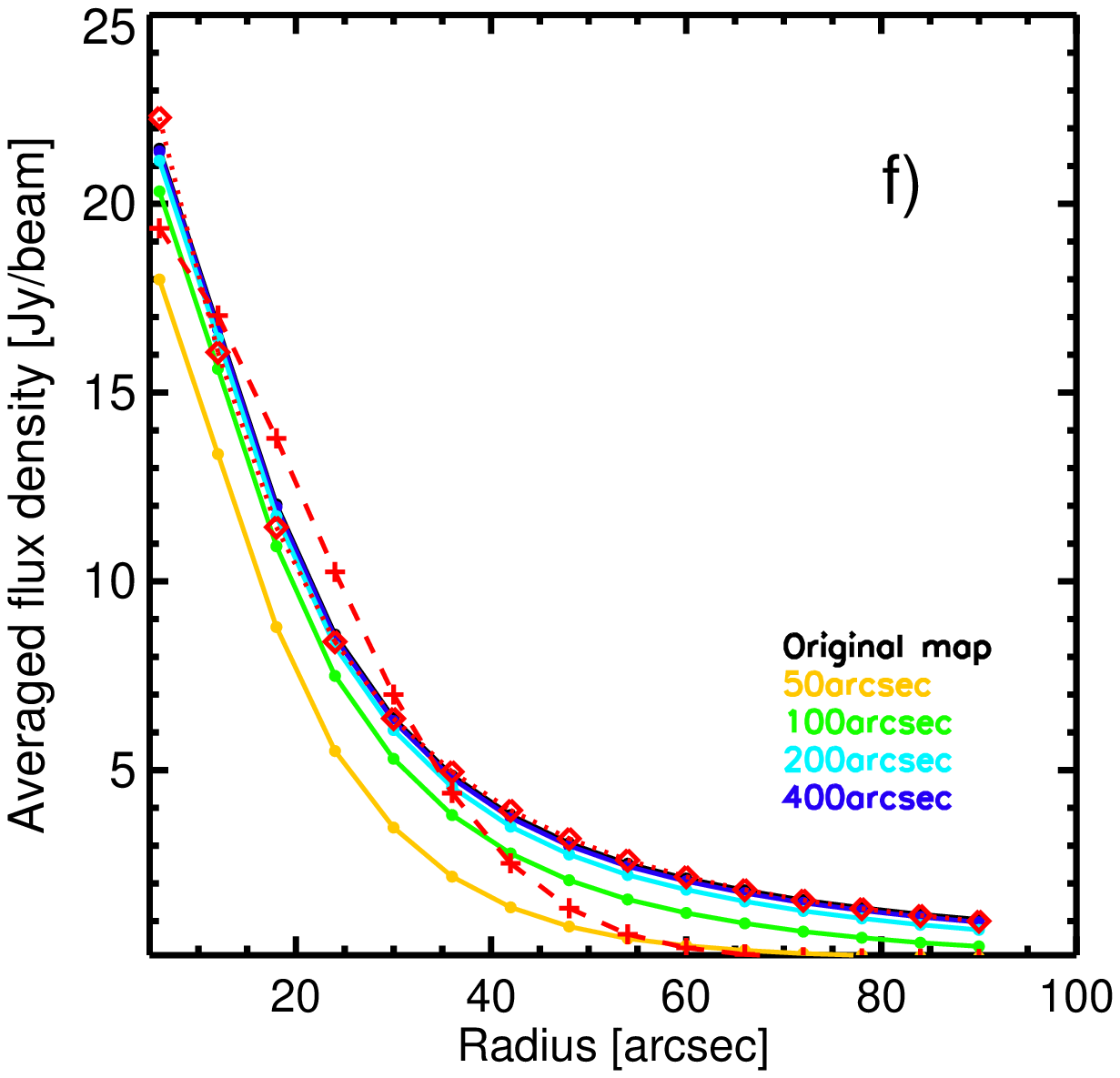}}}
    \caption{
                  One of the brightest sources detected shown with different scales of 
                  filtering, similarly as in Fig.\,\ref{fig:filtering_size}.
                   The peak flux of the object in this case decreases by 20\% from
                  the original images to the most compact one, while the size decreases
                  by only 10\%. Panel {\bf f)} shows {azimuthally averaged} flux density profile
                  with different scales of background emission removed. 
                  The green line corresponds to the
                  filtering used for our catalog, which has 2$\times$50\arcsec\ as 
                  maximum scale.
                  As a comparison a Gaussian profile is indicated in red dashed line with crosses
                  and a dotted line with diamond symbols shows a power-law fit.
                  }
   \label{fig:filtering_size_app}%
   \end{figure*}

{
Since with the filtering some background emission is removed
from the maps, we investigated the impact of this 
on the measured sizes by performing 
source extraction on maps with different scales of background
emission removed.
The change in the azimuthally averaged 
flux density profile by using different scales,
between $50-400$\arcsec, for 
the background is illustrated in Fig.\,\ref{fig:filtering_size_app}.
By comparing the extracted sizes in these measurements, we found 
that the decrease in the measured sizes with respect to that of
the original maps is negligible ($\sim10\%-20\%$)
for the 100~\arcsec\ filtering scale used here.
}

\subsection{Source parameters: the peak flux}\label{app:souparams}

We have performed several tests to estimate the reliability of the
source parameters, such as peak flux density and size extracted by {\gc}.

\begin{figure}
   \centering
   {\rotatebox{90}{\includegraphics[width=6cm]{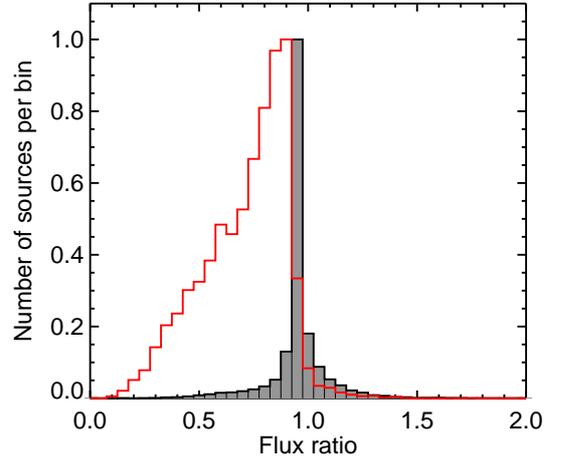}}}
    \caption{
                  The normalized ratio of peak flux density is calculated from the extracted (i.e.\,fitted) 
                  and the actual pixel values at the position of the source. Black histogram shows the 
                  ratio calculated using the filtered maps, while the red line shows the ratio calculated
                  from the original maps with the total emission. (See Sect.\,\ref{app:souparams} for discussion.)
                  }
   \label{fig:filtering_flux}%
   \end{figure}

As a first step we compared the extracted peak fluxes with the fluxes in the filtered and the
original maps. On average we find this difference to be less than 30\% averaged on the individual tiles, 
however locally, in very complex regions this fraction can be higher. Fig.\,\ref{fig:filtering_flux} shows the ratio between the extracted and the measured peak flux density at the position of the source using the both the filtered and the original maps. 
The major fraction of the
sources have peak flux density within 5\% of the pixel values in the filtered maps confirming that the 
algorithm works good finding the peak position of the dust emission. 
On average the measured flux is 97.9\% of the extracted value with a median 
of 98.2\% (Fig.\,\ref{fig:filtering_flux}). The difference comes from two factors: first, fluctuations in the local noise 
may lead to small shift in the determined positions, i.e.\,determining the peak biased towards a higher value
pixel instead of the peak of the Gaussian profile. Hence the measured value in the actual pixel may differ
from the fitted value. We find $\sim$500 sources (5.4\%) of the total sources where this flux ratio differs 
with $>$30\%
compared to unity. We inspected each of these sources and they are dominantly weak and large sources
with no clear Gaussian distribution in which case the peak position is not well defined.
On the other hand, since the algorithm decomposes sources in 
confused regions, we also find flux ratios below 100\%\, which are blended sources.
{
We find that a marginal fraction of the sources is actively
deblended (0.34\% of the total number of sources), where 
within the beam the source is decomposed into two components.
There are, however a larger fraction (24\%) of overlapping sources,
which reflects the clumpy nature of the dust distribution in 
molecular clouds.
}
Altogether we have a good census of the reliability of the extracted peak
flux values of the algorithm.

The extracted peak flux densities compared to the values in the original emission maps show that 
the sources have on average 73.5\% of the pixel value with a median of 78\%. This suggests
that the filtering lowers on average the peak flux values by 20-30\%.

%______________________________________________ 
%______________________________________________ 

%______________________________________________ 
%______________________________________________ 

\section{Color-color plots of star-forming \at\ clumps}

\begin{figure}
   \centering
   {\rotatebox{90}{\includegraphics[width=0.7\linewidth]{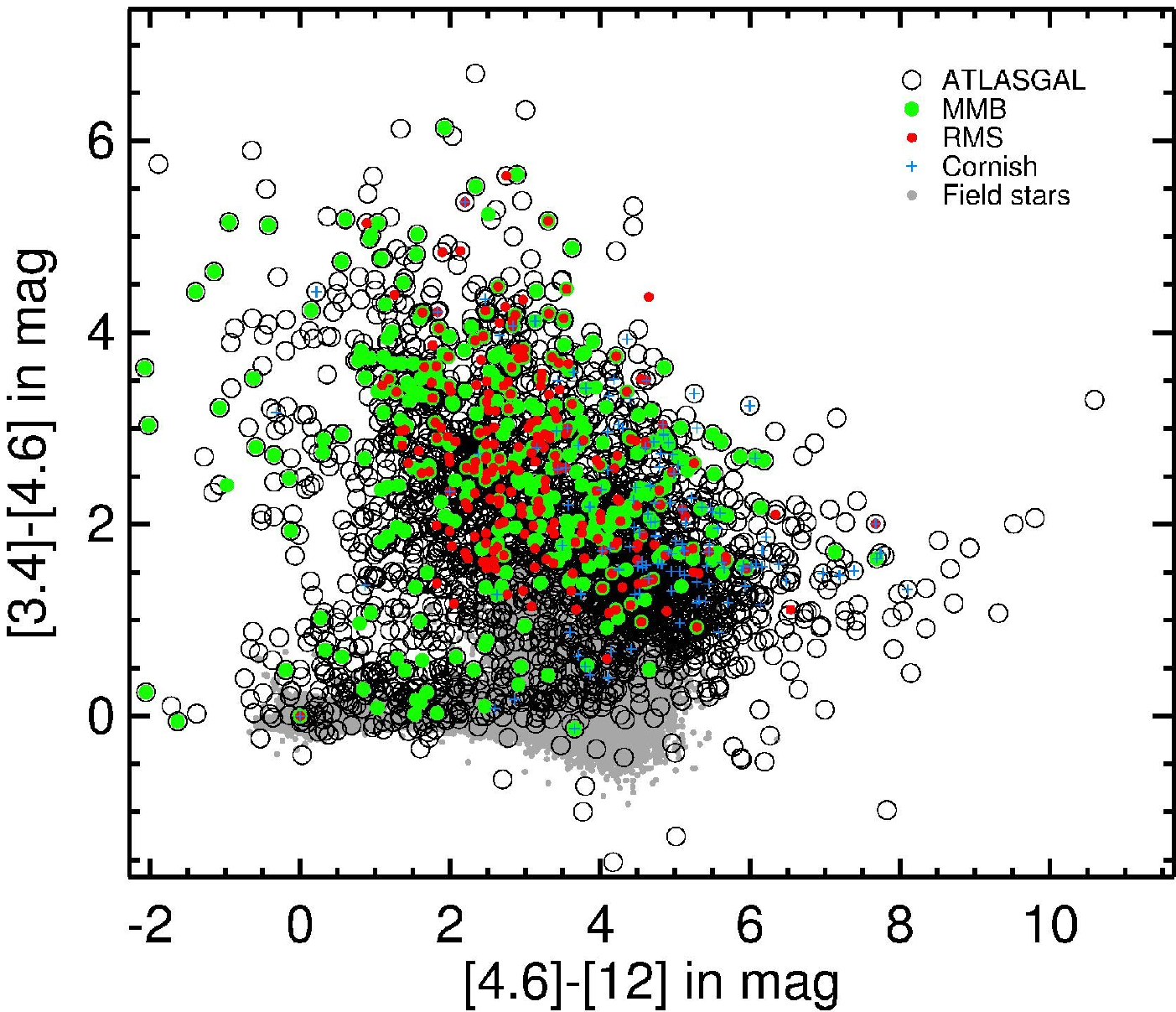}}}
    \caption{
                  Color-color plots of \at\ sources with a WISE source match (black dots).
                  As a comparison, the colors of field stars from a test field are shown in 
                  gray dots.
                  }
   \label{fig:cc1}%
   \end{figure}

\begin{figure}
   \centering
   {\rotatebox{90}{\includegraphics[width=0.7\linewidth]{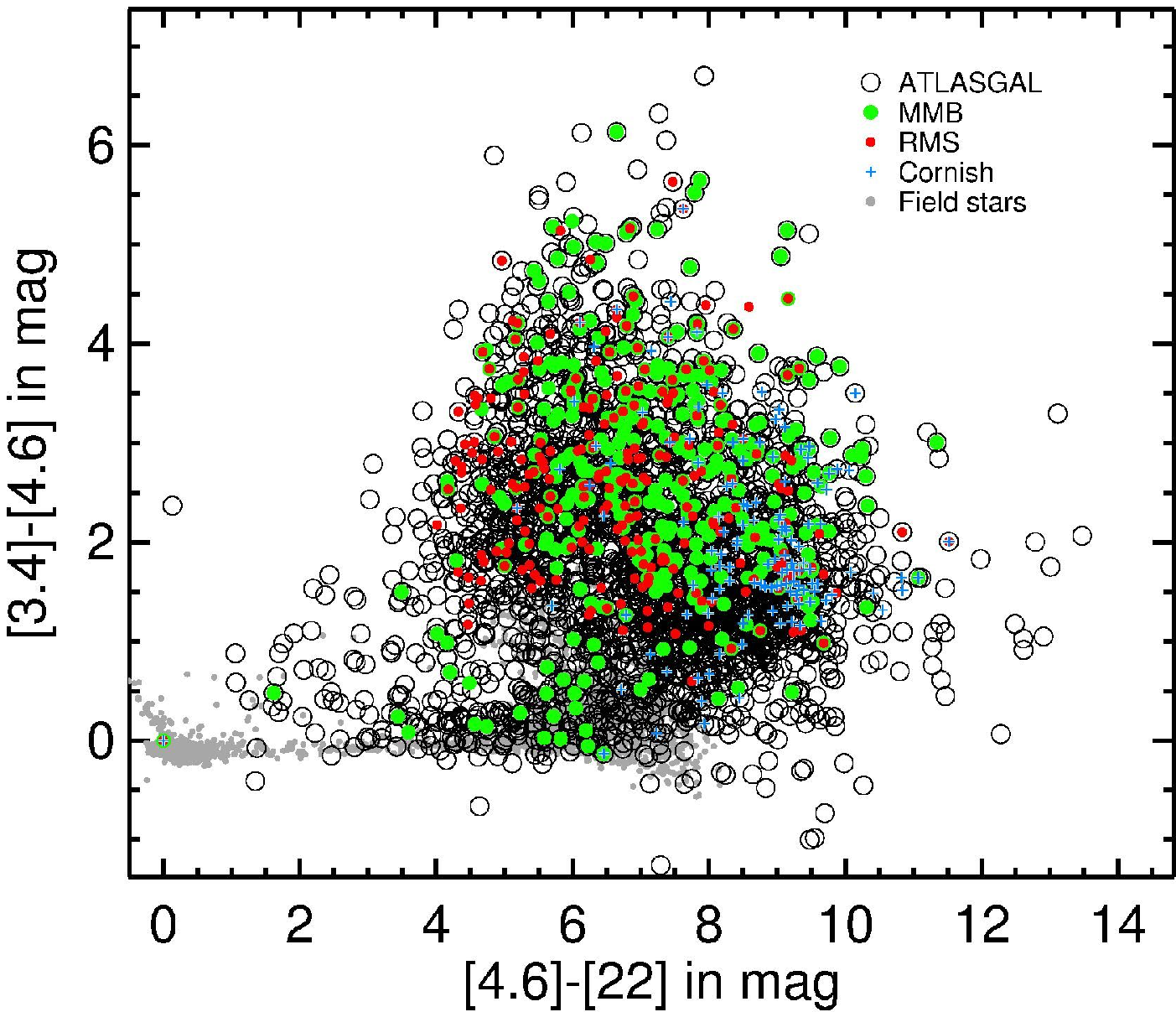}}}
    \caption{
                  Color-color plots of \at\ sources with a WISE source match (black dots).
                  As a comparison, the colors of field stars from a test field are shown in 
                  gray dots.
                  }
   \label{fig:cc2}%
   \end{figure}

\begin{figure}
   \centering
   {\rotatebox{90}{\includegraphics[width=0.7\linewidth]{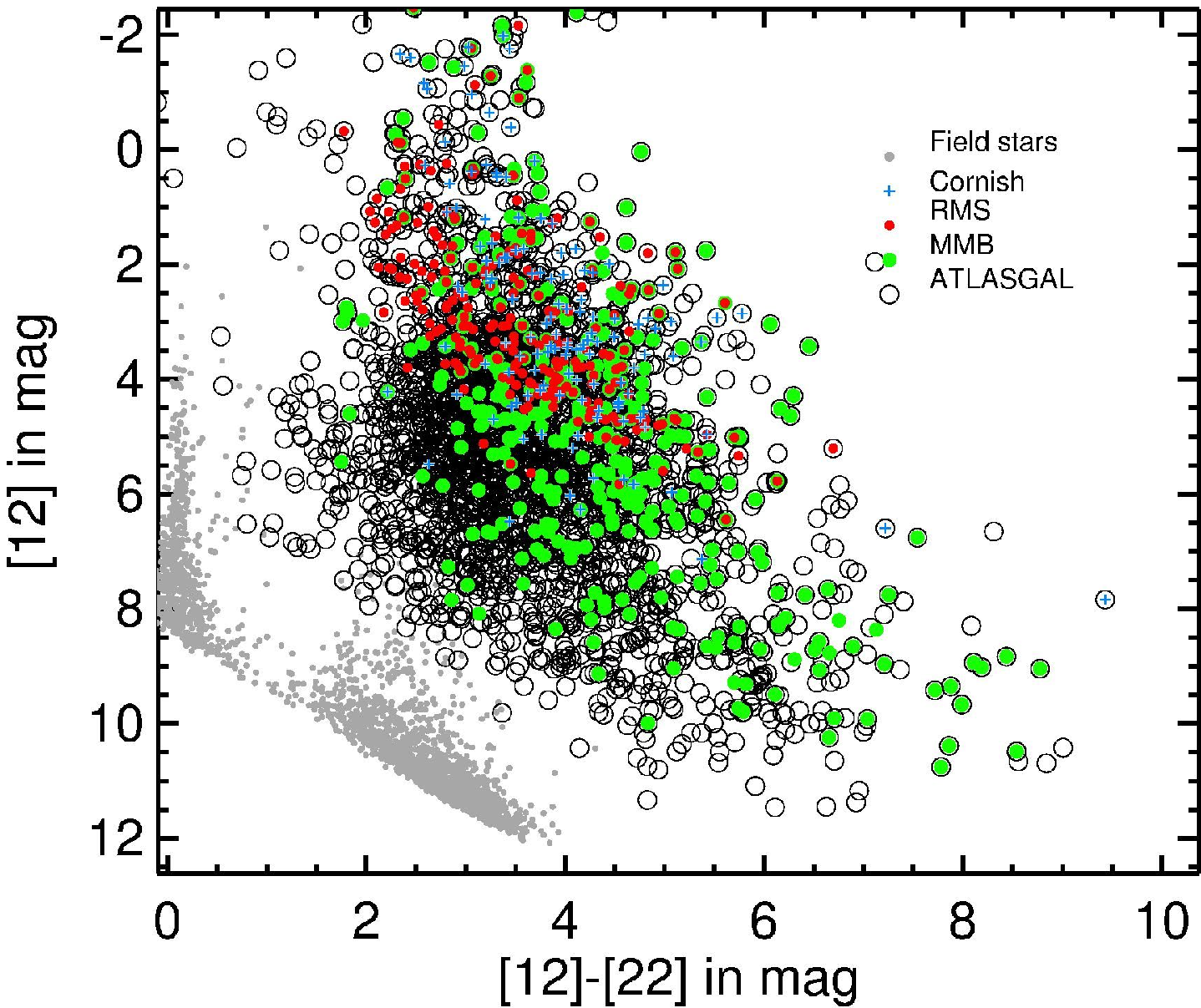}}}
    \caption{
                  Color-magnitude plots of \at\ sources with a WISE source match (black dots).
                  As a comparison, the colors of field stars from a test field are shown in 
                  gray dots.
                  }
   \label{fig:cc3}%
   \end{figure}

\begin{figure}
   \centering
   {\rotatebox{90}{\includegraphics[width=0.7\linewidth]{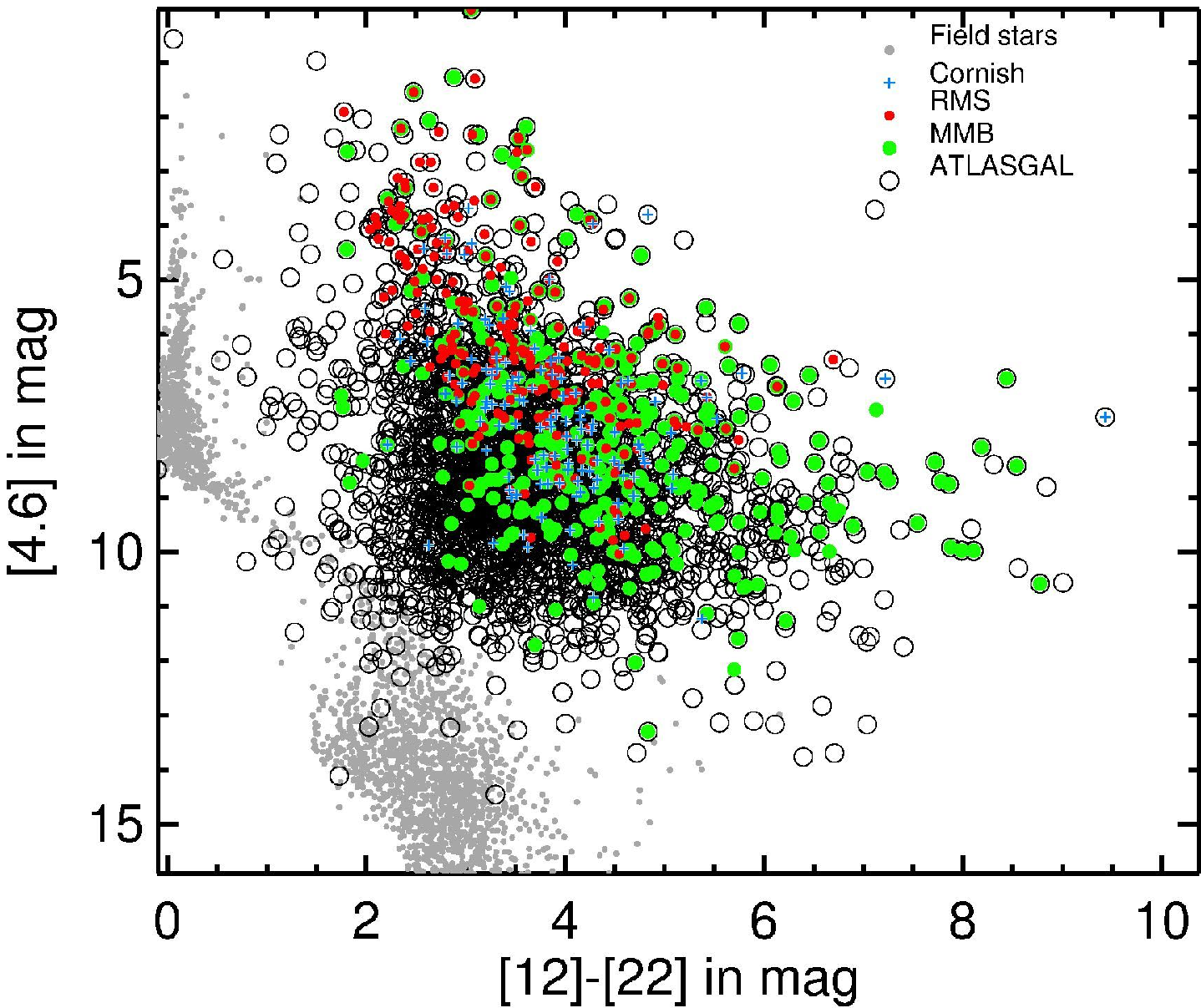}}}
    \caption{
                  Color-magnitude plots of \at\ sources with a WISE source match (black dots).
                  As a comparison, the colors of field stars from a test field are shown in 
                  gray dots.
                  }
   \label{fig:cc4}%
   \end{figure}

Figs.\,\ref{fig:cc1}-\ref{fig:cc4} show color-color plots and a color-magnitude diagrams 
of the \at\ and WISE matched sources.
Comparing
the positions of embedded UC-{\hii} regions, RMS  and MMB sources in Fig.\,\ref{fig:cc3}
we see that especially the former two occupy a well determined region with
bright 12~\mum\ flux and colors between 1.8 $<$ [12]-[22] $<$ 7. These sources
correspond to evolved stages, {\hii} regions and MYSOs. These two samples seem
to have no distinguishable color properties.  

%______________________________________________ 
%______________________________________________ 

\end{appendix}

\end{document}